\documentclass{emulateapj}
\usepackage{epsfig,natbib,multirow} 
\usepackage{amsmath}
\usepackage{graphicx}
\usepackage{verbatim}  
\graphicspath{ {lightcurves/} }

\begin{document}
\pagenumbering{arabic}

\submitted{Accepted by AJ}
\title{detection of stars within $\sim$0.8" of \emph{Kepler} objects of interest}

\author{
Rea~Kolbl\altaffilmark{1},  
Geoffrey~W.~Marcy\altaffilmark{1},  
Howard~Isaacson\altaffilmark{1}, 
Andrew~W.~Howard\altaffilmark{2} 
}   
  
\altaffiltext{1}{University of California, Berkeley, CA 94720}  
\altaffiltext{3}{Institute for Astronomy, University of Hawaii, Honolulu, HI 96822}

\begin{abstract}

We present an algorithm to search for the faint spectrum of a second star mixed with the spectrum of a brighter star in high resolution spectra. We model optical stellar spectra as the sum of two input spectra drawn from a vast library of stars throughout the H-R diagram. From typical spectra having resolution of R=60,000, we are able to detect companions as faint as 1\% relative to the primary star in approximately the V and R bandpasses of photometry. We are also able to find evidence for triple and quadruple systems, given that any additional companions are sufficiently bright. The precise threshold percentage depends on the SNR of the spectrum and the properties of the two stars. For cases of non-detection, we place a limit on the brightness of any potential companions. This algorithm is useful for detecting both faint orbiting companions and background stars that are angularly close to a foreground target star. The size of the entrance slit to the spectrometer, 0.87 x 3 arcsec (typically), sets the angular domain within which the second star can be detected. We analyzed Keck-HIRES spectra of 1160 California \emph{Kepler} Survey objects of interest (KOI) searching for the secondary spectra, with the two goals of alerting the community to two possible host stars of the transiting planet and to dilution of the light curve. We report 63 California \emph{Kepler} Survey objects of interest showing spectroscopic evidence of a secondary star.  

\end{abstract}


\section{Introduction}
\label{sec:intro}

\indent One of the \emph{Kepler} Mission's main goals is to determine the abundance of the terrestrial and larger planets in the habitable zone, to characterize their orbits, and to determine their physical properties. As of February 27, 2012, the catalogue of \emph{Kepler} Planetary Candidates has 2300 entries \citep{Batalha_etal_2013}. Planet candidates are identified by their photometric signals and then confirmed by additional photometric and spectroscopic analysis. However, it is well known that roughly half of all star systems are actually binaries or triples \citep{Raghavan_etal_2010}. Therefore, it is likely that roughly half of the \emph{Kepler} Planetary Candidates have two or more possible host stars. At the typical distances of \emph{Kepler} objects of interest (KOIs) of 0.3 - 1.0 kpc, orbiting stellar companions with typical orbital distances of 5 - 500 AU would be unresolved from the primary stars. This implies that roughly half of the \emph{Kepler} Planetary Candidates have a major ambiguity about the nature of the host star of the transiting: it could be any of the stars in the binary or triple star system. Moreover, even if the planet transits the primary star, the dilution of the \emph{Kepler} photometry from the secondary and tertiary stars causes an underestimate of the radius of the planet. Such an ambiguity can also be caused by a background star, angularly close to the \emph{Kepler} target. The identification of a second set of absorption lines in the spectrum of a KOI highlights cases with such ambiguities and casts appropriate uncertainty on the measured radii of the planets. \\
\indent Besides the dilution of the \emph{Kepler} light curve, there is also a possibility that what appears to be a planet is in fact a false positive. This is a serious concern for the \emph{Kepler} mission, as there are numerous astrophysical phenomena that can produce the dimming indistinguishable from that of the transiting planet. A few such examples are the grazing eclipsing binaries, a giant primary star eclipsed by a dwarf, or a background star. The first two cases can usually be identified by the photometry alone, but the background stars pose further challenges to planet validations, particularly for cases of background eclipsing binaries. There is also another case, especially important to consider for smaller planet candidates. The amount of dimming caused by an Earth-like planet orbiting a Sun-like star can be mimicked by a larger planet orbiting a background star, or a less bright binary system companion. \\
\indent All these alternative situations drastically alter the interpretation of the phenomena that caused the dimming detected by the \emph{Kepler} spacecraft, and significant efforts have been made to characterize the false positive probability (FPP) among the \emph{Kepler} planet candidates. The values in the literature range from FPP $<$ 10\% \citep{Morton_etal_2011, Fressin_etal_2013}, to $\sim$35\% \citep{Santerne_etal_2012} for giant planets. Even the lowest estimated FPP is high enough to raise concerns about individual planet discoveries. In an effort to decrease FPP we developed a method for detecting faint stellar companions in double-lined spectroscopic binaries. \\

\indent In this paper, we present a spectroscopic method of searching for a secondary set of absorption lines in a spectrum. A detection of such lines would reveal a secondary star close enough for its light to reach the aperture and affect interpretation of the phenomena responsible for the dimming. The entrance slit has angular dimensions projected onto the sky of 0.87" x 3.0". The slit is oriented differently for each exposure, with the primary star centered on the slit in both dimensions. A secondary star will be included in the spectrum if it resides anywhere within the slit, which restricts the maximum angular separation between the primary star and the secondary star at 0.43" to 1.5", depending on the orientation of the slit. Most commonly, the secondary star is oriented relative to the primary star at a typical angle of 30 to 60 degrees to the length of the slit, in which case its light will be collected if it resides within 0.5" to 0.8" from the primary star.\\
\indent Our goal is to detect any bound or non-bound secondary star as faint as 1\% relative to the primary target star, with either sufficiently large relative radial velocity to allow for the Doppler separation of the two sets of absorption lines, or with sufficiently different spectral features to allow for the differentiation among the overlapping spectral lines. The presence of such a star does not prove that the transit signal is not a planet. It merely indicates that a second star is located within 0.5" $\sim$ 0.8" of the primary star, causing ambiguity about the host star and uncertainty about the planet radius.\\
\indent A standard tool for spectroscopic binaries is TODCOR \citep{Mazeh_etal_1993}, used extensively to model, self-consistently, the two underlying spectra in a given spectrum. This code has been shown to provide excellent detection and assessment of the properties of the two stars. We, however, wanted to develop our own method for detecting secondary lines with a primary goal of being extremely conservative in our detections. We adopted an "Occam's Razor" approach in which any observed spectrum of just one star would be deemed a sufficient model. We only invoke the spectrum of a second star in the model if absolutely forced to do so by the residuals that show the second star to be statistically significant. Thus, we seek an algorithm that accomplishes a statistically robust identification of secondary lines. We were dedicated to constructing an algorithm that we understand in precise detail, and consider our intimate knowledge of our own algorithm to be vital in assessing its integrity and functionality.

\section{Spectrum Preparation}
\label{sec:spectrum_preparation}
\begin{figure}[h]
	\plotone{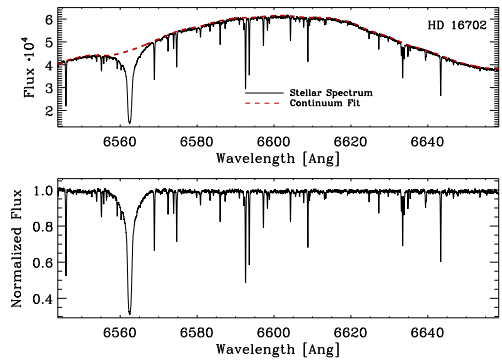}
	\caption{Continuum flattening example using the $H_{\alpha}$ region of the Kepler-16 spectrum. \underline{Above:} Original spectrum with the over plotted $5^{\rm th}$ order polynomial continuum fit, shown as a red dashed line. \underline{Below:} The normalized and flattened spectrum, obtained by dividing the original spectrum shown above by the continuum fit.\\}
	\label{fig:continuum_flattening}
\end{figure}

\indent We worked with spectra from HIRES at Keck Observatory. HIRES has a bandwidth that covers wavelengths from 3643 $\rm\AA$ to 7990 $\rm\AA$. We worked with the wavelength region from 4977 $\rm\AA$ to 7990 $\rm\AA$, avoiding the parts of the spectrum polluted by telluric lines as well as the region of the interstellar sodium D lines. This wavelength domain encompasses approximately the V and R broadbands of classical photometry.\\
\indent Using one of the 0.86"-wide slits, HIRES has a resolving power of 60,000 at 5500 $\rm\AA$. The SNR of the spectra depends on the brightness of the primary star and the duration of the exposure. The typical SNR values range from $\sim$45 to 200 per pixel. \\
\indent First, we continuum-normalize each HIRES spectrum to remove the somewhat variable blaze function from each spectral order. We fit the continuum using the fifth order polynomial, divide the spectrum by the fit and thus obtain a flat continuum at unity, shown in Figure \ref{fig:continuum_flattening}.

\indent Once flattened, we resample the spectrum on a constant ${\Delta}{\lambda}/{\lambda}$ wavelength scale, using interpolation. We are using a wavelength scale with a pixel spacing of $\sim$ 1.3 $\rm{km\,s^{-1}}$. Resampling is done by using a set prescription of converting HIRES pixels into $\Delta\lambda/\lambda$ intervals. This wavelength scale allows for a simpler relationship between the pixel positions of absorption lines and the Doppler shift of the star; with spectrum on a constant ${\Delta}{\lambda}/{\lambda}$ scale, a Doppler shift of $v$ $\rm{km\,s^{-1}}$ corresponds to a simple translation of the spectrum for $\rm{N}_{\rm{pixels}}$ = ($v$ $\rm{km\,s^{-1}}$)/1.3.

\section{The Search Algorithm for a Faint Secondary Spectrum}

\subsection{Fitting the Spectrum of the Primary Star}
\label{sec:primary_star}

\indent Since we are interested in the relative radial velocity between the primary star and the potential companion, we correct for the Doppler shift of the primary star such that the primary star's absorption lines reside at the rest frame wavelengths. We determine the Doppler shift of the primary star using the $\chi^2$ statistic of cross-correlating the corrected NSO Solar spectrum \citep{Wallace_Hinkle_2011} with no Doppler shift. We determine the location of the $\chi^2$ minimum, which corresponds to the radial velocity at which the absorption lines of the Sun's spectrum align with the absorption lines of the primary star - the primary star's Doppler shift. Since we are using a wide range of wavelengths, the discrepancy in the spectral type between the Sun and the star in question does not cause any difficulties in determining the radial velocity of the primary star.

\subsubsection{Finding the Best-Fit}
\label{sec:finding_the_best_fit}

\begin{figure}[h]
	\plotone{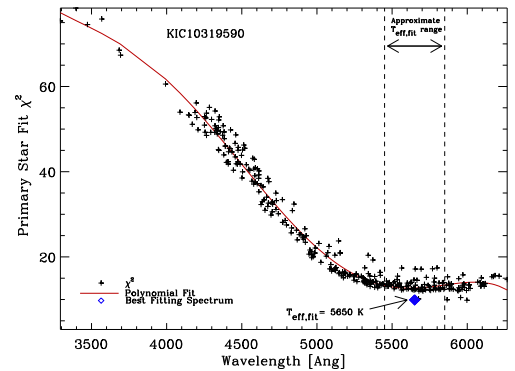}
	\caption{Primary star fit $\chi^2$ as a function of a library star's effective temperature. Value of $\chi^2$ for each library star is shown as a black "+" sign. Due to unknown uncertainties of each point along the spectra, we only look at the $\chi^2$ values of one spectrum relative to another. Approximate $T_{\rm eff}$ range of the primary star is determined based on the minimum of the polynomial fit to the $\chi^2$ versus temperature trend. The fit is shown as a red solid line. Within $\pm$400 K from the minimum of the fit, marked with vertical dashed lines, we then choose the best-fitting library spectrum as the spectrum that produced the lowest $\chi^2$ value. $T_{\rm eff}$ of the best fitting spectrum for KIC 10319590 was 5650 K, shown as the blue diamond.\\}
	\label{fig:allteff_primary_chi2}
\end{figure}

\indent Once the spectrum is normalized and on a constant ${\Delta}{\lambda}/{\lambda}$ wavelength scale, we search through the SpecMatch library of stellar spectra in order to find the best fit for the primary star. The SpecMatch library consists of Keck-HIRES spectra of 640 FGKM stars scattered throughout the H-R Diagram, with a concentration to the main sequence and subgiants. These spectra were obtained as part of the California Planet Search. Each spectrum has SNR=100-200 per pixel between 3800$\rm \AA$-8000$\rm \AA$ with the same spectral resolution as the bulk of the spectra analyzed here, R=60,000. \\
\indent All of the SpecMatch library spectra are available online at the Keck Observatory Archive. Apart from M-dwarfs, each spectrum was analyzed with an advanced version of the spectroscopic analysis package, SME, described in \citet{Valenti_Fischer_2005}. SME was later modified based on the available parallax information, toward a revised, improved version of SME (Brewer et al. 2014, in prep). This analysis yielded the values of $T_{\rm eff}$, log$g$, and [Fe/G] for all FGK stars in the library. Stellar parameters for M-dwarfs were adopted from \citet{Rojas-Ayala_etal_2012}.\\
\indent SpecMatch stellar parameter values range in $T_{\rm{eff}}$=[3250,7260], log$g$=[1.46,5.00], and [Fe/H]=[-1.475,0.558]. We assume that the primary star lies on main sequence and exclude the subgiants from the library as possible best-fit candidates. This choice was made due to the preferential choice of subgiants as the best-fitting stars, even when the observed primary star was on main sequence. Excluding the giants improves the quality of the best-fit and allows for a better subtraction of the primary spectrum, which is explained in Section \ref{sec:primary_subtraction}. This decision was made with an objective of our efforts in mind; since our primary goal is to search for the secondary lines in the spectrum, we sacrifice some of the accuracy of the primary star parameters in order to achieve a better fit to the primary star absorption lines. Nevertheless, if the spectral type of the primary star is known, the algorithm can be modified to include the subgiants in the fitting process as well.\\
\indent We look for the library spectrum that produces the lowest $\chi^2$ value when the absorption lines of the library spectrum are aligned with the studied spectrum. We vary both the rotational broadening of each library spectrum as well as the depth of the absorption lines by diluting the spectrum. The latter is to correct for the finite grid spacing of the metallicity in our library, and is implemented to ensure as complete subtraction of the primary star absorption lines as possible. Due to the inclusion of the constant dilution factor, we are unable to determine the log$g$ of the primary star altogether. However, as mentioned above, the determination of the parameters for the primary star is beyond the scope of this paper, and can be performed using a different method, if so desired. \\
\indent Taking above into account, the $\chi^2$ is computed using
\begin{equation}
	\chi^2 = \sum_{i}{\left(S_i^{\rm lib} - d\cdot (S_{i+p}^{\rm obs}(v{\rm sin}i) - 1) + 1\right)^2}\,,
\end{equation}
where $S^{\rm lib}$ is the SpecMatch library spectrum, $S^{\rm obs}(v{\rm sin}i)$ is the rotationally broadened observed spectrum, $d$ is the dilution factor, and $p$ is the Doppler shift of the primary star in pixels.\\
\indent Discrepancies between each library star and the actual spectrum are both due to Poisson noise as well as intrinsic spectral differences. This is mostly due to the finite grid spacing in parameter space of our library. As a consequence, we cannot establish a good uncertainty estimate for each point along the spectrum, and the actual value of $\chi^2$ bears no significance; rather, it is the relative $\chi^2$ among all of the library spectra that we are concerned with. \\
\indent In order to prevent outliers to be identified as best-fitting stars, we first identify the approximate temperature of the primary star by finding the minimum of the $\chi^2$ versus library star temperature distribution, as shown in Figure \ref{fig:allteff_primary_chi2}. The vertical scatter of the points at the same $T_{\rm eff}$ is due to the discrepancies in other parameters, such as log$g$ and metallicity. Note that we could plot $\chi^2$ as a function of other parameters as well; we choose the temperature as the distribution the has most easily identifiable minimum.\\
\indent Based on the polynomial fit to the $\chi^2$ distribution, we then eliminate the outliers that differ from the polynomial fit for more than 3$\sigma$ at any value of the effective temperature. We then repeat the polynomial fitting, and choose the lowest $\chi^2$ value within the $\pm$200 K range about the minimum of the polynomial fit as our best-fitting library spectrum. While restricting the range of possible $T_{\rm eff}$ to $\pm$200 K around the polynomial minimum might introduce larger uncertainty in the determined $T_{\rm eff}$ of the primary star, it does ensure that the chosen library spectrum is the best achievable fit to the observed spectrum and has approximately correct effective temperature. Restricted range excludes any outliers that have not yet been eliminated at the edges of the effective temperature range ($>$6100K and $<$3400) due to the edge effects on the shape of the polynomial. \\

\indent Since the main goal of the primary star fitting is the subtraction of its absorption lines from the spectrum, we do not report any of the best-fit parameters for the primary star. The accuracy of these results is compromised due to several manipulations of the library spectra to account for the finite grid of our library parameter space. Nevertheless, all of the steps described above do ensure that the primary star absorption lines are fitted almost perfectly and subtracted almost completely. If the parameters for the primary star are known, however, any or all of the values in the parameter space we search over (log$g$, $\rm{[Fe/H]}$, rotational broadening, $T_{\rm eff}$) can be constrained to a known range or fixed to a known value. 

\subsection{Bright Secondary Stars}
\label{sec:bright_stellar_companions}

\begin{figure}[h]
	\plotone{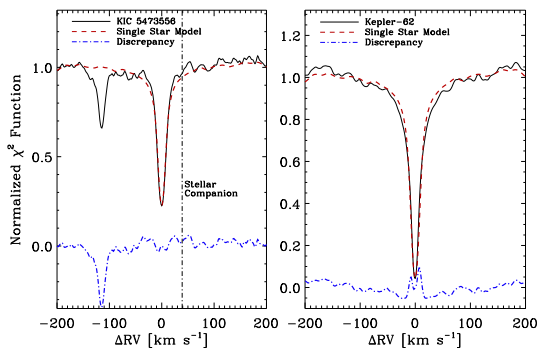}
	\caption{$\chi^2$ as a function of Doppler shift for the fit to the spectrum for two stars, KIC 5473556 (left) and Kepler-62 (right). We compute the $\chi^2$ between the Sun's and a star's spectrum for a range of Doppler shifts. The dip at $\Delta$RV = 0 $\rm{km\,s^{-1}}$ corresponds to the alignment of primary star absorption lines with those of the Sun. The red dashed line corresponds to the $\chi^2$ function of a Sun spectrum with itself, thus representing a characteristic shape for a single star $\chi^2$ function. The blue dot-dashed line shows the difference between the actual $\chi^2$ and the characteristic function. \underline{Left:} KIC 5473556 is a binary system, and we can see the secondary dip at $\Delta$RV $\approx$ 99 $\rm{km\,s^{-1}}$. \underline{Right:} Kepler-62 is a single star, thus its $\chi^2$ function matches closely the shape of a characteristic single star $\chi^2$ function.\\} 
	\label{fig:primary_chi2}
\end{figure}

\indent Brighter stellar companions ($>\,\sim$10\% relative brightness) can be detected using primary $\chi^2$ as a function of Doppler shift. We use the same $\chi^2$ function we used for determining Doppler shift of the studied star, described in Section \ref{sec:primary_star}. As mentioned in Section \ref{sec:finding_the_best_fit}, the actual $\chi^2$ values are of no significance. Therefore, we normalize $\chi^2$ function such that its median value is one when the lines of the NSO Solar spectrum and studied spectrum are misaligned. Deviations from unity are due to the accidental alignments of absorption lines with those corresponding to a different element.\\
\indent We noticed a trend in those accidental alignments that was independent of the spectral type for the primary star, as all of the primary $\chi^2$ functions had the same shape outside the central minimum. We thus created a $\chi^2$ function by cross-correlating the Sun's spectrum with itself, and adopted this shape as the characteristic single star $\chi^2$ function. We then compare the actual $\chi^2$ function to this characteristic single star $\chi^2$ function in order to detect any discrepancies with might indicate the presence of a companion, as shown in Figure \ref{fig:primary_chi2}.

\indent We can see in Figure \ref{fig:primary_chi2} that the primary $\chi^2$ function reveals a secondary star of KIC 5473556 (at left), at a relative RV of +90 $\rm{km\,s^{-1}}$. Such detections are only possible when the secondary star is sufficiently bright with a large enough relative velocity. The $\chi^2$ dip has a FWHM width of approximately 20 $\rm{km\,s^{-1}}$, thus any secondary star with a $\Delta$RV $<$ 10 $\rm{km\,s^{-1}}$ would blend in and remain undetectable. Due to variable rotational broadening of the lines among different spectra, we artificially broaden the characteristic single star $\chi^2$ function to match the width of the actual $\chi^2$ for the analyzed spectrum. Thus, the detection threshold of the relative brightness for the secondary star varies both with $\Delta$RV and the nature of the star's spectrum. \\

\subsection{Faint Secondary Stars}

\subsubsection{Subtraction and the Residuals}
\label{sec:primary_subtraction}

\begin{figure}[h]
	\plotone{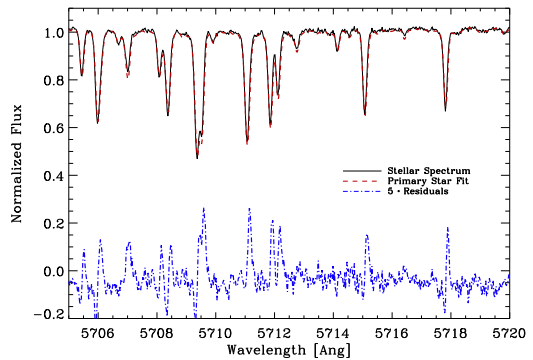}
	\caption{The best-fit to the primary star in the HD 16702 spectrum. We search through $\sim$640 library stars with a range of $T_{\rm{eff}}$, log$g$, and metallicities, as well as vary the Doppler broadening and the dilution of absorption lines to correct for the finite grid spacing of the library spectra's properties. Best-fits are shown as red dashed lines. After the best-fit spectrum has been identified, we subtract it from the spectrum and obtain the residuals, shown as a dot-dashed blue line.\\}
	\label{fig:primary_subtraction}
\end{figure}

\indent Once the best-fit library spectrum is identified and its lines both broadened as well as diluted to match the primary star's absorption lines, we subtract that best-fit library spectrum from the original spectrum, leaving residuals as shown in Figure \ref{fig:primary_subtraction}.

\subsubsection{Evidence of Secondary Lines}
\label{sec:evidence_of_companion}

\begin{figure}[h]
	\plotone{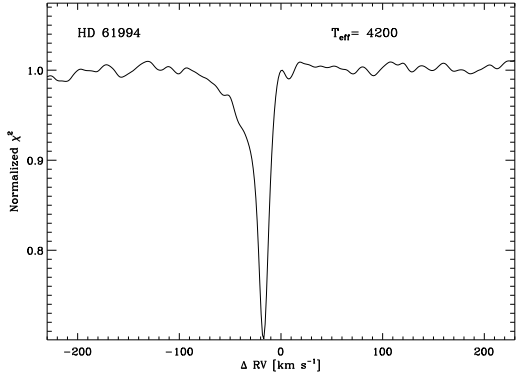}
	\caption{$\chi^2$ as a function of Doppler shift for the residuals of HD 61994 and the median library spectrum with $T_{\rm eff}$ = 4200 K, printed in the upper right corner. There is a significant minimum at $\Delta$RV = -18 $\rm{km\,s^{-1}}$, indicating a presence of a second star.\\} 
	\label{fig:example_chi2_companion}
\end{figure}

\indent We re-normalize the residuals described above back to unity, and search for the secondary set of absorption lines. Since our algorithm is focused mostly on faint secondary stars, the spectral differences among two stars with $\Delta T_{\rm eff}\,<\,100$ K are almost indiscernible. To enhance efficiency, we create 26 median spectra with $T_{\rm eff}$ intervals of 100 K starting at 3300 K and up to 6100 K, rather than utilizing each individual library spectrum. \\
\indent We omit $T_{\rm eff}$ where there are less than three spectra per 100 K temperature interval. As a consequence, not all of the 100 K temperature intervals are represented. In particular, we do not have a representative median library spectrum at effective temperature 3800 K and 3900 K.\\
\indent We then calculate $\chi^2$ as a function of Doppler shift between the residuals and each median library spectrum, and normalize it such that the median is 1. A significant $\chi^2$ minimum occurring at the same $\Delta\rm{RV}$ for several neighboring effective temperatures of the median library spectrum indicates a possible second star in the spectrum. An example of a residual $\chi^2$ function is shown in Figure \ref{fig:example_chi2_companion}.\\
\indent Unless the primary and the secondary spectrum are sufficiently different, we cannot detect secondary stars with $\Delta\rm{RV}\,<$ 10 $\rm{km\,s^{-1}}$ relative to the primary star due to the normal widths of the absorption lines. Since none of the library spectra are the exact match to the studied one, there will always be some imperfections in the best-fitting library star, resulting in larger residuals at the locations of primary star's absorption lines. These residuals may cause either small peak or a small dip at $\Delta\rm{RV}\,$ = 0 $\rm{km\,s^{-1}}$. Thus, if both stars have exactly the same radial velocity, we cannot differentiate between the minimum at $\Delta\rm{RV}$ = 0 $\rm{km\,s^{-1}}$ that is due to the imperfect subtraction of the primary, or the one due to an actual second star. Furthermore, secondary sets of absorption lines with slightly different radial velocities than the primary star have their spectral lines still blended with the primary star's absorption lines. This occurs if the RV separation is less than 10 $\rm{km\,s^{-1}}$, the typical width of an absorption line. This blending causes a fraction of secondary absorption lines to be subtracted away together with the primary star, making a faint second star appear even fainter, and in most cases impossible to detect.\\

\begin{figure}[h]
	\plotone{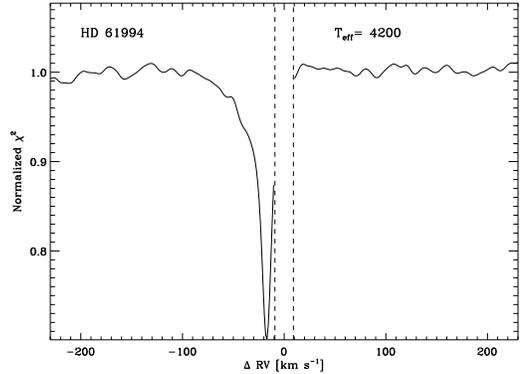}
	\caption{$\chi^2$ as a function of Doppler shift for the residuals of HD 61994 and the median 4200 K library spectrum. We mask out the central region of $\pm$10 $\rm{km\,s^{-1}}$, as the detection of secondary stars with relative radial velocities less than 10 $\rm{km\,s^{-1}}$ is limited. This is both due to the overlap of two sets of lines for $\rm{|RV|}$ separations of less than 10 $\rm{km\,s^{-1}}$, as well as the imperfect subtraction of the primary star. Since the fit for the primary is never exact due to intrinsic spectral differences between the studied spectra and the library spectra, the residual primary star's absorption lines cause a small peak/dip at $\Delta\rm{RV}$ = 0 $\rm{km\,s^{-1}}$. We cannot differentiate between the residual primary minimum and the actual secondary star, thus exclude the central region altogether.\\}
	\label{fig:example_chi2_companion_masked}
\end{figure}

\indent Due to the limitations described above, we mask out the central region with $\Delta\rm{RV}\,<$ 10 $\rm{km\,s^{-1}}$, as shown in Figure \ref{fig:example_chi2_companion_masked}. The only exception is the case where the two spectral types are sufficiently different that primary star spectrum does not interfere with the secondary, regardless of their RV separation. This is the case of a G-type primary and an M-dwarf secondary star, further discussed in Section \ref{sec:G-M_pairs}. 

\subsubsection{Estimating Secondary Star Temperature}
\label{sec:estimating_companion_temperature}

\indent As mentioned in Section \ref{sec:evidence_of_companion}, we obtain 26 distinct residual $\chi^2$ functions, differing in the effective temperature of the library spectrum superimposed with the studied residual spectrum. We expect to find a unique $\Delta\rm{RV}$ location of the secondary star, if one is detectable. Since a secondary set of absorption lines would cause a minimum in $\chi^2$ over a range of library spectral types that are close in $T_{\rm eff}$ to the actual temperature, we assume the secondary star $\Delta\rm{RV}$ to be the mode of the $\chi^2$ minimum locations for a set of all 29 residual $\chi^2$ functions. The exception here is the case where the minimum occurs only for $T_{\rm eff}$ $\sim$ 3500 K. Due to a significantly different M-dwarf spectrum, it is possible that the secondary star is detected only in a few residual $\chi^2$ functions with $T_{\rm eff}\,<\,\sim$3800 K. Those cases are examined and assessed on an individual basis.\\
\indent Using the mode minimum $\Delta\rm{RV}$ location, we plot the $\chi^2$ values at those locations versus the temperature of the median library spectrum used to construct each respective $\chi^2$ function. For most residual $\chi^2$ functions, this $\chi^2$ is indeed the $\chi^2$ minimum. Ideally, the depth of the minimum should only depend on how well the library spectrum correlates with the spectrum of the secondary star, producing the lowest $\chi^2$ minimum when the library spectrum temperature is closest to the effective temperature of the secondary star. However, mostly due to the low SNR of the residuals, we noticed a dependence of the residual $\chi^2$ minimum on the temperature of the library spectrum that was independent of the actual $T_{\rm eff}$ of the secondary star. In order to correct for this, we divide the $\chi^2$ minima plot by the calibration function, construction of which is described in more detail in Section \ref{sec:injection-recovery_companion_calibration}.\\
\indent We then plot the calibrated $\chi^2$ minimum values versus the temperature of the median library spectrum, as shown in Figure \ref{fig:complete_companion_plot} (left). In order to determine approximate temperature of the secondary star, we follow the same procedure as for the primary star, outlined in Section \ref{sec:finding_the_best_fit}. In brief, we fit the points with a polynomial, find a minimum, remove the outliers that differ from the fit by more than 3 $\sigma$, re-fit the remaining points, and restrict the possible range of secondary star $T_{\rm eff}$ to $\pm$ 200 K around the polynomial minimum. Within that range, we then compute the mean $T_{\rm eff}$ of the three median library spectra corresponding to the three lowest $\chi^2$ minima. We adopt this mean as the estimated temperature of the secondary star. Taking the mean of the three lowest $T_{\rm eff}$ values corrects for the missing median templates, as not every 100 K $T_{\rm eff}$ is represented due to varying parameter populations of SpecMatch library. \\
\indent We cannot extrapolate the behavior of the $\chi^2$ minima function to values of $T_{\rm eff}$ higher than 6100 K or lower than 3300 K. Thus, any secondary stars whose $T_{\rm eff}$ is identified at $\sim$6000 K is not guaranteed to be at that temperature; rather this serves as the lower temperature limit, as the secondary star can be hotter than $\sim$6000 K but not cooler. Similarly, secondary stars whose estimated $T_{\rm eff}$ is 3300 K can be cooler than this value, but not hotter. This is discussed more in detail in Section \ref{sec:too_high_teff}.\\

\subsubsection{Estimating Relative Brightness}
\label{sec:estimating_relative_brightness}

\begin{figure}[h]
	\plotone{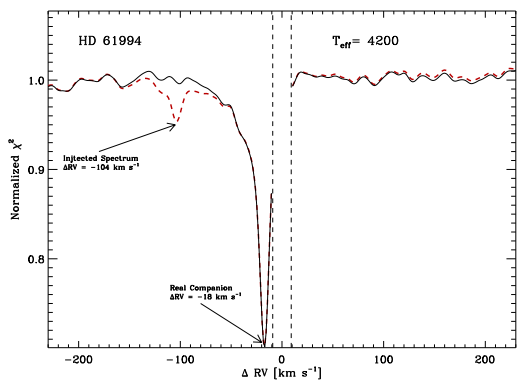}
	\caption{Residual $\chi^2$ function of HD 61994 and the median 4200 K library spectrum. Red line dashed is obtained by injecting a secondary star with $\Delta\rm{RV}$ = $-$104 $\rm{km\,s^{-1}}$ into the original HD 61994 spectrum. $\chi^2$ shown in black is the original residual $\chi^2$, without the injected secondary spectrum (same as Figure \ref{fig:example_chi2_companion}). We can see that there is a minimum at the location of the injected spectrum, annotated with an arrow. The spectrum of the injected secondary star is scaled down such that it contributes 1\% of the total flux.\\}
	\label{fig:example_injected_companion_oneshift}
\end{figure}

\begin{figure}[h]
	\plotone{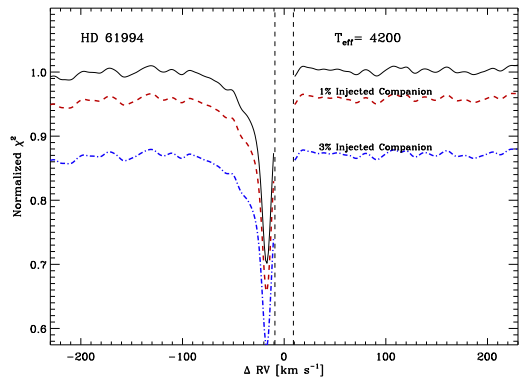}
	\caption{Residual $\chi^2$ functions for HD 61994 and the median 4200 K library spectrum. \underline{Black Curve:} original residual $\chi^2$ function, without any injected secondary star. \underline{Dashed Red Curve:} $\chi^2$ minima values for a secondary star that contributes 1\% of the total flux. \underline{Dot-dashed Blue Curve:} $\chi^2$ minima values for a secondary star that contributes 3\% of the total flux. For both dashed red and dot-dashed blue curves, the value at each $\Delta\rm{RV}$ represents the value that the original residual $\chi^2$ would have if there was an actual secondary star present at that particular $\Delta\rm{RV}$, of a relative brightness indicated on the curve.\\}
	\label{fig:example_injected_companion}
\end{figure}

\indent After determining $T_{\rm eff}$ of the second star, we use the corresponding residual $\chi^2$ function for further analysis. We aim to estimate the brightness of the secondary star relative to the primary star based on the depth of that residual $\chi^2$ function minimum.\\
\indent We do so by synthesizing the effect of secondary spectra. We inject another spectrum into the original studied spectrum. We scale down the injected spectrum such that it contributes either 3\% or 1\% of the total flux, and shift its absorption lines to a known relative radial velocity. The injected spectrum has $T_{\rm eff}$ close to that already estimated for the secondary star temperature, assuring that the superimposed library spectrum fits both the actual secondary star as well as the injected spectrum equally well. The properties of all the possible injected spectra are listed in Table \ref{table:injected_properties}.\\
\indent We then treat this original spectrum, with its synthetic injected secondary star spectrum, as the new studied spectrum, and repeat the whole procedure outlined in Sections \ref{sec:finding_the_best_fit} through \ref{sec:evidence_of_companion}. We record the $\chi^2$ value at the $\Delta\rm{RV}$ location of the injected spectrum, annotated in Figure \ref{fig:example_injected_companion_oneshift}. At that particular $\Delta\rm{RV}$, the difference between the injected spectrum $\chi^2$ and the value of the original $\chi^2$, both shown in Figure \ref{fig:example_injected_companion_oneshift}, represents the minimum depth would be caused by a secondary star of the injected brightness. \\
\indent To obtain a more statistically useful sample of possible secondary spectra, we repeat the injection for several different values $\Delta\rm{RV}$s, and each time record the difference between the value of the $\chi^2$ minimum at the location of the injected secondary spectrum and the original residual $\chi^2$ value at that location. Using those differences, we then calculate the median $\chi^2$ minimum depth relative to the original residual $\chi^2$ caused by a secondary star of that particular brightness. \\
\indent Subtracting the median depth for both 3\% and 1\% injected secondary spectrum from the original residual $\chi^2$ without the injected secondary spectrum, we obtain the plot shown in Figure \ref{fig:example_injected_companion}. At each $\Delta\rm{RV}$, the two colored curves represent the value that the $\chi^2$ function would have if there was a secondary star of the specified brightness located at that particular $\Delta\rm{RV}$. This allows for a visual comparison of the actual residual $\chi^2$ minimum to the characteristic $\chi^2$ minima depths for 3\% and 1\% secondary stars.\\
\indent To estimate the relative brightness of the actual secondary star, we extrapolate the median $\chi^2$ minimum caused by the 3\% and 1\% injected secondary spectrum to the actual $\chi^2$ minimum for the residual function. We use a least-squares linear fit, with a restriction that 0\% secondary star (none present) causes a $\chi^2$ minimum of depth 0. We report both the $\Delta\rm{RV}$ of the secondary star, as well as the estimated relative brightness as a percentage of total flux, on the plot.\\
\indent Since our wavelength domain encompasses approximately the V and R broadbands of the classical photometry, the relative brightness of the two stars can be used to compute the dilution factor of the {\it Kepler} light curves. Thus, if the primary star has any planet candidates, this dilution factor can help determine a more accurate radius of the planet. \\

\subsubsection{Secondary Star Detection Summarized}

\begin{figure}[h]
	\plotone{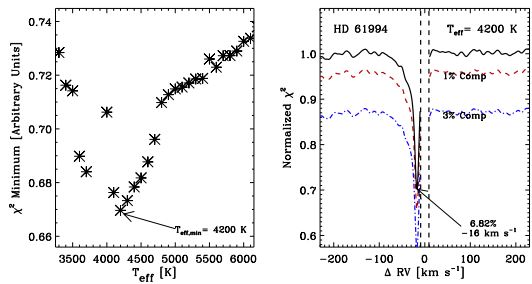}
	\caption{\underline{Left:} residual $\chi^2$ minimum distribution as a function of temperature. Each point on the plot represents the value of the $\chi^2$ minimum, constructed with the residuals of HD 61994 after the primary star was subtracted, and the library spectrum of $T_{\rm eff}$ as indicated on the $x$-axis. The lowest $\chi^2$, annotated with an arrow, corresponds to the estimated secondary star temperature. \underline{Right:} residual $\chi^2$ function constructed with a $T_{\rm eff}$ = 4200 K library spectrum. There is a notable minimum at $\Delta\rm{RV}$ = $-$16 $\rm{km\,s^{-1}}$. The red dashed and the blue dot-dashed lines correspond to the $\chi^2$ minima values that a 1\% and 3\% secondary star would exhibit, respectively. Based on the depth of those two curves, the percentage of the secondary star in HD 61994 spectrum is estimated to contribute 6.82\% of the total flux.\\}
	\label{fig:complete_companion_plot}
\end{figure}

\indent Figure \ref{fig:complete_companion_plot} has two panels, the left showing the calibrated residual $\chi^2$ minimum as a function of $T_{\rm eff}$ and the right showing $\chi^2$ vs. $\Delta$RV. This includes the minima curves for 1\% and 3\% injected secondaries, shown in red and blue respectively (see Section \ref{sec:estimating_relative_brightness}).\\
\indent This final Figure \ref{fig:complete_companion_plot} is intended to summarize our knowledge about the secondary star: estimated $T_{\rm eff}$, estimated percentage contribution to the total flux spectrum in the V and R bands, and the radial velocity of the secondary relative to the primary star. Any additional parameters for the secondary star cannot be determined accurately enough to be published.\\
\indent On the final secondary star plot, we annotate any minimum that contributes at least 0.5\% of the total flux. Depending on the SNR of the original spectrum (typically 50-200 per pixel), the spectral types of the two stars, and their $\Delta\rm{RV}$, secondaries with fluxes between 0.5\% - 1\% relative to the primary star can be detected with our algorithm.\\
\indent We visually inspect each diagnostic plot to assess the probability that the identified minimum is indeed due to a secondary star. When there is no clear $\chi^2$ minimum, or when the $\chi^2$ minimum is due to fluctuations, we establish the threshold limits on the secondary star as follows. Any {\it undetected} secondary must contribute less to the total flux than would have been revealed above the fluctuations in the $\chi^2$ functions. If this percentage is not marked on the plot, the brightness limit is 0.5\% relative to the primary star. Otherwise, the appropriate percentage is marked on the plot. M-dwarf secondaries contributing more than 0.5\% of the total flux can be typically detected in our spectra if present, or ruled out if not present.\\

\section{Assessing the Algorithm: Tests of Synthetic and Real Binaries}
\label{sec:assesing_the_algorithm}

\subsection{Injection-Recovery Experiment}
\label{sec:injection_recovery_experiment}

\indent In order to test and calibrate our algorithm, we performed injection-recovery experiments. The goals were to estimate the uncertainty for the determined secondary star parameters, allow for the correction of any systematic errors, and identify the limitations of the algorithm.\\
\indent We synthesized three sets of 360 binary stars, composed of pairs of our library spectra with added 2\% Poisson noise. The noise level reflects that (2\% per pixel, i.e. SNR$\approx$45) common to many of the Keck-HIRES spectra we analyzed in this paper, namely the 1160 {\it Kepler} Objects of Interest, described in Section \ref{sec:Kepler_survey}. Since the spectra used to synthesize the binaries are real, observed spectra with Keck-HIRES, they already contain a certain amount of other types of noise (such as red noise), and thus the synthetic binaries accurately capture the nature of any other spectra analyzed in this paper. \\
\indent We chose library stars in $T_{\rm eff}$ increments of 500 K, ranging from 3500 K to 6000 K. We then created pairs of spectra for all possible permutations of $T_{\rm eff}$, beginning at 3500 K for both the primary and the secondary star. We considered all possible combinations, including those where the primary star is cooler than the secondary, to account for both bound systems and background companions. The secondary set of absorption lines was shifted from the primary by $\Delta \rm RV$ of $+$50 $\rm km\,s^{-1}$, where the two sets of lines are well separated. Tests have shown that for $\Delta \rm RV$ separations larger than $\pm$10 $\rm km\,s^{-1}$ the detectability of the second star is independent of the relative radial velocity for the two stars. For smaller $\Delta \rm RV$ separations, the two sets of absorption lines overlap due to the natural broadening. At such small $\Delta \rm RV$ separations, we can only detect M-dwarf secondary to a G-type primary star. For those cases, additional tests have been carried out with the secondary set of absorption lines shifted from the primary by only $+$5 $\rm km\,s^{-1}$. \\
\indent Each of the three sets of 360 spectra differs by the relative brightness of the secondary star. We constructed cases in which the second star contributed 10\%, 3\%, and 1\% of the total flux in our optical spectra (V and R bandpasses). For the cases of G-type primary and an M-dwarf secondary star we performed additional tests with even higher flux ratios, where the M-dwarf secondary star contributed as low as 0.05\%, 0.1\$ and 0.5\% of the total flux. To obtain statistically significant sample, we created 10 different binary spectra for each such permutation, each with a different realizations of Poisson noise. We then blindly executed our search algorithm for secondary spectra on each synthetic binary. In the search, we removed the spectra used to construct that particular case from the library, preventing those to be chosen as best fitting stars.\\

\subsubsection{Secondary Star Recovery Rate}
\label{sec:inject-recovery_comapnion_recrate}

\indent We used the synthetic binary spectra with added noise to determine the uncertainties in the deduced parameters for the secondary star. We investigated each temperature combination, as well as each flux ratio, separately. We examined the recovery rate by counting all the cases where the secondary star was identified, disregarding the accuracy of the estimated parameters. Results are summarized in Table \ref{table:companion_recovery_rate}. The Table can be read as follows: $T_{\rm eff}$ values running horizontally denote the effective temperature of the secondary star, and the values running vertically down the first column denote $T_{\rm eff}$ of the primary star. For example, 60\% rate quoted in the last column of the first data row means that the secondary star in the binary spectrum consisting 99\% of a 3500 K star and 1\% 3500 K star was successfully identified in 6 out of 10 cases.\\
\indent For all synthetic spectra, we added a Poisson noise at the level of 2\%. Since we subtract the primary star absorption lines, a 1\% secondary implies an SNR for the residual spectrum of only 0.5, 3\% has an SNR of 1.5, and 10\% secondary yields a residual spectrum with SNR of 5. Of course, such low SNR values per pixel are overcome by the thousands of absorption lines, detected in thousands of pixels, each contributing to the detectability of the secondary star. Thus the effective signal-to-noise of the secondary star is much higher than the per-pixel SNR.\\

\subsubsection{Calibration of the Secondary Star Temperature}
\label{sec:injection-recovery_companion_calibration}

\begin{figure}[h]
	\plotone{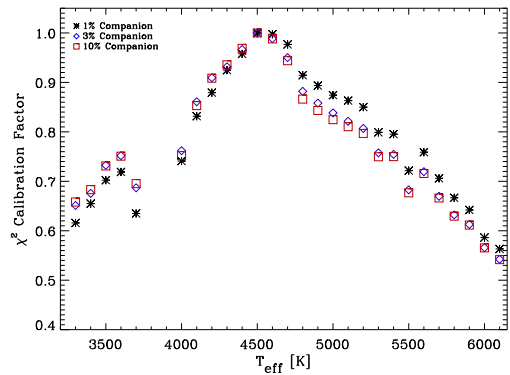}
	\caption{Calibration functions for a 1\%, 3\%, and a 10\% secondary star, as shown on the plot. These calibration functions were obtained by calculating the mean residual $\chi^2$ minimum distribution functions over all possible $T_{\rm eff}$ secondaries. If the $\chi^2$ was unbiased against the temperature of the superimposed spectrum, the calibration function would be constant.\\}
	\label{fig:calibration_functions}
\end{figure}

\indent We isolated the synthetic binary cases where the binarity of the spectrum was established, and examined the predicted secondary star temperature as opposed to the actual values. As mentioned in Section \ref{sec:estimating_companion_temperature}, we noticed a dependence of the depth of the residual $\chi^2$ minimum on the temperature of the library spectrum, which was superimposed with the studied spectrum to construct the residual $\chi^2$. Except when the secondary star was an M-dwarf, the residual $\chi^2$ minimum was always the deepest for $T_{\rm eff}\,\sim$ 4500 K, regardless of whether or not that was the actual $T_{\rm eff}$ of the secondary star. We suspect this bias is caused by the nature of the 4500 K spectrum itself, as it contains an abundance of deep and sharp absorption lines from neutral metals that dominate FGK spectra. With the primary star subtracted from the spectrum, the residuals are dominated by Poisson noise at a 2\% level, especially so when the secondary star contributes only a few percent to the total flux. These neutral metal absorption lines tend to accidentally align with the noise much more frequently due to their deep, sharp features. Such accidental alignments cause a deeper $\chi^2$ minimum for the 4500 K library star with the studied spectrum of any $T_{\rm eff}$, particularly so for the residuals with SNR values near unity.\\
\indent This bias was strong enough to influence the estimated temperature of the second star and make our deduced raw derived secondary star parameters inaccurate. In order to correct for this temperature bias, we calculated the average residual $\chi^2$ minimum distribution for all of the test cases of a particular flux ratio, excluding non-detections. If the distribution was unbiased against library spectrum's $T_{\rm eff}$, the mean distribution should be a approximately constant with respect to temperature, as we assured that there were an equal number of cases at each secondary star $T_{\rm eff}$. \\
\indent We found, however, that this was not the case; the mean distribution peaked at 4500 K, as expected. Since the shape of any true $\chi^2$ minimum distribution will be affected by this peak, we use the mean distribution, normalized such that the maximum occurs at $\chi^2$ = 1, as our calibration function. Calibration functions for a 1\%, 3\%, and 10\% secondary star are shown in Figure \ref{fig:calibration_functions}.\\
\indent As we can see from Figure \ref{fig:calibration_functions}, calibration functions are almost identical among different relative brightnesses of the two stars. Therefore, there will be negligible errors resulting from dividing a residual $\chi^2$ minimum distribution by these calibration functions for cases where the relative brightness of the studied star is not 1\%, 3\%, or 10\% exactly.\\

\subsubsection{Secondary Star Parameters}

\indent Using calibrated $\chi^2$ minimum distribution, we estimated the second star temperature and relative brightness, as described in Sections \ref{sec:estimating_companion_temperature} and \ref{sec:estimating_relative_brightness}. We then compared the deduced values to the known parameters, and the discrepancies are shown in Tables \ref{table:injection-recovery_companion_temperature} and \ref{table:injection-recovery_companion_brightness}, respectively.\\
\indent Table \ref{table:injection-recovery_companion_temperature} lists the discrepancies in the SpecMatch $T_{\rm eff}$ value and the one derived from our algorithm, ($T_{\rm SpecMatch}\,-\,T_{\rm deduced}$). Table \ref{table:injection-recovery_companion_brightness} lists the discrepancies in the set secondary star percentage contribution to the total flux and the value derived by our algorithm. We know the actual flux contribution precisely, as we artificially scaled down the secondary spectra when constructing synthetic binaries. The values are listed in terms of the percentage of the total flux, ($\%_{known}\,-\%_{deduced}$). Each primary-secondary temperature combination consisted of 10 trial cases, with different realizations of added Poisson noise.\\
\indent We report both the systematic difference (the mean discrepancy) as well as the associated standard deviation. Since the systematic differences are not constant over all temperature combinations or relative brightnesses, we cannot apply a single correction to obtain a more accurate values. As such, Tables \ref{table:injection-recovery_companion_temperature} and \ref{table:injection-recovery_companion_brightness} serve more as illustrations of the injection-recovery results.\\
\indent We noticed larger $T_{\rm eff}$ systematic differences for 4000 K secondary stars. This is mostly due to the scarce SpecMatch library populations in that region, with some 100 K $T_{\rm eff}$ intervals missing altogether. On the other hand, the algorithm did consistently identify correctly the M dwarf secondaries, owing their detectability to their distinct spectral features. As expected, the systematic differences in $T_{\rm eff}$ as well as the standard deviations decreased for brighter secondary stars, as the secondary spectrum becomes brighter than the 2\% noise. \\
\indent On the other hand, absolute errors in percentage increased when the secondary star was brighter. This was expected, since the errors in linear extrapolations of $\chi^2$ minimum depth become larger for brighter stars. Nevertheless, the errors were still mostly around 1\% only, and the relative errors rarely exceed 50\%. We also noticed a pattern that relative brightness of hotter secondary stars was mostly overestimated, while cooler stars were underestimated. These trends can be useful when establishing limits on brightnesses of any real secondaries.\\
\indent We combine both the systematic error and the uncertainty to form more conservative uncertainties to be used with the secondary star parameters deduced in the analysis of actual spectra. These are shown in Table \ref{table:uncertainties_companion}. Columns two through seven show the absolute uncertainty in the effective temperature of the secondary star for several possible primary-secondary star pairs; columns eight through 13 show the relative uncertainty in the flux ratio. Based on the effective temperatures of the primary-secondary pair and their flux ratio, appropriate uncertainty for both the effective temperature of the secondary star and its relative brightness can be read from Table \ref{table:uncertainties_companion}.

\subsubsection{G-type Primary and M-dwarf Secondary Star}
\label{sec:G-M_pairs}

\indent As a part of the injection-recovery experiment, we also examined more closely the case of a G-type primary star with a temperature of 5500 K and an M-dwarf secondary at 3500 K. Due to its low temperature, M-dwarf spectrum contains many molecular lines in the visible spectrum, quite distinct from the spectrum of FGK-type stars. As such, it is much easier to detect an M-dwarf secondary spectra, even when the $\Delta\rm{RV}$ for the two stars is low.\\
\indent We considered a situation where the two stars are separated by only 5 $\rm{km\,s^{-1}}$, placing the secondary set of absorption lines into the masked-out region on our residual $\chi^2$ versus $\Delta\rm{RV}$ plot. We constructed three sets 20 different synthetic binary stars with a 5500 K primary and a 3500 K secondary star, the secondary contributing either 1\%, 3\%, or 5\% of the total flux. We then executed the algorithm on each spectrum, determined the recovery rate, and calculated the discrepancies in the deduced $T_{\rm eff}$ and percentage flux. Results are shown in Table \ref{table:injection-recovery_closein_G-M_pairs}. Each companion percentage contribution sample consisted of 20 test cases.\\
\indent These results quantify the ability of our algorithm to detect M-dwarf secondary stars even when the two sets of absorption lines overlap. Nevertheless, some of the M-dwarf absorption lines still get subtracted away together with the primary star, thus the relative brightness of the secondary was consistently underestimated for all cases. This suggests an important limitation on our algorithm; when the secondary stars are detected in the $\Delta\rm{RV}\,<$ 10 $\rm{km\,s^{-1}}$ region, the estimated relative brightness of the second star is more of a lower limit rather than an actual value.\\
\indent Besides considering M-dwarf secondary stars with low $\Delta \rm RV$ separation, we also explored the detectability limit in terms of relative brightness of the two stars. We constructed synthetic binaries where the M-dwarf secondary contributed 0.05\%, 0.1\%, and 0.5\% of the total flux. Recovery rate for the 0.5\% M-dwarf secondary star was 40\%, while none of the fainter companions were detected. Since all of the spectra had a 2\% percent noise, this places the M-dwarf detectability to a SNR $\sim$ 0.25.\\

\subsection{Known Real Spectroscopic Binaries}

\indent We executed our algorithm on several cases of known spectroscopic binaries. We then compared our results to those available in the literature. Results are summarized in Table \ref{table:results_table_primary} for the primary star parameters, and in Table \ref{table:results_table_companion} for the secondary. The uncertainties for the secondary star parameter are based on the Table \ref{table:uncertainties_companion}. Even though we normally do not report the parameters of the primary star, we include them in this experiment. This is to demonstrate that even though our primary star parameters might carry larger errors and uncertainties than those derived using alternative methods, they still poses a sufficient accuracy for the purposes of the subtraction of the primary star absorption lines.\\

\subsubsection{KIC 10319590}

\begin{figure}
	\plotone{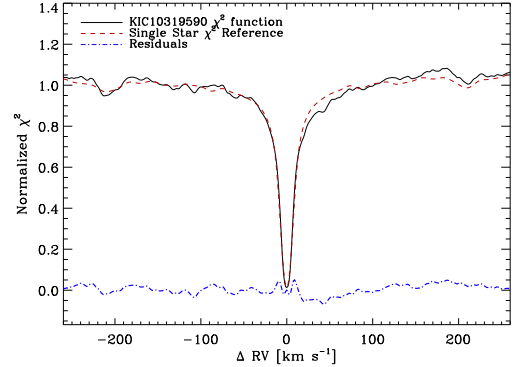}
	\caption{Primary star $\chi^2$ function for KIC 10319590. We can see a slight deviation from the characteristic single star $\chi^2$ function at $\Delta\rm RV\,\approx\,$+45 $\rm{km\,s^{-1}}$, indicating a possible secondary star. This deviation alone, however, is not convincing enough to make any conclusive decisions.\\}
	\label{KIC10319590_primary}
\end{figure}

\begin{figure}
	\plotone{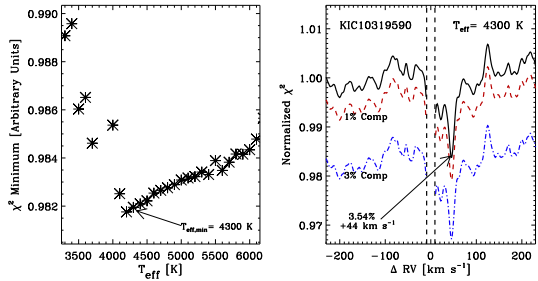}
	\caption{Final secondary star plot for KIC 103194950. Same as Figure \ref{fig:complete_companion_plot}. As foreshadowed by the Figure \ref{KIC10319590_primary}, a secondary star is detected at +44 $\rm{km\,s^{-1}}$ $\Delta\rm{RV}$, with an estimated $T_{\rm eff}$ of 4300 K and contributing 3.54\% to the total flux.\\}
	\label{KIC10319590}
\end{figure}

\indent KIC 10319590 is a precessing eclipsing binary, with time-varying eclipse depths \citep{Rappaport_etal_2013, Slawson_etal_2011}. There is very little evidence of the secondary star in the $\chi^2$ function for the primary star shown in Figure \ref{KIC10319590_primary}, indicating that the secondary is faint. \\
\indent The residual $\chi^2$ function shown in Figure \ref{KIC10319590} shows an evidence of a secondary set of absorption lines at $\Delta$RV = +44 $\rm{km\,s^{-1}}$. Based on the residual $\chi^2$ minimum distribution, we estimate the secondary star $T_{\rm eff}$ = 4300 $\pm$ 500 K, with uncertainty based on Table \ref{table:uncertainties_companion}. \\
\indent Deduced percentage of the total flux for the secondary star is 3.54\%, yielding the flux ratio $\rm{F_{B}/F_{A}}$ = 0.036 $\pm$ 0.07. The comparison of our results to known parameters is limited by the lack of the available information.\\

\subsubsection{KIC 5473556} 

\begin{figure}
	\plotone{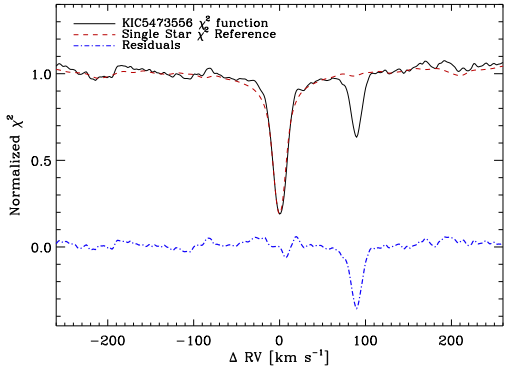}
	\caption{Primary star $\chi^2$ function for KIC 5473556. We can see a clear evidence of secondary star at $\Delta\rm RV\,\approx\,$+90 $\rm km\,s^{-1}$. While this is a fairly convincing argument for the spectral binarity, residual $\chi^2$ function provides a greater insight into parameters for the secondary.\\}
	\label{KIC5473556_primary}
\end{figure}

\begin{figure}
	\plotone{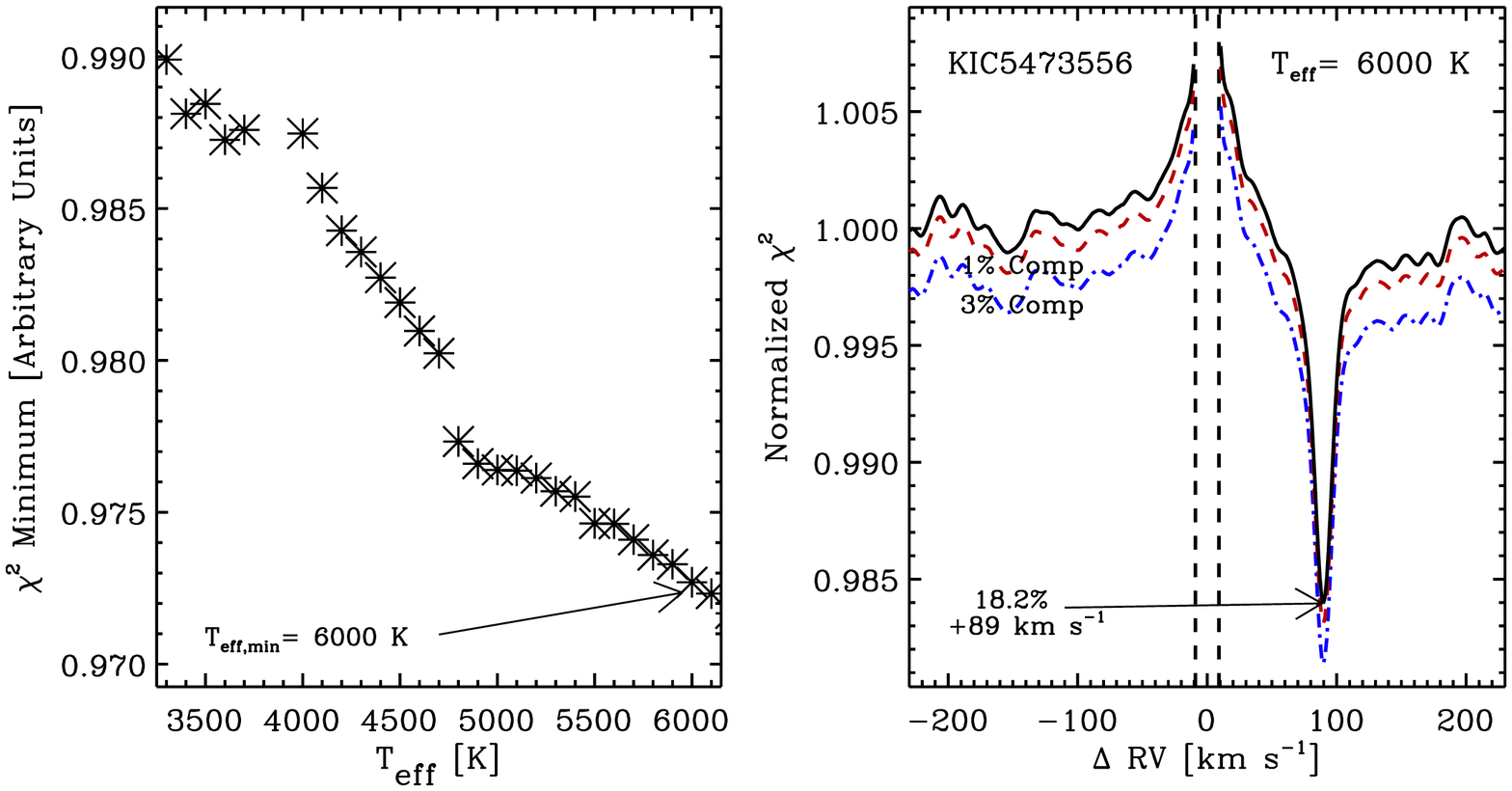}
	\caption{Final secondary star plot for KIC 5473556. Same as Figure \ref{fig:complete_companion_plot}. As indicated by the Figure \ref{KIC5473556_primary}, a secondary star is detected at +89 $\rm{km\,s^{-1}}$ $\Delta\rm RV$, with an estimated $T_{\rm eff}$ of 6000 K, contributing 18.2\% to the total flux.\\}
	\label{KIC5473556}
\end{figure}

\indent KIC 5473556 is another example of a SB2 eclipsing binary \citep{Ruth_Angus}. For this system, the companion is sufficiently bright to cause a secondary minimum in the $\chi^2$ function for the primary star, shown in Figure \ref{KIC5473556_primary}. \\
\indent There is a clear detection shown in the residual $\chi^2$ (Figure \ref{KIC5473556}) at $\Delta$RV = +89 $\rm{km\,s^{-1}}$. We estimate the secondary star $T_{\rm eff}$ at 6000 $\pm$ 100 K. At the relative flux of $\rm F_{B}/F_{A}$ = 0.22 $\pm $ 0.07, this could either be a bound companion or a background star, as the primary star $T_{\rm eff}$ is similar to the estimated secondary star value. \\
\indent In order to resolve the ambiguity, we executed our algorithm using a different observation taken at a different time. If the two stars indeed orbit each other, a different orbital phase should yield a different $\Delta\rm RV$ value. This was indeed the case, as another observation shows a companion at $\Delta\rm{RV}\,\approx$ -115 $\rm{km\,s^{-1}}$. \\
\indent The flux ratio of 0.22 $\pm$ 0.07 and an approximately the same $T_{\rm eff}$ for the two stars indicates that our deduced companion temperature might be an overestimate, and the actual companion $T_{\rm eff}$ most likely lies at the lower uncertainty limit, $T_{\rm eff}\,\sim$ 5900 K. Moreover, it is possible that only a fraction of the flux from the companion actually entered the slit during the observation. Since the parameters for the companion of KIC 5473556 are not known, we cannot compare our derived flux ratio, $\rm{F_{B}/F_{A}}$ = 0.22 $\pm$ 0.07, to a known value. \\

\subsubsection{KIC 8572936}

\indent KIC 8572936 is an SB2 binary system, also known for its circumbinary planet, Kepler-34 \citep{Welsh_etal_2012}. The two stars are both Sun-like stars, and their fluxes are comparable; $\rm{F_{B}/F_{A}}$=0.8475$\pm$0.005 \citep{Welsh_etal_2012}. The effective temperatures of the two stars are similar, $T_{\rm{eff,A}}$ = 5932$\pm$130 K and $T_{\rm{eff,B}}$ = 5867$\pm$130 K. \\
\begin{figure}
	\plotone{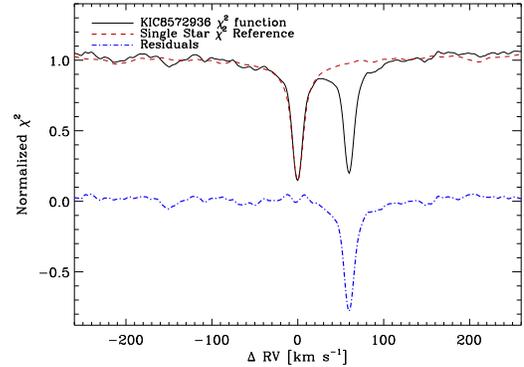}
	\caption{Primary star $\chi^2$ function for KIC 8572936. There is an almost equally bright secondary star detected at $\Delta\rm RV\,\approx\,$+60 $\rm{km\,s^{-1}}$. Again, while this is a fairly convincing argument for the spectral binarity, residual $\chi^2$ function provides a greater insight into parameters for the secondary star.\\}
	\label{KIC8572936_primary}
\end{figure}

\begin{figure}
	\plotone{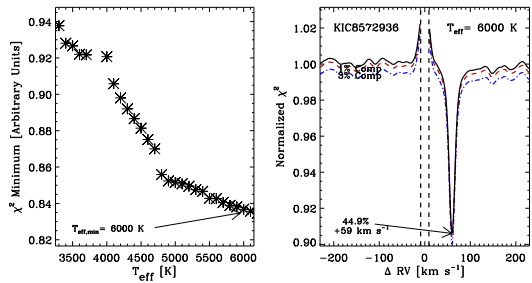}
	\caption{Final secondary star plot for KIC 8572936. Same as Figure \ref{fig:complete_companion_plot}. As already revealed by the Figure \ref{KIC5473556_primary}, a secondary star is detected at +59 $\rm{km\,s^{-1}}$ $\Delta\rm{RV}$, with an estimated $T_{\rm eff}$ of 6000 K, contributing 44.9\% to the total flux.\\}
	\label{KIC8572936}
\end{figure}
\indent This is an interesting case due to the similarity between the two stars, both in their spectral types as well as in the fractional contribution to the total flux. The main problem with such cases is that the two sets of absorption lines interfere with the best-fit for the primary star, thus we inevitably subtract some of the secondary lines together with the primary star. This leads to a somewhat larger uncertainty in the derived flux ratio for the stars using our method; however, the detection of both stars is clear and unambiguous, as seen in Figure \ref{KIC8572936_primary} and \ref{KIC8572936}. \\
\indent The estimated temperature of the companion of $T_{\rm{eff}}$ = 6000 $\pm$ 100 K, is in agreement with $T_{\rm{eff}}$ = 5812 $\pm$ 150 K quoted by \citet{Welsh_etal_2012}. We calculated the flux ratio of $\rm{F_{B}/F_{A}}$ = 0.82 $\pm$ 0.25, which, in the uncertainty limit, also matches the \citet{Welsh_etal_2012} value of 0.8475 $\pm$ 0.005.\\

\subsubsection{KIC 9837578}

\begin{figure}
	\plotone{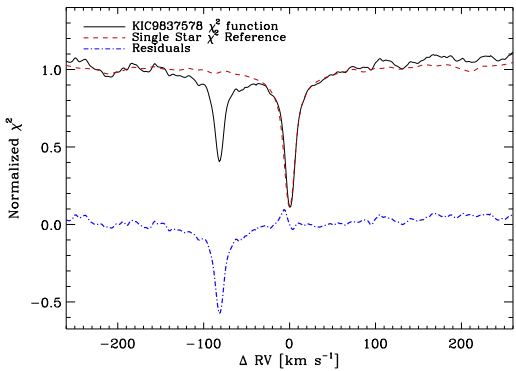}
	\caption{Primary star $\chi^2$ function for KIC 9837578. A bright secondary star is detected at $\Delta\rm RV\,\approx\,$-80 $\rm{km\,s^{-1}}$.\\}
	\label{KIC9837578_primary}
\end{figure}

\begin{figure}
	\plotone{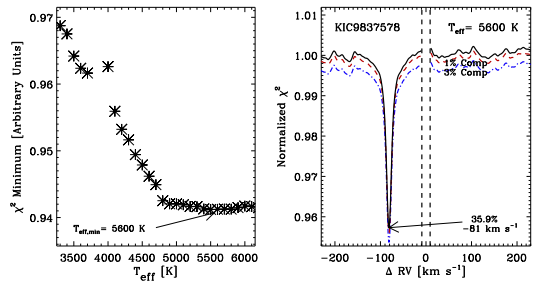}
	\caption{Final secondary star plot for KIC 9837578. Same as Figure \ref{fig:complete_companion_plot}. A secondary star is detected at -81 $\rm{km\,s^{-1}}$ $\Delta\rm{RV}$, with an estimated $T_{\rm eff}$ of 5600 $\pm$ 400 K, and contributing 35.9\% to the total flux.\\}
	\label{KIC9837578}
\end{figure}

\indent The case of KIC 9837578 is a similar to KIC 8572936. It is also an SB2 binary system, with the two stars of similar $T_{\rm{eff}}$, but the flux ratio in the \emph{Kepler} bandpass is somewhat lower, $\rm{F_{B}/F_{A}}$ = 0.3941 \citep{Welsh_etal_2012}. \\
\indent We find a secondary star at $\Delta$RV = -81 $\rm{km\,s^{-1}}$, with the estimated effective temperature of 5600 $\pm$ 400 K (Figures \ref{KIC9837578_primary} and \ref{KIC9837578}). Taking into account the uncertainty, $T_{\rm{eff}}$ agrees with \citet{Welsh_etal_2012} value of 5202 $\pm$ 100 K. The estimated flux ratio for the two stars of 0.56 $\pm$ 0.14 is larger than the \citet{Welsh_etal_2012} value by 7\%. Some of this discrepancy can be attributed to slightly different wavelength bandpass of our spectra as compared to \citet{Welsh_etal_2012}, and some due to the mixing of the two sets of absorption lines as both stars are of relatively similar spectral types with the secondary star contributing a significant amount of the total flux.\\

\subsubsection{HD 61994}

\begin{figure}
	\plotone{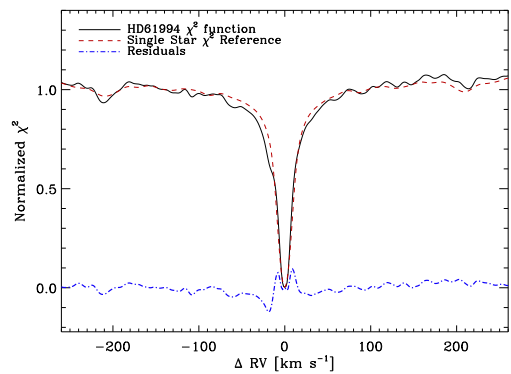}
	\caption{Primary star $\chi^2$ function for HD 61994. There is a slight asymmetry in the shape of the $\chi^2$ minimum, indicating a possible secondary star at $-$20 $\rm{km\,s^{-1}}$. The asymmetry alone, however, is not a conclusive evidence for the presence of another star, and we need to refer to the residual $\chi^2$ function for more details on a potential secondary.\\}
	\label{HD61994_primary}
\end{figure}

\indent HD 61994 is a double-lined spectroscopic binary with a period of 552.8 days \citep{Strassmeier_etal_2012}. When the spectrum was taken, the relative radial velocity for the two stars was only +18 $\rm{km\,s^{-1}}$, an interesting example for cases where the relative radial velocity is low and the secondary star is something other than an M-dwarf. Furthermore, the companion is faint - the visible brightness ratio for the two stars $\rm{F_{B}/F_{A}}$ is only 0.069 \citep{Strassmeier_etal_2012}. \\
\indent The primary $\chi^2$ deviates slightly from the characteristic single-star $\chi^2$ function at $\Delta$RV $\approx$ -20 $\rm{km\,s^{-1}}$, shown in Figure \ref{HD61994_primary}. However, more convincing evidence for the presence of a secondary star comes from the residual $\chi^2$ function shown in Figure \ref{fig:complete_companion_plot}, indicating a secondary star at $\Delta$RV = $-$16 $\rm{km\,s^{-1}}$. Its residual $\chi^2$ minimum distribution function sets the secondary star $T_{\rm eff}$ at 4200 $\pm$ 650 K. Within the uncertainty limit, this agrees with the \citet{Strassmeier_etal_2012} value of $T_{\rm{eff}}$ = 4775 $\pm$ 150 K. The calculated flux ratio, $\rm{F_{B}/F_{A}}$ = 0.055 $\pm$ 0.022, also matches the flux ratio of $\rm{F_{B}/F_{A}}$ = 0.069, quoted by \citet{Strassmeier_etal_2012}.\\

\subsubsection{KOI-54}
\label{ex:koi-54}

\begin{figure}
	\plotone{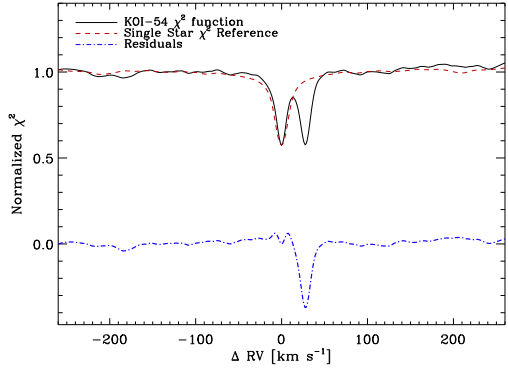}
	\caption{Primary star $\chi^2$ function for KOI-54. $\chi^2$ minimum is shallow compared to other examples, because the NSO Stellar spectrum used to construct the $\chi^2$ functions differs significantly from a spectrum of a 8800 K star. The companion detected at $\approx$ +30 $\rm{km\,s^{-1}}$ appears equally bright.\\}
	\label{KOI-54_primary}
\end{figure}

\begin{figure}
	\plotone{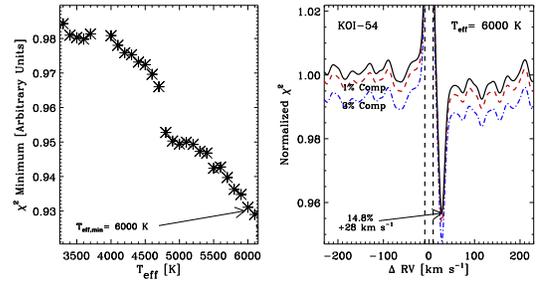}
	\caption{Final secondary star plot for KOI-54. Same as Figure \ref{fig:complete_companion_plot}. A secondary star is detected at +28 $\rm{km\,s^{-1}}$ $\Delta\rm{RV}$, with an estimated $T_{\rm eff}$ of 6200 K, and contributing 17.4\% to the total flux. This low relative brightness is surprising, as the two minima appeared almost equal in Figure \ref{KOI-54_primary}. For this system, however, both constituent stars are $\sim$ 2000 K hotter than the hottest star in the SpecMatch library, thus our derived parameters are highly inaccurate.\\}
	\label{KOI-54}
\end{figure}

\indent KOI-54 is a non-eclipsing, SB2 binary system. The two stars are in an almost face-on, highly eccentric orbit \citep{Welsh_etal_2011}. \\
\indent This example shows the ability to detect the presence of both stars even when their effective temperatures are outside the range of our library and above the highest SpecMatch library spectrum temperature used to construct the residual $\chi^2$ panels. The primary star has a temperature $T_{\rm{eff}}$ = 8800$\pm$200 K, and the companion lags behind this value only slightly with $T_{\rm{eff}}$ = 8500$\pm$200 K \citep{Welsh_etal_2011}. This is over 2000 K higher than any of our library spectra. \\
\indent Nonetheless, we are able to see the minimum in the $\chi^2$ function shown in Figure \ref{KOI-54_primary} due to each of the two stars, as well as detect a clear minimum indicating the presence of the companion after the primary star was subtracted, as seen in Figure \ref{KOI-54}. We don't state the parameters obtained for KOI-54 system, since they are incorrect due to the large discrepancy between the stellar spectra for the KOI-54 stars and the available library spectra; more on this issue is explained in Section \ref{sec:too_high_teff}.\\

\subsubsection{HD 16702}

\begin{figure}
	\plotone{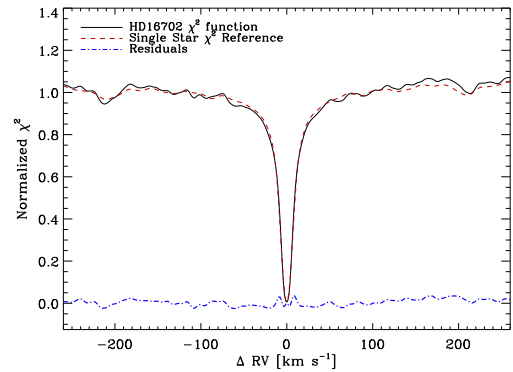}
	\caption{Primary star $\chi^2$ function for HD 16702. Since the companion is really faint and separated only by 5 $\rm{km\,s^{-1}}$ from the primary star, the primary $\chi^2$ function alone shows no evidence of spectrum binarity.}
	\label{HD16702_primary}
\end{figure}

\begin{figure}
	\plotone{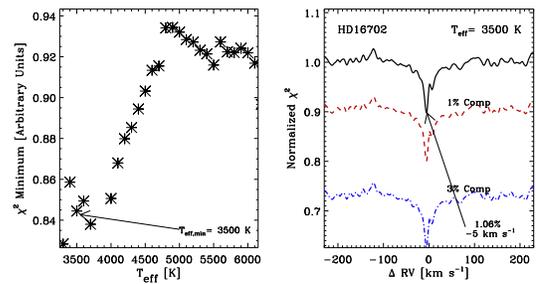}
	\caption{Final secondary star plot for HD 16702. Same as Figure \ref{fig:complete_companion_plot}. Despite its low relative brightness and small RV separation, the companion is detected at $\Delta\rm{RV}$ = $-$5 $\rm{km\,s^{-1}}$. The estimated companion temperature is 3500 $\pm$ 250 K, and it contributes 1.06\% of the total flux.}
	\label{HD16702}
\end{figure}

\indent HD 16702 is identified as a binary star in \citet{Diaz_etal_2012}, with the mass ratio for the companion and the primary star is $\rm{M_{B}/M_{A}}$ = 0.35$-$0.41 \citep{Diaz_etal_2012}. We detect a secondary star at $\Delta\rm{RV}$ = $-$5 $\rm{km\,s^{-1}}$, with an estimated $T_{\rm eff}$ = 3500 $\pm$ 250 K and contributing 1.06\% of the total flux, as shown in Figures \ref{HD16702_primary} and \ref{HD16702}.\\
\indent Since the only known companion parameter is its mass, we performed some analysis to confirm that our detection agrees with \citet{Diaz_etal_2012} value of 0.40 $\pm$ 0.05 $\rm{M_{\odot}}$. Since the spectrum was taken when the velocity of both stars was almost exclusively along the line of sight, we can predict the mass of the companion based on their radial velocity separation. Using 9 precise RV measurements \citep{Marcy_etal_1992}, we determine the primary star velocity relative to the center of mass of the binary system to be at its maximum at the time of observation, $v_{p}$ = -2.2 $\rm{km\,s^{-1}}$. Using the primary mass of 0.98 $\rm{M_{\odot}}$ \citep{Diaz_etal_2012}, and $\Delta\rm{RV}$ of $-$5 $\rm{km\,s^{-1}}$, as determined by our algorithm, we predict the mass ratio $\rm{M_{B}/M_{A}}$ = 0.30. This indicates a companion mass of $\sim$0.3 $\rm{km\,s^{-1}}$, only slightly below the \citet{Diaz_etal_2012} value. Using known calibrations, both the flux ratio of $\sim$0.01 and the companion temperature of $\sim$3500 K are consistent with a star whose mass is approximately 0.4 $\rm{M_{\odot}}$.\\

\section{Application: 1160 {\it Kepler} Objects}
\label{sec:Kepler_survey}

\indent We have carried out a spectroscopic survey of the 1160 {\it Kepler} Objects of Interest (KOI) that are brighter than {\it Kepler} magnitude of 14.2, announced as of September 2013 by the {\it Kepler} Project and listed in the Exoplanet Archive\footnote[1]{http://exoplanetarchive.ipac.caltech.edu}. A complete description of the 1160 KOIs and our spectra of them are provided by Howard et al. (2014, in prep). We obtained a Keck-HIRES spectrum with a SNR of $\sim$45 per pixel and a resolution of R = 60,000 from 3800 $\rm\AA$ to 8000 $\rm\AA$ for each one. All spectra are available to the public on the Keck Observatory Archive.\\
\indent In Table \ref{table:CK0s_results} we identify the KOIs where a secondary star was detected by our algorithm here, and provide the deduced parameters of both the primary star and the companion. The uncertainties are determined using Table \ref{table:uncertainties_companion}, which were computed using the injection-recovery tests. Companion plots for identified KOI binary systems are shown in Appendix, Figures \ref{fig:KOI-5} through \ref{fig:KOI-4871}.\\
\indent Out of 1160 KOIs, 60 KOIs show clear evidence of companion stars using our spectroscopic algorithm. Additionally, we report three detections that were marginal. Among all of them, we found 41 KOIs that have one planetary candidate (one validated as a planet), three KOIs that have two planetary candidates, and one KOI with three planetary candidates. In addition, 15 of the 60 KOIs had been marked as false positives in the Exoplanet Archive. \\
\indent Of course, these spectroscopic detections of secondary stars do not necessarily imply that the planet candidate is false. Instead, each detection merely indicates that there is either a background star or a bound companion to the target KOI. In either case, the transit lightcurve could be caused by a planet transiting the primary star, albeit with a diluted light curve, or an orbiting object that blocks the light of the secondary star, either bound or background.\\
\indent The relative brightness of the secondary and the primary star can be used to compute the dilution factor of the {\it Kepler} light curves. Our wavelength domain encompasses approximately the V and R bandpasses of the classical photometry, and the relative brightness can thus be converted to the corresponding value in the {\it Kepler} bandpass, and used to determine a more accurate radius for the planet.\\
\indent Some of the KOIs had secondary stars hotter than the primary star, but contributed only a few percent of the total flux. There are three possible explanations for such cases. First, the primary star might actually have already evolved off main sequence, and became a red giant. In this case, it would be possible that it is in fact cooler than the secondary, if the secondary is still on main sequence. Second possible explanation is that only a fraction of the flux from the secondary actually entered the spectrometer slit, thus not all of the flux from the secondary star is collected by the spectrometer. A third alternative is that the secondary star is simply a background star much further away.\\

\subsection{KOI Companion Detections}

This section is intended to be used in conjunction with the list of detected companions in Table \ref{table:CK0s_results}, providing additional information relevant for selected KOIs. We also list the number of confirmed planets, planetary candidates, or indicate whether the KOI was identified as a false positive.

\subsubsection{Confirmed Planets}
\paragraph{KOI-1361}
Also known as Kepler-61, it has one confirmed planet \citep{Ballard_etal_2013}. We determined the primary star temperature to be $T_{\rm eff}$ = 4265 $\pm$ 145, which is $\sim$100 K larger compared to \citet{Ballard_etal_2013}. We identified an even cooler companion, with $T_{\rm eff}$ = 3600 $\pm$ 130 K, at a radial velocity of + 40 $\rm{km\,s^{-1}}$ relative to the primary star. \citet{Ballard_etal_2013} reports a companion 0.94" north of the primary star, detected using adaptive optics. The magnitude difference between the primary star and the companion, as reported by \citet{Ballard_etal_2013}, is 2.98 mag in {\it Kepler} bandpass. This corresponds to the flux ratio of $\rm{F_{A}/F_{B}}$ = 0.064, which is only slightly above the value deduced by our algorithm, $\rm{F_{A}/F_{B}}$ = 0.022 $\pm$ 0.003. Some of this difference can be attributed to the difference in the {\it Kepler} bandpass regions as compared to the regions used in our algorithm. It is also possible that the spectrometer slit was not necessarily aligned to capture all of the light from the companion, thus the flux ratio is correspondingly lower. 

\subsection{Secondary Stars to KOIs with Three Planet Candidates}

\paragraph{KOI-2311}
Due to a low RV separation of +11 $\rm km\,s^{-1}$, the flux ratio of 0.276 $\pm$ 0.069 is a lower limit. The final secondary star plot is shown in Figure \ref{fig:KOI-2311}.

\subsubsection{Secondary Stars to KOIs with Two Planet Candidates}

\paragraph{KOI-5}
The observed RV trend indicates a possibility of a stellar companion \citep{Wang_etal_2014}. Low $\rm{\Delta{RV}}$ of +11 $\rm km\,s^{-1}$ might have caused some of the secondary absorption lines to be subtracted away with the primary star, thus the quoted flux ratio of 0.066 $\pm$ 0.020 is a lower limit rather than an actual value. We analyzed several different observations of the same star, and the RV separation varied with observation time. This indicates that the secondary and the primary star form a bound system. The final companion plot is shown in Figure \ref{fig:KOI-5}.
\paragraph{KOI-354}
Detection of the secondary star to the KOI-354 is marginal, thus we don't list in in the Table \ref{table:CK0s_results}. The system consists of a 5936 $\pm$ 102 primary star (cfop.ipac.caltech.edu) and an M-dwarf secondary with $T_{\rm eff}$ = 3500 $\pm$ 250 K. The secondary star has a low relative radial velocity of +11 $\rm{km\,s^{-1}}$, and thus the flux ratio of $\rm{F_{A}/F_{B}}$ = 0.005 $\pm$ 001 is a lower limit. The secondary star plot for KOI-354 is shown in Figure \ref{fig:KOI-354}.
\paragraph{KOI-1613}
Due to a low RV separation of +10 $\rm km\,s^{-1}$, the flux ratio of 0.044 $\pm$ 0.013 is a lower limit. \citet{Law_etal_2013} detected a secondary star at a separation of 0.22", 1.3 magnitudes fainter than the primary star. 1.3 magnitude difference indicates a flux ratio of 0.30, much higher than our lower limit. Since they observed the system in a different bandpass (roughly 7000-8000$\rm{\AA}$), this flux discrepancy can be somewhat due to a different flux ratios in different wavelength regions. However, we suspect that most of the discrepancy can be attributed to the low $\Delta\rm{RV}$ separation of the two stars of similar spectral types. At 10 $\Delta\rm{RV}$ = $\rm{km\,s^{-1}}$, the two sets of absorption lines are barely resolved, and most likely a large fraction of the secondary star was subtracted away together with the primary, leaving only a weak signal in the residuals. The final secondary star plot is shown in Figure \ref{fig:KOI-1613}.
\paragraph{KOI-2059}
M-dwarf secondary star has a small RV separation of only +5 $\rm km\,s^{-1}$, making the flux ratio of 0.016 $\pm$ 0.008 a lower limit. Nevertheless, due to a significantly different secondary spectral type of the secondary compared to the primary with $T_{\rm eff}$ of 5996 $\pm$ 103 K (cfop.ipac.caltech.edu), we don't expect the actual flux ratio to deviate significantly from this lower limit. Using AO, a companion was detected at a separation of 0.38", 1.1 magnitude fainter \citep{Law_etal_2013}. A magnitude difference of 1.1 predicts the flux ratio of about 0.36, which is again larger than the ratio predicted by our algorithm. Again, since the \citet{Law_etal_2013} observation was made in the near-infrared region, where the flux of the M-dwarfs is considerably larger, we do expect that the flux ratio of an M-dwarf and a $\sim$5000 K primary star is higher as compared to the ratio in the V and R bandpasses. Therefore, our flux ratio is still consistent with the \citet{Law_etal_2013} value, especially when accounting for the fact that the ratio predicted by our algorithm as a lower limit. The final secondary star plot is shown in Figure \ref{fig:KOI-2059}.

\subsubsection{Companions to KOIs with One Planet Candidate}

\indent Here we provide notes about selected KOIs that have one planet candidate, and an indication from our spectroscopic analysis of a secondary set of absorption lines in the spectrum. A complete listing of all such detections is in Table \ref{table:CK0s_results}.\\
\paragraph{KOI-151}
Due to a low $\Delta{\rm{RV}}$ separation of +14 $\rm km\,s^{-1}$, the flux ratio of 0.012 $\pm$ 0.006 represents a lower limit for KOI-151 and its secondary star. The final secondary star plot is shown in Figure \ref{fig:KOI-151}.
\paragraph{KOI-219}
Both the flux ratio of 0.330 $\pm$ 0.033 as well as the secondary star $T_{\rm eff}$ of 6000 $\pm$ 100 K represent lower limits due to a restricted $T_{\rm eff}$ library range and low RV separation, respectively. The final secondary star plot is shown in Figure \ref{fig:KOI-219}.
\paragraph{KOI-652}
KOI-652 is at least a triple star system, with a possible marginal detection of the fourth star. All three companions are M-dwarfs, each contributing only a few percent to the total flux. The final companion plot is shown in Figure \ref{fig:KOI-652}. In Table \ref{table:CK0s_results}, we don't list the faintest companion at $-$44 $\rm{km\,s^{-1}}$, as its detection is only marginal.
\paragraph{KOI-698}
This system is analyzed in \citet{Santerne_etal_2012}. Assuming a main sequence companion, our $T_{\rm eff}$ estimate for the secondary star of 4800 $\pm$ 600 K indicates a secondary mass of 0.7 - 0.8 $\rm{M_{\odot}}$, which is below the predicted value of 1.1 $\pm$ 0.1 $\rm{M_{\odot}}$ by \citet{Santerne_etal_2012}. The final secondary star plot is shown in Figure \ref{fig:KOI-698}.
\paragraph{KOI-716}
The detection of KOI-716 is marginal, and we don't include it in the Table \ref{table:CK0s_results}. KOI-716 secondary star is extremely faint and at the threshold of detection, with a flux ratio of $\rm{F_{A}/F_{B}}$ = 0.004 $\pm$ 0.001. The primary star is has a $T_{\rm eff}$ = 6096 $\pm$ 150 K (cfop.ipac.caltech.edu), and the secondary star is an M-dwarf with $T_{\rm eff}$ = 3500 $\pm$ 250 K. Due to a relatively low RV separation of +18 $\rm{km\,s^{-1}}$, the flux ratio might be underestimated, yet we don't expect the secondary star contribution to be larger than 1\% of the total flux. KOI-716 has 1 planetary candidate, and the secondary star plot is shown in Figure \ref{fig:KOI-716}. 
\paragraph{KOI-984}
AO imaging of KOI-984 revealed a companion of approximately equal magnitude 1.5 arcseconds apart \citep{Law_etal_2013}. Our algorithm, however, detects a marginal secondary star at $\Delta\rm{RV}$ = +33 $\rm{km\,s^{-1}}$ with $T_{\rm eff}$ = 3500 $\pm$ 150 K, contributing less than 1\% to the total flux. This suggests that only a small fraction of the light from secondary star star actually entered the slit, and while we can confirm the possibility of the companion, its estimated parameters are highly inaccurate. The final secondary star plot is shown in Figure \ref{fig:KOI-984}. Since the detection is marginal, we don't list it in Table \ref{table:CK0s_results}.
\paragraph{KOI-1020}
The existence of the companion already speculated by \citep{Hirano_etal_2012}. The final secondary star plot is shown in Figure \ref{fig:KOI-1020}.
\paragraph{KOI-1137}
Despite low RV separation of +11 $\rm km\,s^{-1}$, the two stars are of sufficiently distinct spectral types that we don't expect the flux ratio to be significantly higher than the calculated lower limit of 0.020 $\pm$ 0.004. The final secondary star plot is shown in Figure \ref{fig:KOI-1137}.
\paragraph{KOI-1452}
The primary star has $T_{\rm eff}$ = 7172 $\pm$ 211 K (cfop.ipac.caltech.edu), which is outside our library range. While the binarity of the spectrum is clear, the parameters for the secondary star cannot be determined accurately enough to be published. This is mostly due to a high rotational broadening of the secondary absorption lines, indicating a secondary star with a spectral type outside the range of our SpecMatch library. The final secondary star plot is shown in Figure \ref{fig:KOI-1452}.
\paragraph{KOI-1784}
Due to a low RV separation of $-$13 $\rm km\,s^{-1}$, the flux ratio of 0.192 $\pm$ 0.058 represents a lower limit. The final secondary star plot is shown in Figure \ref{fig:KOI-1784}.
\paragraph{KOI-2075}
The secondary star is faint, contributing less than 1\% to the total flux. While the flux ratio of 0.008 $\pm$ 0.004 is a lower limit due to a low RV separation of $-$13 $\rm km\,s^{-1}$, we don't expect that a large fraction of the secondary absorption lines got subtracted away together with the primary star, as the two spectral types differ significantly from each other. The final secondary star plot is shown in Figure \ref{fig:KOI-2075}.
\paragraph{KOI-2215}
This is a triple-lined spectrum, with both secondary and tertiary stars clearly detectable. Both additional stars are treated separately, but the uncertainties on the parameters might be slightly higher due to the blending of spectral lines as the spectral types of all three stars are relatively similar. The flux ratio for the third star of 0.155 $\pm$ 0.047 is a lower limit due to a low RV separation of $-$15 $\rm km\,s^{-1}$. The final secondary star plot is shown in Figure \ref{fig:KOI-2215}.
\paragraph{KOI-2457}
The primary star effective temperature of 6728 $\pm$ 155 (cfop.ipac.caltech.edu), exceeds our temperature range of the SpecMatch library. This hinders our ability to determine accurate parameters for the secondary star at a low $\Delta\rm RV$ separation of +14 $\rm km\,s^{-1}$. Nevertheless, our analysis indicates a secondary star with a lower limit flux ratio of 0.076 $\pm$ 0.023. The final secondary star plot is shown in Figure \ref{fig:KOI-2457}.
\paragraph{KOI-3471}
The spectrum of KOI-3471 contains four stars, all but one of comparable brightness. We provide estimates for the parameters for each of the constituent stars, but we expect larger errors due to significant blending of their spectral lines. Since all of the stars are contained within less than 80 $\rm{km\,s^{-1}}$ $\Delta\rm{RV}$, we expect that the temperatures as well as the flux ratios for additional stars are affected by the presence of the other all stars. The final plot is shown in Figure \ref{fig:KOI-3471}.
\paragraph{KOI-3573}
Low RV separation of $-$15 $\rm km\,s^{-1}$ together with similar spectral types of the two stars suggest that the actual flux ratio is probably higher than the calculated lower limit of 0.109 $|pm$ 0.033. As the two sets of absorption lines overlap slightly, it is likely that a fraction of the secondary absorptions lines got subtracted away together with the primary star, thus making the secondary star appear fainter. The final secondary star plot is shown in Figure \ref{fig:KOI-3573}.
\paragraph{KOI-3583}
The secondary star $\chi^2$ minimum is significantly broadened, indicating rotational broadening of secondary star absorption lines. Thus, despite the RV separation of $-$23 $\rm{km\,s^{-1}}$, there is a possibility for the blending of the two sets of absorption lines, and the calculated flux ratio of 0.109 $\pm$ 0.033 is a lower limit. The final secondary star plot is shown in Figure \ref{fig:KOI-3583}. 
\paragraph{KOI-3606}
Both the secondary and the primary $\chi^2$ minimum are broadened, indicating a significant rotational broadening of the spectral lines. As this is characteristic of hot stars, we assume that the actual $T_{\rm eff}$ for the secondary star is above the quoted lower limit of 6000 $\pm$ 100 K. The final secondary star plot is shown in Figure \ref{fig:KOI-3606}.
\paragraph{KOI-3721}
The spectrum of KOI-3721 contained three sets of absorption lines, with the faintest star contributing only $\sim$1\% to the total flux. Brighter secondary star $\chi^2$ is broad, indicating significant rotational broadening. This is consistent with the estimate of the secondary star $T_{\rm eff}$ above 6000 $\pm$ 100 K. The final plot is shown in Figure \ref{fig:KOI-3721}.
\paragraph{KOI-3782}
The calculated flux ratio is a lower limit due to a low RV separation of $-$7 $\rm km\,s^{-1}$, but we don't expect a significant deviation from the stated flux ratio due to significantly different spectral types of the two stars. The final secondary star plot is shown in Figure \ref{fig:KOI-3782}.
\paragraph{KOI-4713}
Due to similar spectral types of the two stars as well as the small RV separation of $-$15$\rm km\,s^{-1}$, the calculated flux ratio of 0.344 $\pm$ 103 is a lower limit. The final secondary star plot is shown in Figure \ref{fig:KOI-4713}.
\paragraph{KOI-4871}
Despite the RV separation of $-$23 $\rm{km\,s^{-1}}$, we state the flux ratio of 0.012 $\pm$ 0.003 as the lower limit due to the similarity of the primary and secondary star spectral type. This similarity allows for a chance of overlap of some of the absorption lines, and some of the secondary absorption lines might have been subtracted away with the primary star. The final secondary star plot is shown in Figure \ref{fig:KOI-4871}.

\subsubsection{Confirmed False Positives}

\paragraph{KOI-1152}
KOI-1152 has a known companion, previously detected using AO \citep{Law_etal_2013}. The planet candidate identified as a false positive (cfop.ipac.caltech.edu). We analyzed several different observations of the same star, and the RV separation varied with different observation times. This indicates that the secondary and the primary star form a bound system. The final secondary star plot is shown in Figure \ref{fig:KOI-1152}.
\paragraph{KOI-3000}
Flux ratio of 0.055 $\pm$ 0.019 is a lower limit rather than an actual value due to a low RV separation of +10 $\rm km\,s^{-1}$. Nevertheless, the two stars have sufficiently different spectral types that we do not expect significant departure from the stated flux ratio. The final secondary plot is shown in Figure \ref{fig:KOI-3000}.
\paragraph{KOI-3035}
Despite moderate RV separation of +20 $\rm{km\,s^{-1}}$ both sets of spectral lines are rotationally broadened, causing the two sets of absorption lines to overlap slightly. Therefore, the flux ratio of 0.212 $\pm$ 0.064 is a lower limit rather than an actual value. The final secondary star plot is shown in Figure \ref{fig:KOI-3035}.
\paragraph{KOI-3162}
Flux ratio of 0.553 $\pm$ 0.166 is a lower limit due to a low RV separation of $-$13 $\rm km\,s^{-1}$. The final secondary star plot is shown in Figure \ref{fig:KOI-3162}.
\paragraph{KOI-3216}
Flux ratio of 0.326 $\pm$ 0.033 is a lower limit due to a low RV separation of +13 $\rm km\,s^{-1}$. The final secondary star plot is shown in Figure \ref{fig:KOI-3216}.
\paragraph{KOI-4355}
The secondary star to KOI-4355 exhibited a broad minimum, indicating significant rotational broadening. Due to shallower spectral lines, the uncertainties on the secondary star parameters might be somewhat larger. The final secondary star plot is shown in Figure \ref{fig:KOI-4355}.
\paragraph{KOI-4457}
The calculated flux ratio of 0.012 $\pm$ 0.003 is a lower limit due to a low RV separation of $-$15 $\rm km\,s^{-1}$, but we don't expect significant deviation from the stated value due to significantly different spectral types of the two stars. The final secondary star plot is shown in Figure \ref{fig:KOI-4457}. 

\section{Limitations}

\subsection{Relative Radial Velocity for the Secondary Star}

\indent Our code cannot detect secondary stars with $|\Delta\rm{RV}|\,<$ 10 $\rm{km\,s^{-1}}$, with the exception of M-dwarfs. For bound systems consisting of two solar type stars, this corresponds to the physical separation of at most $\sim$ 5 AU, assuming the orbital plane is viewed edge-on and that the spectrum was taken when the two stars were coming directly towards/away from us. On the other hand, if the phase of the two stars is such that their momentary velocity is close to tangential, our code will not be able to detect the presence of the secondary star, regardless of their relative brightness. Somewhat less strictly, we are also limited to the orbits with periods larger than $\sim$2.5 days, as the maximum detectable Doppler shift for the secondary star is at $\pm$200 ${\rm km\,s^{-1}}$. This limit is determined assuming that the two stars in the binary system are solar type, that the spectrum was taken when the two stars were coming directly towards/away from us, and that the orbit is viewed edge-on. Shorter orbital periods can thus be detected for binaries with lower mass companions, for orbits viewed at an angle, or when the spectrum was taken at a different orbital phase.\\
\indent For both bound companions and background secondary stars, whenever the spectral types of the two stars are similar and their relative radial velocity low ($\lesssim$20 $\rm km\,s^{-1}$, with the exact number depending upon the amount of the absorption line broadening in the spectrum), there is a possibility that some of the absorption lines of the secondary spectrum get subtracted away together with the primary star. In those cases, the flux of the secondary star is somewhat underestimated, and its parameters carry somewhat larger uncertainties.

\subsection{Effective Temperature}
\label{sec:too_high_teff}

\indent The SpecMatch library used for our algorithm contains stars with $T_{\rm{eff}}$ from 3200 K to 6550 K. For detecting the secondary star, the $T_{\rm eff}$ range is slightly narrower, ranging from 3300 K to 6100 K. This is due to the fact that we only construct the median library spectrum for a particular 100 K $T_{\rm eff}$ range when there are at least three different SpecMatch spectra within that range. For any primary star with $T_{\rm{eff}}$ outside the SpecMatch library range, the fit will not be exact, and the subtraction will be imperfect - this decreases the detectability of the secondary stars with $\Delta\rm RV\,\lesssim\,10\,km\,s^{-1}$. Furthermore, for any secondary star with $T_{\rm eff}$ much lower than 3300 K or much higher than 6100 K we cannot accurately estimate the percentage flux as the injected spectrum cannot match the actual secondary star spectrum. Due to these reasons, deduced $T_{\rm eff}$ and flux contribution for stars outside the SpecMatch parameter range carry larger uncertainties. Nevertheless the binarity of the spectrum can still be established with a sufficient degree of certainty (see KOI-54 example in Section \ref{ex:koi-54}). \\
\indent Since we cannot extrapolate the $\chi^2$ minima distribution function to temperatures lower than 3300 K, we cannot predict whether a secondary star whose effective temperature is estimated at $\sim$ 3300 K is actually cooler than this limiting temperature. Thus, the 3300 K secondary star $T_{\rm eff}$ is always more of an upper limit than an actual value. Similarly, we cannot extrapolate the behavior of the $\chi^2$ minima distribution to effective temperatures higher than $\sim$6000 K, and any $\sim$6100 K estimated secondary star $T_{\rm eff}$ is a thus a lower limit rather than the actual estimate. \\

\subsection{Rotational Broadening}
\label{sec:rotbro}

\indent Rotational broadening does not pose a major challenge to the identification of the primary star, as we artificially broaden the library spectrum's absorption lines to match the amount of broadening present in the spectrum. We can fit and subtract the primary absorption lines successfully regardless of the star's rotation. For the secondary star, however, any significant broadening of absorption lines impairs our ability to detect it. As the lines become broader, they become shallower, thus less distinct from the noise. This also causes a less sharp and distinct residual $\chi^2$ minimum, that cannot be as easily distinguished from accidental alignments.\\
\indent For cases where the secondary star has a significant rotational broadening but is sufficiently bright to enable detection, the estimated secondary temperature and its flux contribution become more uncertain. Since we do not artificially broaden the lines of the median library spectra used for the search of the secondary star, even the spectrum with $T_{\rm eff}$ matching that of the actual secondary star is a poor match. Thus, since none of the median library spectra match the secondary star well, the differences in the $\chi^2$ minima are smaller among the $T_{\rm eff}$ values for the median library spectra, and the residual $\chi^2$ minimum distribution's minimum is harder to identify correctly. Furthermore, shallower absorption lines cause the underestimation of the flux contribution, and the flux ratio for the two stars is underestimated as well. \\ 

\subsection{Fractional Flux Contribution from the Secondary Star}
\label{sec:ffcont}

\indent For the secondary star to be detectable, its fractional contribution most often needs to reach a value that is at least one half of the noise; for example, a 1\% flux from the secondary star for a spectrum with 2\% can be detected. This brings the SNR of the residuals after the subtraction of the primary star to 0.5. The exact value also depends on the spectral differences between the two stars, as well as their relative radial velocity. \\
\indent At the other end, secondary stars with fractional contributions comparable to that of their primary stars pose a challenge as well. While those stars are easily detectable, the determination of the properties for the secondary star is limited by the pollution of one set of absorption lines by the other. This effect is especially apparent when their relative radial velocity is low, with one set of lines encroaching on the other significantly. In those cases, both the primary star fit as well as the secondary star parameters suffer by this mutual pollution of their respective absorption lines. Since some of the secondary absorption lines get subtracted away together with the primary star, this also causes the calculated flux ratio for the two stars to be below the true value.\\

\section{Conclusion}
\label{sec:conclusion}

\indent Using the algorithm described in this paper we were able to detect secondary stars in the Keck-HIRES spectra that contributed as low as 1\% of the total flux in the spectrum with a 2\% Poisson noise. These secondary stars can be angularly separated from the primary star by at most 0.43" to 1.5", depending on the orientation of the aperture slit. For detection, the relative radial velocity for the two stars needs to exceed 10 $\rm{km\,s^{-1}}$, except for the case of the G-type primary star and an M-dwarf. For G-M pairs, there is no restriction on the radial velocity separation for the two stars, given that the two sets of absorption lines are not overlapping exactly; for these, we were able to detect secondaries as faint as 0.5\% relative to the primary star in a spectrum with a 2\% Poisson noise. Our results suggests the detectability threshold SNR for the secondary star of 0.5 for any spectral type combination of the two stars, and 0.25 for M-dwarf secondaries to G-type primary stars. \\
\indent For cases where the secondary star is detected, its effective temperature and the fractional flux contribution can be determined. The uncertainties in the derived parameters depend on the relative brightness of the secondary star as well as on the spectral types of the two stars in the spectrum. These values are summarized in Table \ref{table:uncertainties_companion}. For the cases of non-detections, we place an upper limit for the brightness of any potential hidden companions.\\
\indent Currently, we are limited to stars with the effective temperature above 3300 K and below 6500 K; for the stars outsides this temperature range, the presence of the sufficiently bright secondary star can still be evaluated, but the derived parameters carry larger uncertainties. The parameter errors also increase when the two stars have comparable flux contributions, or when their spectral types are similar. Our results are most accurate for the cases where the relative radial velocity of the two stars exceeds 10 $\rm{km\,s^{-1}}$, both stars have $T_{\rm{eff}}$ between 3000 K and 6000 K, and the secondary star contributes less than 20\% of the total flux for the system. The recovery rate for a given radial velocity separation and relative brightness of the secondary star is the highest for M-dwarfs, given that the primary stars is of a different spectral type.\\
\indent This spectroscopic algorithm can be useful for transiting planets towards recognizing multiple potential host stars for the planet and for computing photometric dilution in the analysis of the planet. As an application of the algorithm, we analyzed 1160 {\it Kepler} objects of interest and found and evidence of at least one additional star in 63 KOIs. A secondary star does not imply that the planet is false, but merely indicates that there is another star present angularly close to the primary star, either bound or background. Our estimate of the fractional flux contribution from the secondary star for each of these 63 KOIs can be used to compute the dilution factor for their {\it Kepler} light curves, and help to determine a more accurate radius of the planet.\\

\section*{Acknowledgments}

\indent We thank John Johnson for the CalTech Keck Telescope time to obtain many of the spectra of the KOIs. We also thank John Brewer and Debra Fischer for their preliminary estimates of the stellar parameters for the Specmatch library stars, as their parameters are an improvement over the values in \citet{Valenti_Fischer_2005}, with a published paper in preparation. G. Marcy, the Alberts Chair at UC Berkeley, would like to thank the Marilyn and Watson Alberts for funding that made this research possible. We also thank the John Templeton Foundation and NASA grant NNX11AK04A for funding this research. We thank the government of the Republic of Slovenia and its Slovene Human Resources and Scholarship Fund and its Ad futura programme that helps fund R. Kolbl.\\
\indent {\it Kepler} was competitively selected as the tenth NASA Discovery mission. Funding for this mission is provided by the NASA Science Mission Directorate. Some of the data presented herein were obtained at the W. M. Keck Observatory, which is operated as a scientific partnership among the California Institute of Technology, the University of California, and the National Aeronautics and Space Administration. The Keck Observatory was made possible by the generous financial support of the W. M. Keck Foundation. \\
\indent The authors  would like to thank the many people who gave so generously of their time to make NASA {\it Kepler} Mission a success. All spectra used in this paper are available at the Keck Observatory Archive. All {\it Kepler} data products are available to the public at the Mikulski Archive for Space Telescopes {\url{http://stdatu.stsci.edu/kepler}} and the spectra and their products are made available at the NExSci Exoplanet Archive and its CFOP website: {\url{http://exoplanetarchive.ipac.caltech.edu}} We thank the many observers who contributed to the measurements reported here. We gratefully acknowledge the efforts and dedication of the Keck Observatory staff, especially Scott Dahm, Hien Tran, and Grant Hill for support of HIRES and Greg Wirth for support of remote observing. This work made use of the SIMBAD database (operated at CDS, Strasbourg, France) and NASA's Astrophysics Data System Bibliographic Services. This research has made use of the NASA Exoplanet Archive, which is operated by the California Institute of Technology, under contract with the National Aeronautics and Space Administration under the Exoplanet Exploration Program. Finally, the authors wish to extend special thanks to those of Hawai`ian ancestry on whose sacred mountain of Mauna Kea we are privileged to be guests. Without their generous hospitality, the Keck observations presented herein would not have been possible.

\begin{deluxetable*}{ccccccc}[H]
	\tabletypesize{\footnotesize}
	\tablecaption{Stellar parameters for injected SpecMatch library spectra
	\label{table:injected_properties}}
	\tablewidth{0pt}
	\tablehead{\colhead{Name} & \colhead{$T_{\rm eff}$ [K]} & \colhead{log$g$ [mag]} & \colhead{M [$\rm{M_{\odot}}$]} &  \colhead{R [$\rm{R_{\odot}}$]} & \colhead{$[\rm{Fe/H}]$ [mag]} & \colhead{Source}}
	\startdata
	GL 273 & 3293 & \nodata & 0.29 & 0.31 & $-$0.17 & \citet{Rojas-Ayala_etal_2012} \\
	GL 687 & 3395 & \nodata & 0.40 & 0.40 & $-$0.09 & \citet{Rojas-Ayala_etal_2012} \\
	GL 408 & 3526 & \nodata & 0.38 & 0.38 & $-$0.09 & \citet{Rojas-Ayala_etal_2012} \\
	GL 250B & 3569 & \nodata & 0.45 & 0.43 & 0.01 & \citet{Rojas-Ayala_etal_2012} \\
	GL 686 & 3693 & \nodata & 0.45 & 0.43 & $-$0.28 & \citet{Rojas-Ayala_etal_2012} \\
	HIP 24284 & 3992 & 4.88 & 0.41 & 0.36 & $-$0.49 & SME Analysis \\
	HIP 41689 & 4088 & 4.83 & 0.47 & 0.39 & $-$0.45 & SME Analysis \\
	HD 217357 & 4200 & 4.74 & 0.55 & 0.46 & $-$0.17 & SME Analysis \\
	HIP 105341 & 4298 & 4.72 & 0.58 & 0.55 & $-$0.05 & SME Analysis \\
	HIP 99205 & 4397 & 4.70 & 0.63 & 0.59 & $-$0.18 & SME Analysis \\
	HIP 36551 & 4501 & 4.69 & 0.65 & 0.60 & $-$0.30 & SME Analysis \\
	HIP 103650 & 4602 & 4.68 & 0.69 & 0.63 & $-$0.04 & SME Analysis \\
	HIP 63762 & 4701 & 4.62 & 0.74 & 0.69 & 0.09 & SME Analysis \\
	HD 220221 & 4797 & 4.58 & 0.79 & 0.75 & 0.18 & SME Analysis \\
	HD 51866 & 4906 & 4.59 & 0.80 & 0.75 & 0.13 & SME Analysis \\
	HD 23356 & 4988 & 4.60 & 0.78 & 0.73 & $-$0.04 & SME Analysis \\
	HD 216520 & 5097 & 4.55 & 0.78 & 0.77 & $-$0.19 & SME Analysis \\
	HD 205855 & 5204 & 4.53 & 0.86 & 0.83 & 0.05 & SME Analysis \\
	HD 75732 & 5295 & 4.49 & 0.97 & 0.92 & 0.39 & SME Analysis \\
	HD 58727 & 5399 & 4.52 & 0.97 & 0.90 & 0.24 & SME Analysis \\
	HD 147750 & 5496 & 4.52 & 0.90 & 0.85 & $-$0.11 & SME Analysis \\
	HD 20619 & 5600 & 4.42 & 0.87 & 0.94 & $-$0.25 & SME Analysis \\
	HD 12661 & 5699 & 4.38 & 1.09 & 1.11 & 0.35 & SME Analysis \\
	HD 148284 & 5799 & 4.39 & 1.11 & 1.10 & 0.29 & SME Analysis \\
	HD 205351 & 5896 & 4.29 & 1.11 & 1.23 & 0.09 & SME Analysis \\
	HD 27859 & 5997 & 4.41 & 1.14 & 1.09 & 0.14 & SME Analysis \\
	HD 48682 & 6104 & 4.36 & 1.18 & 1.18 & 0.11 & SME Analysis \\
	\enddata
	\tablecomments{No precise log$g$ values are available for M-dwarfs.}
\end{deluxetable*}

\begin{deluxetable*}{ccccccc}[H]
	\tabletypesize{\footnotesize}
	\tablecaption{Percentage recovery rates for injection-recovery experiment
	\label{table:companion_recovery_rate}}
	\tablewidth{0pt}
        \tablehead{
	Primary $T_{\rm eff}$ & \multicolumn{6}{c}{Secondary $T_{\rm eff}$ [K]} \\
	 & \colhead{3500} & \colhead{4000} & \colhead{4500} & \colhead{5000} & \colhead{5500} & \colhead{6000}}
	\startdata
	 & \multicolumn{6}{c}{1\% Secondary}\\
	3500 & 50\% & 40\% & 40\% & 60\% & 60\% & 60\% \\
	4000 & 70\% & 70\% & 70\% & 70\% & 70\% & 70\% \\
	4500 & 80\% & 70\% & 80\% & 80\% & 70\% & 70\% \\
	5000 & 80\% & 80\% & 70\% & 70\% & 80\% & 70\% \\
	5500 & 100\% & 90\% & 90\% & 100\% & 100\% & 80\% \\
	6000 & 80\% & 70\% & 80\% & 70\% & 80\% & 80\% \\
	 & \multicolumn{6}{c}{3\% Secondary} \\
	3500 & 100\% & 90\% & 100\% & 100\% & 100\% & 100\% \\
	4000 & 90\% & 90\% & 90\% & 90\% & 90\% & 90\% \\
	4500 & 100\% & 100\% & 100\% & 100\% & 100\% & 90\% \\
	5000 & 90\% & 100\% & 100\% & 100\% & 90\% & 90\% \\
	5500 & 100\% & 100\% & 100\% & 100\% & 100\% & 100\% \\
	6000 & 90\% & 100\% & 100\% & 100\% & 90\% & 90\% \\
	 & \multicolumn{6}{c}{10\% Secondary} \\
	3500 & 100\% & 100\% & 100\% & 100\% & 100\% & 100\% \\
	4000 & 100\% & 100\% & 100\% & 100\% & 100\% & 100\% \\
	4500 & 100\% & 100\% & 100\% & 100\% & 100\% & 100\% \\
	5000 & 100\% & 100\% & 100\% & 100\% & 100\% & 100\% \\
	5500 & 100\% & 100\% & 100\% & 100\% & 100\% & 100\% \\
	6000 & 100\% & 100\% & 100\% & 100\% & 100\% & 100\% \\
	\enddata
\end{deluxetable*}

\begin{deluxetable*}{ccccccc}[H]
	\tabletypesize{\footnotesize}
	\tablecaption{Systematic errors and uncertainties for the estimated secondary star temperature 
	\label{table:injection-recovery_companion_temperature}}
	\tablewidth{0pt}
        \tablehead{
	Primary $T_{\rm eff}$ [K] & \multicolumn{6}{c}{Secondary $T_{\rm eff}$ [K]} \\
	 & \colhead{3500} & \colhead{4000} & \colhead{4500} & \colhead{5000} & \colhead{5500} & \colhead{6000}}
	\startdata
	 & \multicolumn{6}{c}{1\% Secondary} \\
	3500 & 75 $\pm$ 70 & 565 $\pm$ 200 & 30 $\pm$ 695 & $-$245 $\pm$ 415 & 45 $\pm$ 465 & 500 $\pm$ 465 \\
	4000 & $-$45 $\pm$ 155 & 175 $\pm$ 440 & $-$55 $\pm$ 530 & $-$470 $\pm$ 440 & $-$230 $\pm$ 320 & 180 $\pm$ 330 \\
	4500 & 0 $\pm$ 125 & 195 $\pm$ 455 & 80 $\pm$ 220 & $-$285 $\pm$ 260 & $-$355 $\pm$ 70 & 115 $\pm$ 180\\
	5000 & $-$85 $\pm$ 155 & 500 $\pm $ 125 & 230 $\pm$ 135 & $-$205 $\pm$ 230 & $-$14- $\pm$ 390 & 50 $\pm$ 20 \\
	5500 & 5 $\pm$ 155 & 230 $\pm$ 440 & 140 $\pm$ 235 & $-$260 $\pm$ 235 & $-$375 $\pm$ 95 & 60 $\pm$ 25\\
	6000 & $-$85 $\pm$ 155 & 500 $\pm$ 125 & 230 $\pm$ 125 & $-$205 $\pm$ 230 & $-$140 $\pm$ 390 & 50 $\pm$ 20 \\
	 & \multicolumn{6}{c}{3\% Secondary} \\
	3500 & $-$60 $\pm$ 185 & 470 $\pm$ 340 & 200 $\pm$ 355 & $-$85 $\pm$ 410 & $-$180 $\pm$ 360 & 85 $\pm$ 65 \\
	4000 & $-$5 $\pm$ 130 & 380 $\pm$ 420 & 195 $\pm$ 325 & $-$80 $\pm$ 480 & $-$255 $\pm$ 440 & 205 $\pm$ 450 \\
	4500 & 25 $\pm$ 135 & 385 $\pm$ 370 & 250 $\pm$ 350 & $-$195 $\pm$ 295 & $-$355 $\pm$ 145 & 50 $\pm$ 150 \\
	5000 & 0 $\pm$ 135 & 80 $\pm$ 370 & 250 $\pm$ 250 & $-$210 $\pm$ 385 & 225 $\pm$ 435 & 330 $\pm$ 495 \\
	5500 & $-$5 $\pm$ 160 & 375 $\pm$ 255 & 225 $\pm$ 255 & $-$220 $\pm$ 320 & $-$375 $\pm$ 105 & 50 $\pm$ 150 \\
	6000 & 0 $\pm$ 135 & 80 $\pm$ 375 & 250 $\pm$ 250 & $-$210 $\pm$ 370 & 205 $\pm$ 435 & 330 $\pm$495 \\
	 & \multicolumn{6}{c}{10\% Secondary} \\
	3500 & $-$15 $\pm$ 145 & 295 $\pm$ 170 & 22 $\pm$ 185 & 70 $\pm$ 245 & $-$155 $\pm$ 345 & $-$50 $\pm$ 50\\
	4000 & $-$5 $\pm$ 130 & 250 $\pm$ 120 & $-$50 $\pm$ 220 & $-$15 $\pm$ 275 & $-$230 $\pm$ 280 & $-$50 $\pm$ 50 \\
	4500 & $-$10 $\pm$ 130 & 205 $\pm$ 135 & 100 $\pm$ 50 & $-$20 $\pm$ 315 & $-$195 $\pm$ 320 & 100 $\pm$ 14 \\
	5000 & 0 $\pm$ 150 & 205 $\pm$ 85 & 25 $\pm$ 150 & 30 $\pm$ 255 & $-$145 $\pm$ 245 & $-$20 $\pm$ 50 \\
	5500 & $-$5 $\pm$ 150 & 190 $\pm$ 160 & 130 $\pm$ 65 & $-$95 $\pm$ 345 & $-$320 $\pm$ 125 & $-$50 $\pm$ 20 \\
	6000 & 0 $\pm$ 150 & 300 $\pm$ 80 & 150 $\pm$ 100 & 30 $\pm$ 255 & $-$145 $\pm$ 245 & $-$40 $\pm$ 50	 
	\enddata
\end{deluxetable*}

\begin{deluxetable*}{ccccccc}[H]
	\tabletypesize{\footnotesize}
	\tablecaption{Systematic errors and uncertainties for the estimated percentage flux of the secondary star
	\label{table:injection-recovery_companion_brightness}}
	\tablewidth{0pt}
        \tablehead{
	Primary $T_{\rm eff}$ [K] & \multicolumn{6}{c}{Secondary $T_{\rm eff}$ [K]} \\
	 & \colhead{3500} & \colhead{4000} & \colhead{4500} & \colhead{5000} & \colhead{5500} & \colhead{6000}}
	\startdata
	& \multicolumn{6}{c}{1\% Secondary} \\
	3500 & 0.25 $\pm$ 0.10 & $-$0.28 $\pm$ 0.89 & 0.03 $\pm$ 0.33 & $-$0.30 $\pm$ 0.24 & $-$0.21 $\pm$ 0.29 & $-$0.15 $\pm$ 0.34 \\
	4000 & 0.21 $\pm$ 0.11 & 0.20 $\pm$ 0.26 & $-$ 0.01 $\pm$ 0.11 & $-$0.18 $\pm$ 0.24 & $-$0.10 $\pm$ 0.18 & $-$0.04 $\pm$ 0.21\\
	4500 & 0.10 $\pm$ 0.26 & 0.11 $\pm$ 0.26 & $-$0.07 $\pm$ 0.12 & $-$0.25 $\pm$ 0.34 & $-$0.16 $\pm$ 0.11 & $-$0.14 $\pm$ 0.16 \\
	5000 & 0.32 $\pm$ 0.19 & 0.32 $\pm$ 0.19 & 0.03 $\pm$ 0.19 & $-$0.04 $\pm$ 0.25 & $-$0.02 $\pm$ 0.17 & $-$0.05 $\pm$ 0.22 \\
	5500 & 0.15 $\pm$ 0.18 & 0.11 $\pm$ 0.26 & $-$0.11 $\pm$ 0.17 & $-$0.12 $\pm$ 0.20 & $-$0.23 $\pm$ 0.31 & $-$0.07 $\pm$ 0.14\\
	6000 & 0.32 $\pm$ 0.18 & 0.32 $\pm$ 0.19 & 0.03 $\pm$ 0.19 & $-$0.04 $\pm$ 0.25 & $-$0.02 $\pm$ 0.17 & $-$0.05 $\pm$ 0.22 \\
	 & \multicolumn{6}{c}{3\% Secondary} \\
	 3500 & 0.34 $\pm$ 0.90 & 0.67 $\pm$ 0.45 & 0.00 $\pm$ 0.40 & $-$0.21 $\pm$ 0.66 & $-$0.34 $\pm$ 0.70 & $-$0.35 $\pm$ 0.74 \\
	 4000 & 0.32 $\pm$ 0.43 & 0.71 $\pm$ 0.65 & $-$0.08 $\pm$ 0.32 & $-$0.23 $\pm$ 0.57 & $-$0.24 $\pm$ 0.78 & 0.19 $\pm$ 0.69 \\
	 4500 & 0.38 $\pm$ 0.43 & 0.62 $\pm$ 0.50 & 0.07 $\pm$ 0.69 & $-$0.25 $\pm$ 0.50 & $-$0.31 $\pm$ 0.52 & $-$0.02 $\pm$ 0.34 \\
	 5000 & 0.31 $\pm$ 0.86 & 0.11 $\pm$ 0.79 & $-$0.12 $\pm$ 0.23 & $-$0.18 $\pm$ 0.18 & $-$0.11 $\pm$ 0.14 & $-$0.49 $\pm$ 0.16 \\
	 5500 & 0.25 $\pm$ 0.49 & 0.67 $\pm$ 0.49 & $-$0.21 $\pm$ 0.38 & $-$0.18 $\pm$ 0.42 & $-$0.45 $\pm$ 0.53 & 0.04 $\pm$ 0.29 \\
	 6000 & 0.18 $\pm$ 0.87 & 0.11 $\pm$ 0.79 & $-$0.12 $\pm$ 0.23 & 0.18 $\pm$ 0.18 & 0.11 $\pm $0.14 & 0.69 $\pm$ 0.16 \\
	 & \multicolumn{6}{c}{10\% Secondary} \\
	 3500 & 1.17 $\pm$ 1.34 & 3.09 $\pm$ 1.21 & 0.35 $\pm$ 1.85 & 0.06 $\pm$ 2.25 & $-$2.63 $\pm$ 2.50 & $-$1.17 $\pm$ 1.42 \\
	 4000 & 1.26 $\pm$ 1.40 & 3.43 $\pm$ 1.09 & 0.23 $\pm$ 1.40 & $-$1.92 $\pm$ 1.44 & $-$1.84 $\pm$ 0.77 & $-$0.68 $\pm$ 1.44 \\
	 4500 & 1.32 $\pm$ 1.55 & 2.95 $\pm$ 0.77 & $-$0.12 $\pm$ 1.05 & $-$1.26 $\pm$ 0.83 & $-$1.97 $\pm$ 1.08 & $-$1.07 $\pm$ 1.97 \\
	 5000 & $-$0.07 $\pm$ 0.15 & 2.23 $\pm$ 1.43 & 0.29 $\pm$ 0.70 & $-$1.25 $\pm$ 0.93 & $-$1.76 $\pm$ 0.83 & $-$1.55 $\pm$ 1.53 \\
	 5500 & 0.55 $\pm$ 1.56 & 2.00 $\pm$ 1.00 & $-$0.84 $\pm$ 0.78 & $-$1.71 $\pm$ 0.41 & $-$1.71 $\pm$ 0.53 & $-$0.47 $\pm$ 0.50 \\
	 6000 & $-$0.07 $\pm$ 0.16 & 2.23 $\pm$ 1.43 & 0.29 $\pm$ 0.70 & $-$0.42 $\pm$ 0.58 & $-$1.76 $\pm$ 0.83 & $-$1.55 $\pm$ 1.53
	\enddata
\end{deluxetable*}

\begin{deluxetable*}{ccccccccccccc}[H]
	\tabletypesize{\footnotesize}
	\tablecaption{Uncertainties in the effective temperature of the secondary star 
	\label{table:uncertainties_companion}}
	\tablewidth{0pt}
        \tablehead{
	Primary $T_{\rm eff}$ [K] & \multicolumn{12}{c}{Secondary $T_{\rm eff}$ [K]} \\
	 & \colhead{3500} & \colhead{4000} & \colhead{4500} & \colhead{5000} & \colhead{5500} & \colhead{6000} &  \colhead{3500} & \colhead{4000} & \colhead{4500} & \colhead{5000} & \colhead{5500} & \colhead{6000}}
	\startdata
	 & \multicolumn{6}{c}{Absolute $T_{\rm eff}$ Uncertainty, $\sigma_{T_{\rm eff}}$ [K]}  & \multicolumn{6}{c}{Relative Flux Ratio Uncertainty${}^{*}$, $\sigma_{f}/f$}\\
	 & \multicolumn{6}{c}{1\% Secondary} & \multicolumn{6}{c}{1\% Secondary} \\
	3500 & 150 & 750 & 750 & 650 & 500 & 950 & 0.35 & 1.15 & 0.35 & 0.55 & 0.50 & 0.50 \\
	4000 & 200 & 600 & 600 & 900 & 550 & 500 & 0.30 & 0.45 & 0.10 & 0.40 & 0.30 & 0.25 \\
	4500 & 150 & 650 & 300 & 550 & 450 & 300 & 0.35 & 0.35 & 0.20 & 0.60 & 0.25 & 0.30 \\
	5000 & 250 & 650 & 350 & 450 & 400 & 100 & 0.50 & 0.50 & 0.20 & 0.30 & 0.20 & 0.30 \\
	5500 & 150 & 650 & 400 & 500 & 450 & 100 & 0.35 & 0.35 & 0.30 & 0.30 & 0.55 & 0.20 \\
	6000 & 250 & 650 & 350 & 450 & 550 & 100 & 0.50 & 0.50 & 0.20 & 0.30 & 0.20 & 0.25 \\
	 & \multicolumn{6}{c}{3\% Secondary}  & \multicolumn{6}{c}{3\% Secondary}\\
	3500 & 250 & 800 & 550 & 500 & 550 & 150 & 0.40 & 0.35 & 0.15 & 0.30 & 0.35 & 0.35 \\
	4000 & 150 & 800 & 500 & 550 & 700 & 650 & 0.25 & 0.45 & 0.15 & 0.25 & 0.35 & 0.30 \\
	4500 & 150 & 755 & 600 & 500 & 500 & 200 & 0.25 & 0.35 & 0.25 & 0.25 & 0.25 & 0.10 \\
	5000 & 150 & 450 & 500 & 600 & 650 & 850 & 0.40 & 0.30 & 0.10 & 0.10 & 0.10 & 0.20 \\
	5500 & 150 & 650 & 500 & 550 & 500 & 200 & 0.25 & 0.40 & 0.20 & 0.20 & 0.30 & 0.10 \\
	6000 & 150 & 450 & 500 & 600 & 650 & 850 & 0.35 & 0.30 & 0.10 & 0.10 & 0.10 & 0.30 \\
	 & \multicolumn{6}{c}{10\% Secondary} & \multicolumn{6}{c}{10\% Secondary}\\
	3500 & 150 & 450 & 200 & 300 & 550 & 100 & 0.25 & 0.45 & 0.20 & 0.25 & 0.50 & 0.25 \\
	4000 & 150 & 350 & 250 & 350 & 500 & 100 & 0.25 & 0.45 & 0.15 & 0.35 & 0.25 & 0.20 \\
	4500 & 150 & 350 & 150 & 350 & 500 & 100 & 0.30 & 0.35 & 0.10 & 0.20 & 0.30 & 0.30 \\
	5000 & 150 & 300 & 200 & 300 & 400 & 100 & 0.10 & 0.35 & 0.10 & 0.20 & 0.25 & 0.30 \\
	5500 & 150 & 350 & 200 & 450 & 450 & 100 & 0.20 & 0.30 & 0.15 & 0.20 & 0.20 & 0.10 \\
	6000 & 150 & 400 & 250 & 300 & 400 & 100 & 0.05 & 0.35 & 0.10 & 0.10 & 0.25 & 0.30 	 
	\enddata
	\tablecomments{${}^{*}$ The flux ratio of the secondary and the primary star is here denoted by $f$, where $f\,=\,F_{B}/F_{A}$.}
\end{deluxetable*}

\begin{deluxetable*}{cccc}[H]
	\tabletypesize{\footnotesize}
	\tablecaption{Injection-recovery experiment: recovery rates and parameter uncertainties for a G-type primary star and an M-dwarf secondary at $+$5 $\rm km\,s^{-1}\,\Delta RV$
	\label{table:injection-recovery_closein_G-M_pairs}}
	\tablewidth{0pt}
        \tablehead{
	\colhead{Secondary star} & \colhead{Recovery} & \colhead{$T_{\rm eff}$ [K]} & \colhead{Relative Flux [\%]} \\
	\colhead{Brightness} & \colhead{Rate} & \colhead{$T_{\rm actual}\,-\,T_{\rm deduced}$} &  \colhead{$\%_{\rm actual}\,-\,\%_{\rm deduced}$}} 
	\startdata
	1\% & 40\% & 35 $\pm$ 170 & 0.01 $\pm$ 0.22 \\
	3\% & 90\% & 10 $\pm$ 130 & 1.18 $\pm$ 0.56 \\
	5\% & 90\% & -5 $\pm$ 140 & 2.05 $\pm$ 0.70
	\enddata
\end{deluxetable*}

\begin{deluxetable*}{ccccc}[H]
	\tabletypesize{\footnotesize}
	\tablecaption{Parameters for the Test Cases of Binary Stars; Primary Star
	\label{table:results_table_primary}}
	\tablewidth{0pt}
	\tablehead{
	\colhead{Parameter} & \colhead{Literature} & \colhead{Our Results} & \colhead{Discrepancy} & \colhead{Notes}}
	\startdata
	\multicolumn{5}{c}{\bf KIC 10319590}\\
	$T_{\rm{eff}}$ $[K]$ & 5518 $\pm$ 200  & 5650 $\pm$ 200 & / &  MAST Online Catalogue \\
	Mass [$\rm{M}_{\odot}$] & \nodata &  1.0 $\pm$ 0.5 & \nodata & MAST Online Catalogue \\
	Radius [$\rm{R}_{\odot}$] & \nodata & 0.9 $\pm$ 0.5 & \nodata & MAST Online Catalogue \\
	log$g$ [mag] & 4.4 $\pm$ 0.5 & 4.5 $\pm$ 0.5 & / & MAST Online Catalogue \\
	\multicolumn{5}{c}{\bf KIC 5473556}\\
	$T_{\rm{eff}}$ [K] & 5932 $\pm$ 200  & 5800 $\pm$ 200 & / &  MAST Online Catalogue \\
	Mass [$\rm{M}_{\odot}$] & \nodata &  1.0 $\pm$ 0.5 & \nodata & MAST Online Catalogue \\
	Radius [$\rm{R}_{\odot}$] & \nodata & 0.9 $\pm$ 0.5 & \nodata & MAST Online Catalogue \\
	log$g$ [mag] & 4.028 $\pm$ 0.5 & 4.5 $\pm$ 0.5 & / & MAST Online Catalogue \\
	\multicolumn{5}{c}{\bf KIC 8572936}\\
	$T_{\rm{eff}}$ [K] & 5913 $\pm$ 130  & 6000 $\pm$ 200 & / &  \citet{Welsh_etal_2011} \\
	Mass [$\rm{M}_{\odot}$] & 1.0479 &  1.0 $\pm$ 0.5 & / & \citet{Welsh_etal_2011} \\
	Radius [$\rm{R}_{\odot}$] &1.1618 & 1.1 $\pm$ 0.5 & / & \citet{Welsh_etal_2011} \\
	log$g$ [mag] & 4.3284$\pm$0.5 & 4.3 $\pm$ 0.5 & / & \citet{Welsh_etal_2011} \\
	\multicolumn{5}{c}{\bf KIC 9837578}\\
	$T_{\rm{eff}}$ [K] & 5606 $\pm$ 150  & 5800 $\pm$ 200 & / &  \citet{Welsh_etal_2011} \\
	Mass [$\rm{M}_{\odot}$] & 0.8877 &  0.9 $\pm$ 0.5 & / & \citet{Welsh_etal_2011} \\
	Radius [$\rm{R}_{\odot}$] & 1.0284 & 0.9 $\pm$ 0.5 & / & \citet{Welsh_etal_2011} \\
	log$g$ [mag] & 4.3623$\pm$0.5 & 4.5 $\pm$ 0.5 & / & \citet{Welsh_etal_2011} \\
	\multicolumn{5}{c}{\bf HD 61994}\\
	$T_{\rm eff}$ [K] & 5630 $\pm$ 150  & 5750 $\pm$ 200 & / &  \citet{Strassmeier_etal_2012} \\
	Mass [$\rm{M}_{\odot}$] & \nodata &  1.1 $\pm$ 0.5 & \nodata & \nodata \\
	Radius [$\rm{R}_{\odot}$] & \nodata & 1.3 $\pm$ 0.5 & \nodata & \nodata \\
	log$g$ [mag] & 4.13 $\pm$ 0.11 & 4.2 $\pm$ 0.5 & / & \citet{Strassmeier_etal_2012} \\
	\multicolumn{5}{c}{\bf HD 16702} \\
	$T_{\rm eff}$ [K] & 5908 $\pm$ 25 & 5800 $\pm$ 200 & / & \citet{Diaz_etal_2012} \\
	Mass [$\rm{M_{\odot}}$] & 0.98 $\pm$ 0.04 & 1.1 $\pm$ 0.5 & / & \citet{Diaz_etal_2012}\\
	Radius [$\rm{R_{\odot}}$] & \nodata & 1.0 $\pm$ 0.5 & \nodata & \nodata \\
	log$g$ [mag] & 4.46 $\pm$ 0.03 & 4.4 $\pm$ 0.5 & / & \citet{Diaz_etal_2012}
	\enddata 
\end{deluxetable*}

\begin{deluxetable*}{ccccc}[H]
	\tabletypesize{\footnotesize}
	\tablecaption{Parameters for the Test Cases of Binary Stars; Stellar Companion
	\label{table:results_table_companion}}
	\tablewidth{0pt}
	\tablehead{
	\colhead{Parameter} & \colhead{Literature} & \colhead{Our Results} & \colhead{Discrepancy} & \colhead{Notes}}
	\startdata
	\multicolumn{5}{c}{\bf KIC 10319590}\\
	$T_{\rm{eff}}$ [K] & \nodata & 4300 $\pm$ 500 & \nodata & MAST Online Catalogue \\
	Flux Ratio, $\rm{F_{B}/F_{A}}$ & \nodata & 0.036 $\pm$ 0.007 & \nodata & \nodata\\
	\multicolumn{5}{c}{\bf KIC 5473556}\\
	$T_{\rm{eff}}$ [K] & \nodata & 6000 $\pm$ 100 & \nodata & MAST Online Catalogue \\
	Flux Ratio, $\rm{F_{B}/F_{A}}$ & \nodata & 0.22 $\pm$0.07 & \nodata & \nodata\\
	\multicolumn{5}{c}{\bf KIC 8572936}\\
	$T_{\rm{eff}}$ [K] & 5867 $\pm$ 130 & 6000 $\pm$ 250 & / & \citet{Welsh_etal_2011} \\
	Flux Ratio, $\rm{F_{B}/F_{A}}$ & 0.8475 $\pm$ 0.005 & 0.82 $\pm$ 0.25 & / & \citet{Welsh_etal_2011}\\
        \multicolumn{5}{c}{\bf KIC 9837578}\\
        $T_{\rm{eff}}$ [K] & 5202 $\pm$ 100 & 5600 $\pm$ 400 & / & \citet{Strassmeier_etal_2012} \\
        Flux Ratio, $\rm{F_{B}/F_{A}}$ & 0.3941 &  0.56 $\pm$ 0.14 & 7\% & \citet{Strassmeier_etal_2012}\\
	\multicolumn{5}{c}{\bf HD 61994}\\
	$T_{\rm{eff}}$ [K] & 4775 $\pm$ 150 & 4200 $\pm$ 650 & / & \citet{Strassmeier_etal_2012} \\
	Flux Ratio, $\rm{F_{B}/F_{A}}$ & 0.069 & 0.055 $\pm$ 0.022 & / & \citet{Strassmeier_etal_2012}\\
	\multicolumn{5}{c}{\bf HD 16702}\\
	$T_{\rm eff}$ [K] & \nodata & 3500 $\pm$ 250 & \nodata & \nodata\\
	Flux Ratio, $\rm{F_{B}/F_{A}}$ & \nodata & 0.016 $\pm$ 0.008 & \nodata & \nodata
	\enddata 
\end{deluxetable*}

\begin{deluxetable*}{cccccc}[!h]
	\tabletypesize{\footnotesize}
	\tablecaption{California \emph{Kepler} Search: Binary Systems
	\label{table:CK0s_results}}
	\tablewidth{0pt}
	\tablehead{
	\multirow{2}{*}{KOI} & \colhead{Primary Star} & \multicolumn{3}{c}{Companion Parameters} & \colhead{Planetary}\\
	 & \colhead{$T_{\rm eff,A}$ [K]} & \colhead{$T_{\rm eff,B}$ [K]} & \colhead{$\rm{F_{B}/F_{A}}$} & \colhead{$\rm{\Delta{RV}}$ $[\rm{km\,s^{-1}}]$}& \colhead{Data${}^{*}$}}
	\startdata
	   5 & 5753 $\pm$ 75 & 5900 $\pm$ 850 & $\geq$0.066 $\pm$ 0.020 & +11 & 2 PC\\
	 151 & 6276 $\pm$ 163 & 3500 $\pm$ 250 & $\geq$0.012 $\pm$ 0.006 & +14 & 1 PC\\ 
	 219 & 5513 $\pm$ 184 & $\geq$6000 $\pm$ 100 & $\geq$0.330 $\pm$ 0.033 & +13 & 1 PC\\
	 652 & 4700 $\pm$ 128 & 3700 $\pm$ 150 & 0.092 $\pm$ 0.028 & +22 & 1 PC\\
	 & & 4000 $\pm$ 350 & 0.020 $\pm$ 0.007 & +46\\
	 & & 3500 $\pm$ 150 & 0.006 $\pm$ 0.002 &$-$44 \\
	 698 & 6120 $\pm$ 196 & 4800 $\pm$ 600 & 0.048 $\pm$ 0.005  & +29& 1 PC\\
	 969 & 6224 $\pm$ 186 & $\geq$6000 $\pm$ 100 & 0.821 $\pm$ 0.246 & $-$42 & 1 PC\\
	1020 & 6058 $\pm$ 158 & 6000 $\pm$ 100 & 0.292 $\pm$ 0.088 & +39 & 1 PC\\
	1121 & 5671 $\pm$ 156 & 6000 $\pm$ 100 & $\geq$0.077 $\pm$ 0.008 & $-$12& 1 PC\\
	1137 & 5324 $\pm$ 178 & 3600 $\pm$ 150 & $\geq$0.020 $\pm$ 0.004 & +11 & 1 PC\\
	1152 & 3806 $\pm$ 50 & 4200 $\pm$ 350 & 0.307 $\pm$ 0.138 & +27 & 1 FP\\
	1227 & 5658 $\pm$ 159 & $\geq$6000 $\pm$ 100 & 0.821 $\pm$ 0.082 & $-$123 & 1 PC\\
	1326 & 5378 $\pm$ 173 & $\geq$6000 $\pm$ 100 & 0.395 $\pm$ 0.040 & +42 & 1 PC \\
	1361 & 4014 $\pm$ 76 & 3600 $\pm$ 200 & 0.022 $\pm$ 0.007 & +40 & 1 CP \\
	1452 & 7172 $\pm$ 211 & \nodata & \nodata & +81& 1 PC\\
	1613 & 6044 $\pm$ 117 & $\geq$6000 $\pm$ 850 & $\geq$0.044 $\pm$ 0.013 & +10 & 2 PC \\
	1645 & 5193 $\pm$ 170 & 4500 $\pm$ 200 & 0.287 $\pm$ 0.029 & +55 & 1 PC\\
	1684 & 6428 $\pm$ 134 & $\geq$6000 $\pm$ 100 & 0.300 $\pm$ 0.090 & $-$ 50& 1 PC\\
	1784 & 5936 $\pm$ 150 & $\geq$6000 $\pm$ 100 & $\geq$0.192 $\pm$ 0.058 & $-$13 & 1 PC \\
	1796 & 6065 $\pm$ 205 & 3500 $\pm$ 250 & 0.012 $\pm$ 0.006 & +36 & 1 PC\\
	2059 & 4996 $\pm$ 103 & 3600 $\pm$ 250 & 0.016 $\pm$ 0.008 & +5 & 2 PC\\
	2075 & 6403 $\pm$ 159 & 3400 $\pm$ 250 & $\geq$0.008 $\pm$ 0.004 & $-$13 & 1 PC\\
	2215 & 5974 $\pm$ 187 & $\geq$6000 $\pm$ 100 & 0.185 $\pm$ 0.056 & +32 & 1 PC\\
	 & & $\geq$6000 $\pm$ 100 & $\geq$0.155 $\pm$ 0.047 & $-$15 & \\
	2311 & 5754 $\pm$ 147 & 5600 $\pm$ 400 & $\geq$0.276 $\pm$ 0.069 & +11 & 3 PC\\
	2457 & 6728 $\pm$ 155 & $\geq$6000 $\pm$ 100 & $\geq$0.076 $\pm$ 0.023 & +14& 1 PC\\
	2787 & 6335 $\pm$ 177 & $\geq$6000 $\pm$ 100 & 0.222 $\pm$ 0.067 & +16 & 1 FP \\
	2813 & 5143 $\pm$ 164 & $\geq$6000 $\pm$ 100 & 0.195 $\pm$ 0.058 & +26 & 1 PC\\
	2867 & 4865 $\pm$ 194 & 4700 $\pm$ 200 & 0.389 $\pm $ 0.136 & +44 & 1 PC\\
	2965 & 6468 $\pm$ 159 & $\geq$6000 $\pm$ 850 & 0.022 $\pm$ 0.007 & +20 & 1 FP\\
	3000 & 5919 $\pm$ 157 & 4100 $\pm$ 450 & $\geq$0.055 $\pm$ 0.019 & +10 & 1 FP\\
	3002 & 6060 $\pm$ 152 & $\geq$6000 $\pm$ 100 & 0.475 $\pm$ 0.143 & $-$15 & 1 FP\\
	3035 & 6324 $\pm$ 174 & $\geq$6000 $\pm$ 100 & $\geq$0.212 $\pm$ 0.064 & +18 & 1 FP\\
	3161 & 6790 $\pm$ 195 & $\geq$6000 $\pm$ 100 & 0.305 $\pm$ 0.092 & $-$167 & 1 PC\\
	3162 & 6109 $\pm$ 165 & 5900 $\pm$ 100 & $\geq$0.553 $\pm$ 0.166 & $-$13 & 1 FP\\
	3216 & 5615 $\pm$ 166 & $\geq$6000 $\pm$ 100 & $\geq$0.326 $\pm$ 0.033 & +13 & 1 FP\\
	3231 & 6195 $\pm$ 188 & $\geq$6000 $\pm$ 100 & 0.175 $\pm$ 0.053 & $-$22 & 1 FP\\
	3243 & 6348 $\pm$ 146 & $\geq$6000 $\pm$ 100 & 0.590 $\pm$ 0.177 & +45 & 1 FP\\
	3415 & 5619 $\pm$ 178 & 5600 $\pm$ 200 & 0.786 $\pm$ 0.157 &$-$71& 1 PC\\
	3471 & 4847 $\pm$ 118 & 4600 $\pm$ 200 & 0.746 $\pm$ 0.079 & $-$33 & 1 FP\\
	 & & $\geq$6000 $\pm$ 100 & 0.202 $\pm$ 0.061 & +26 \\
	 & & 4100 $\pm$ 300 & 0.117 $\pm$ 0.041 & $-$52 \\
	3506 & 6878 $\pm$ 194 & $\geq$6000 $\pm$ 100 & 0.206 $\pm$ 0.062 & +137 & 1 PC\\
	3515 & 6395 $\pm$ 155 & $\geq$6000 $\pm$ 100 & 0.134 $\pm$ 0.040 & +48 & 1 FP\\
	3527 & 5668 $\pm$ 153 & $\geq$6000 $\pm$ 100 & 0.880 $\pm$ 0.264 & $-$16 & 1 PC\\
	3528 & 6061 $\pm$ 200 & $\geq$6000 $\pm$ 100 & 0.182 $\pm$ 0.055 & +66 & 1 PC\\
	3557 & 6462 $\pm$ 161 & $\geq$6000 $\pm$ 100 & 0.393 $\pm$ 0.118 & +55 & 1 FP\\
	3573 & 6403 $\pm$ 158 & $\geq$6000 $\pm$ 100 & $\geq$0.109 $\pm$ 0.033 & $-$15 & 1 PC\\
	3583 & 7591 $\pm$ 230 & $\geq$6000 $\pm$ 100 & $\geq$0.109 $\pm$ 0.033 & $-$23 & 1 PC\\
	3602 & 5940 $\pm$ 157 & $\geq$6000 $\pm$ 100 & 0.253 $\pm$ 0.076 & +24 & 1 PC\\
	3605 & 5126 $\pm$ 152 & 3600 $\pm$ 150 & 0.044 $\pm$ 0.018 & $-$57 & 1 PC\\
	3606 & 6570 $\pm$ 177 & $\geq$6000 $\pm$ 100 & 0.399 $\pm$ 0.120 & +162 & 1 PC\\
	3721 & 6261 $\pm$ 185 & $\geq$6000 $\pm$ 100 & 0.143 $\pm$ 0.043 & +51 & 1 PC\\
	 & & 4500 $\pm$ 350 & 0.016 $\pm$ 0.003 & $-$61 & \\
	3782 & 5018 $\pm$ 250 & 3400 $\pm$ 135 & $\geq$0.022 $\pm$ 0.011 & $-$7 & 1 PC\\ 
	3837 & 6540 $\pm$ 181 & 4500 $\pm$ 500 & 0.048 $\pm$ 0.005 & +37 & 1 PC\\
	3853 & 5081 $\pm$ 118 & 3400 $\pm$ 250 & 0.009 $\pm$ 0.001 & +22 & 1 PC\\
	3875 & 6022 $\pm$ 198 & $\geq$6000 $\pm$ 100 & 0.623 $\pm$ 0.187 & $-$85& 1 PC\\
	4345 & 5776 $\pm$ 185 & 3600 $\pm$ 250 & 0.012 $\pm$ 0.006 & $-$48 & 1 PC\\
	4355 & 5780 $\pm$ \nodata & $\geq$6000 $\pm$ 850 & 0.044 $\pm$ 0.013 & $-$21 & 1 FP\\
	4457 & 6384 $\pm$ 159 & 3600 $\pm$ 250 & $\geq$0.012 $\pm$ 0.006 & +11 & 1 FP\\
	4713 & 5780 $\pm$ \nodata & $\geq$6000 $\pm$ 100 & $\geq$0.344 $\pm$ 0.103 & $-$15 & 1 PC\\
	4871 & 6532 $\pm$ 159 & $\geq$6000 $\pm$ 100 & $\geq$0.012 $\pm$ 0.003 & $-$23 & 1 PC\\
	\enddata
	\tablecomments{Primary star $T_{\rm eff}$ and the associated uncertainties are taken from cfop.ipac.caltech.edu. Companion parameters are derived using our algorithm, and the uncertainties are determined using Table \ref{table:uncertainties_companion}. Unless stated otherwise, we only have one observation of each spectrum listed above; therefore, we cannot test whether the secondary star is a part of the bound system or a background star.\\
	${}^{*}$From cfop.ipac.caltech.edu. PC - planet candidate, CP - confirmed planet, FP - false positive}
\end{deluxetable*}

\clearpage	
\appendix

\indent Graphical supporting evidence for KOIs that showed an evidence of secondary star, listed in Table \ref{table:CK0s_results}.
\begin{figure}[h]
	\plotone{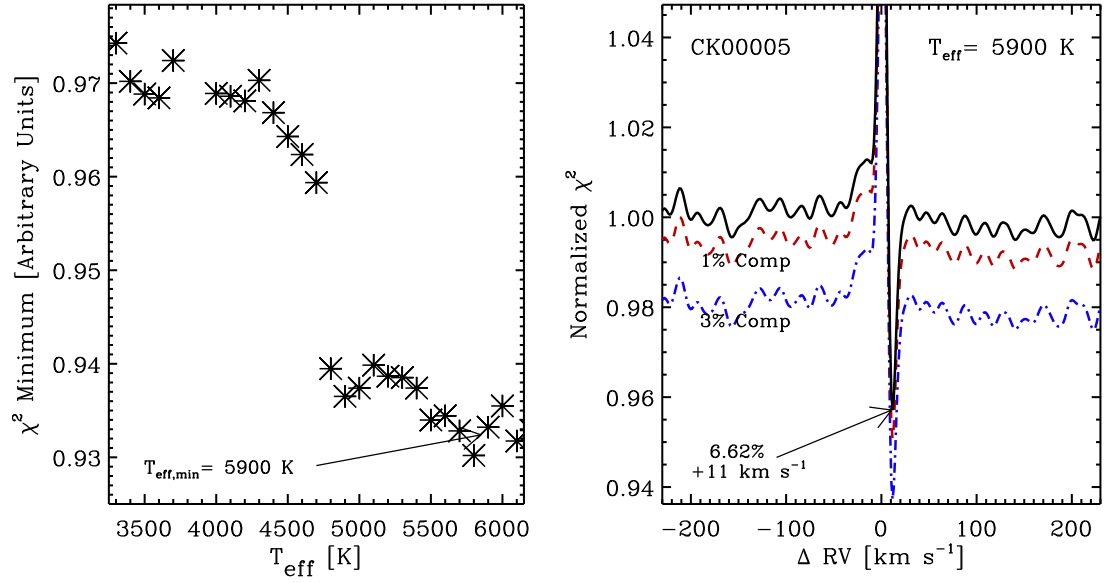}
	\caption{Final secondary star plot for KOI-5. Same as Figure \ref{fig:complete_companion_plot}.}
	\label{fig:KOI-5}
\end{figure}

\begin{figure}[h]
	\plotone{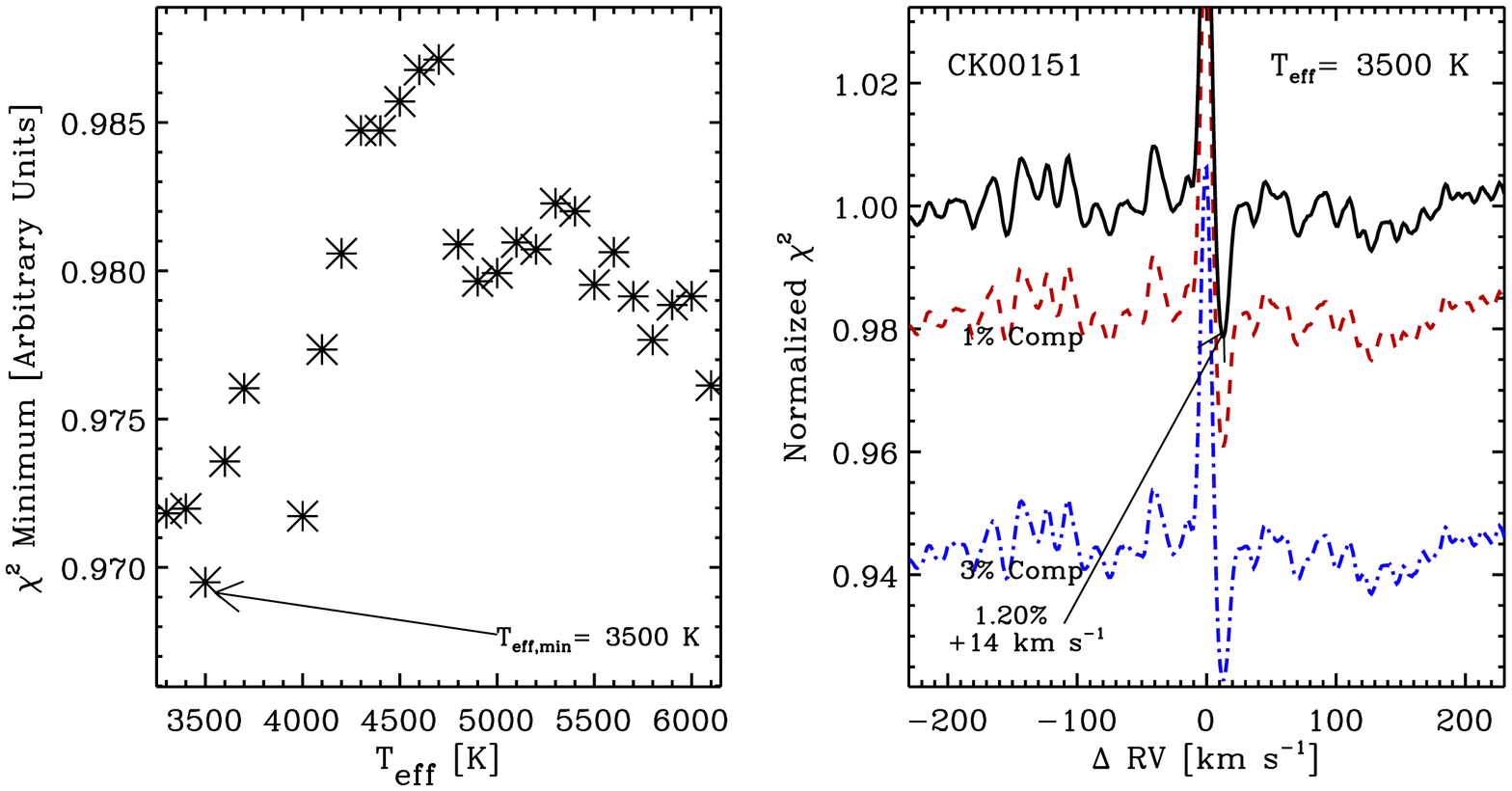}
	\caption{Final secondary star plot for KOI-151. Same as Figure \ref{fig:complete_companion_plot}.}
	\label{fig:KOI-151}
\end{figure}
\begin{figure}[h]
	\plotone{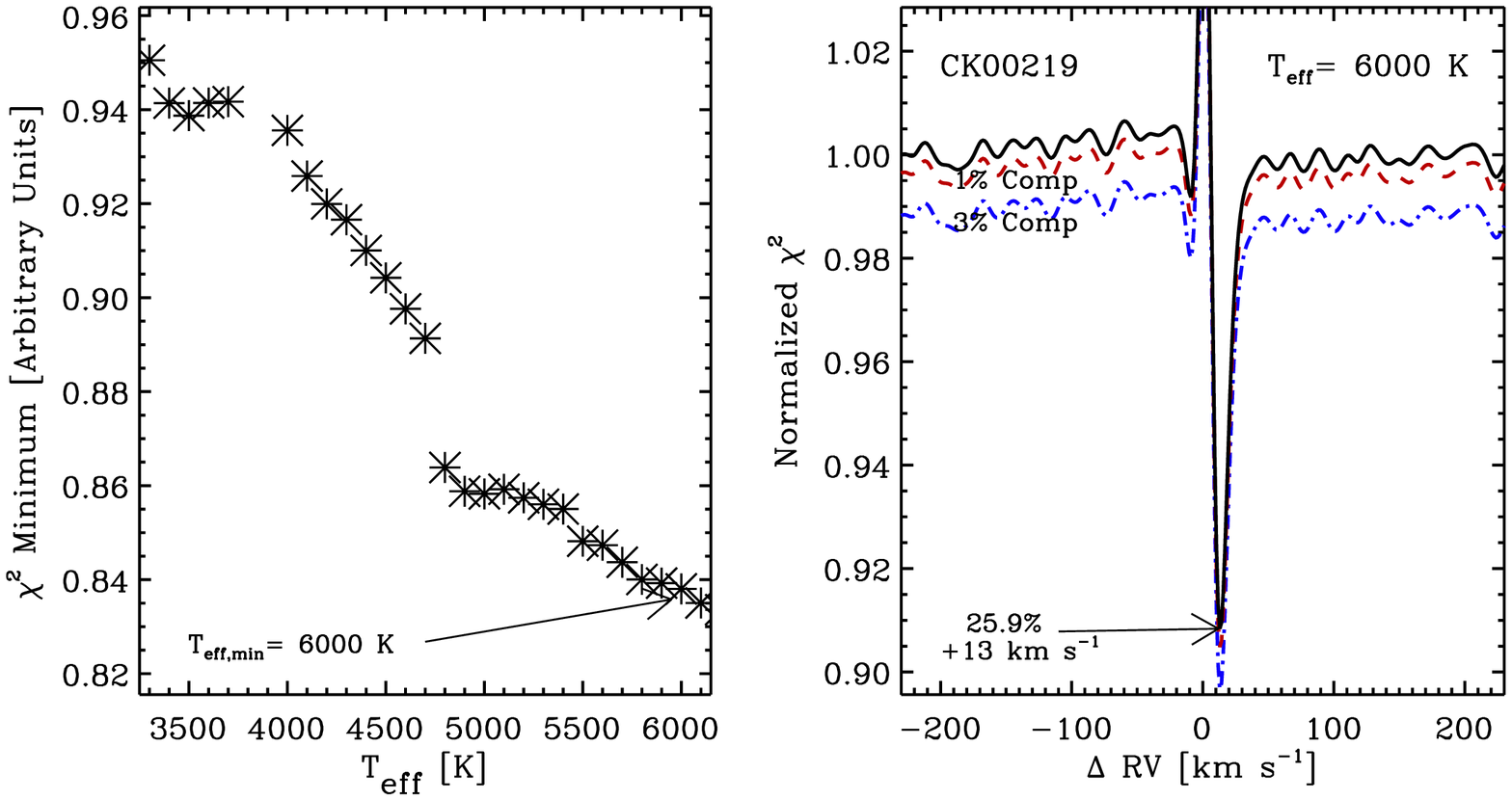}
	\caption{Final secondary star plot for KOI-219. Same as Figure \ref{fig:complete_companion_plot}.}
	\label{fig:KOI-219}
\end{figure}

\begin{figure}[h]
	\plotone{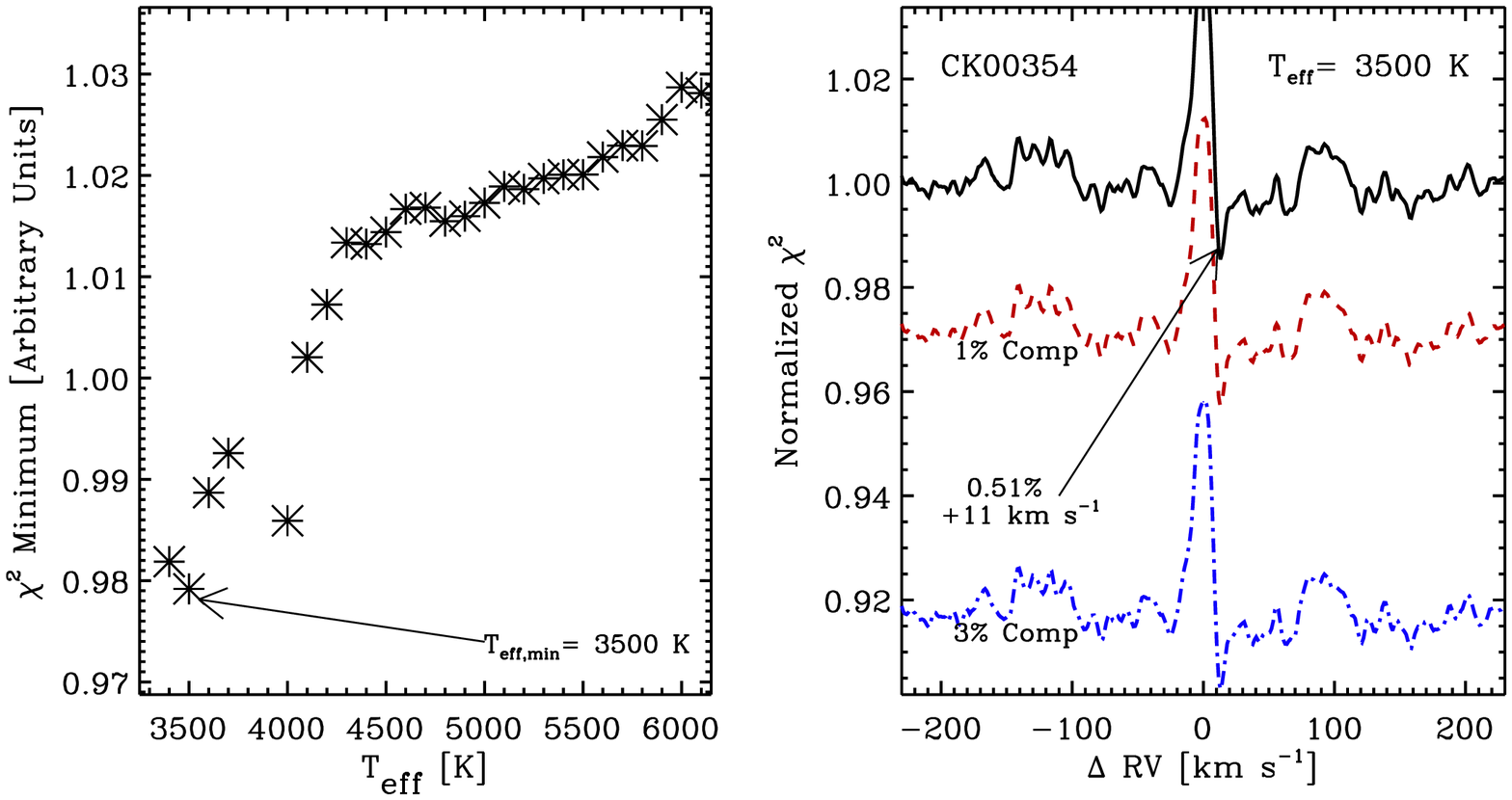}
	\caption{Final secondary star plot for KOI-354. Marginal detection. Same as Figure \ref{fig:complete_companion_plot}.}
	\label{fig:KOI-354}
\end{figure}

\begin{figure}[h]
	\plotone{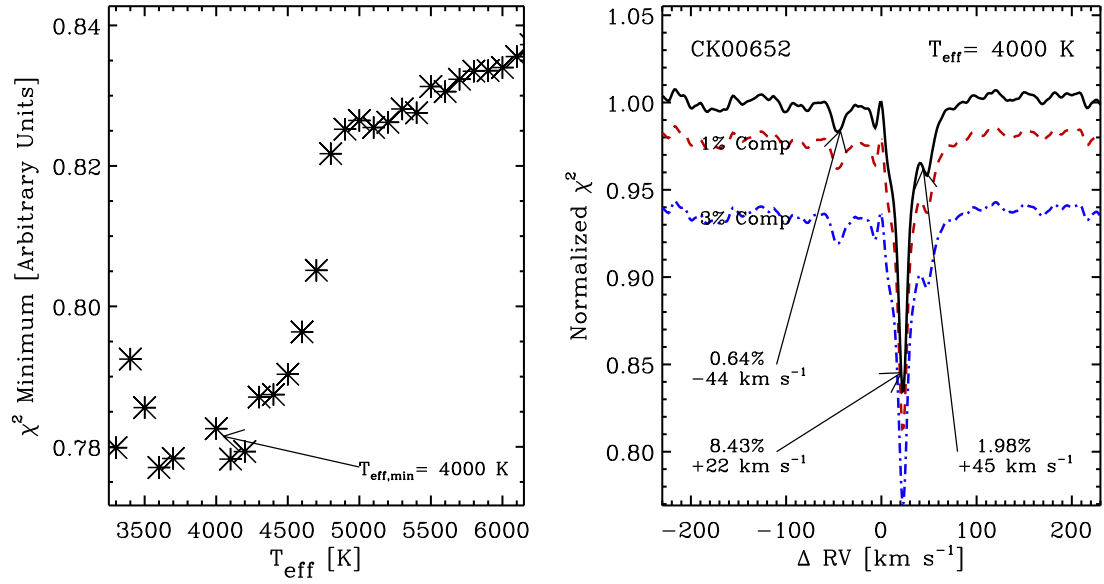}
	\caption{Final secondary star plot for KOI-652. Same as Figure \ref{fig:complete_companion_plot}. At left, we only show the $\chi^2$ minimum function for brightest companion. At right, all three companions are annotated with an arrow. The detection of the faintest companion, contributing 0.6\% to the total flux, is marginal thus not listed in Table \ref{table:CK0s_results}.}
	\label{fig:KOI-652}
\end{figure}

\begin{figure}[h]
	\plotone{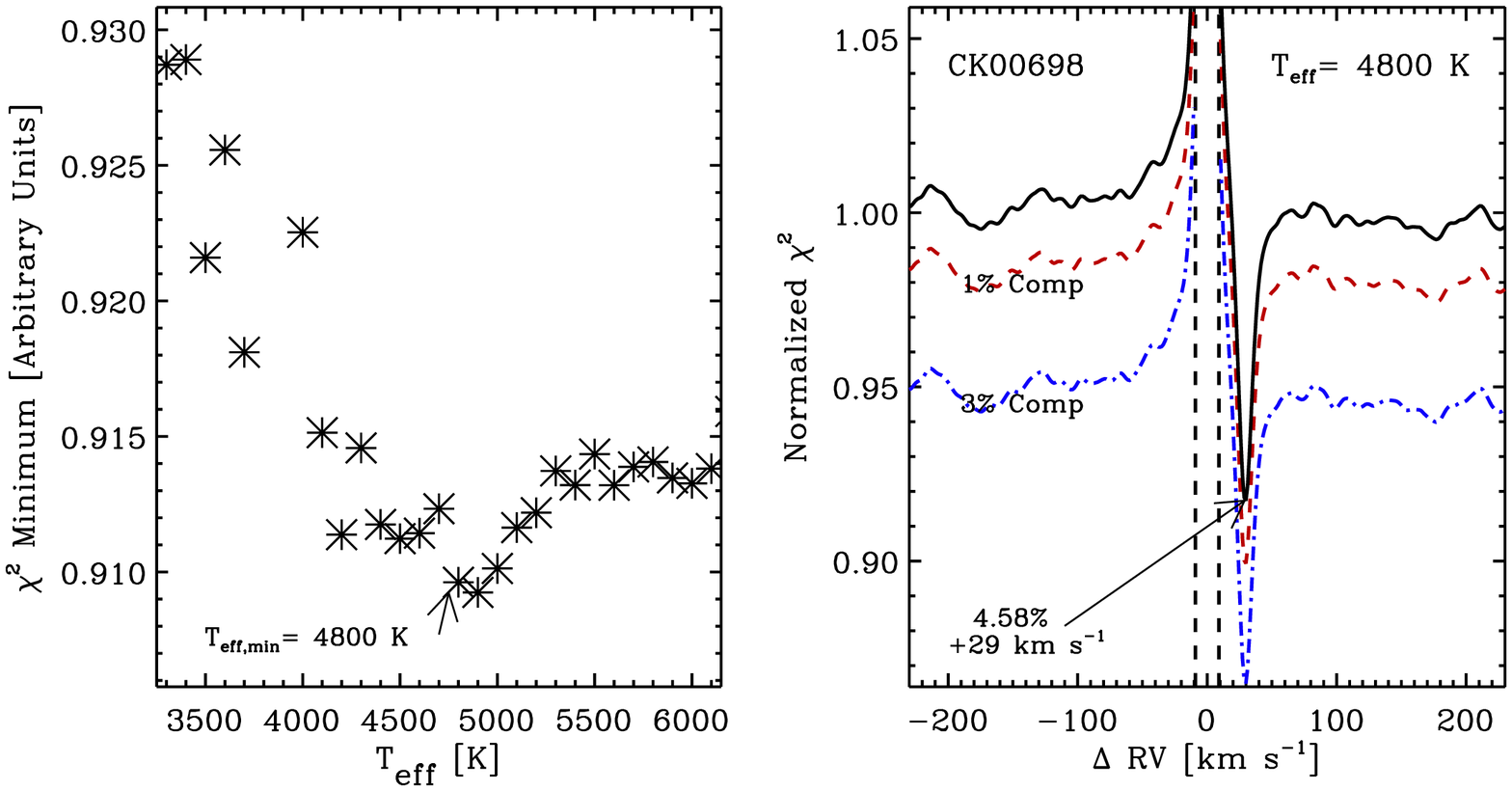}
	\caption{Final secondary star plot for KOI-698. Same as Figure \ref{fig:complete_companion_plot}.}
	\label{fig:KOI-698}
\end{figure}

\begin{figure}[h]
	\plotone{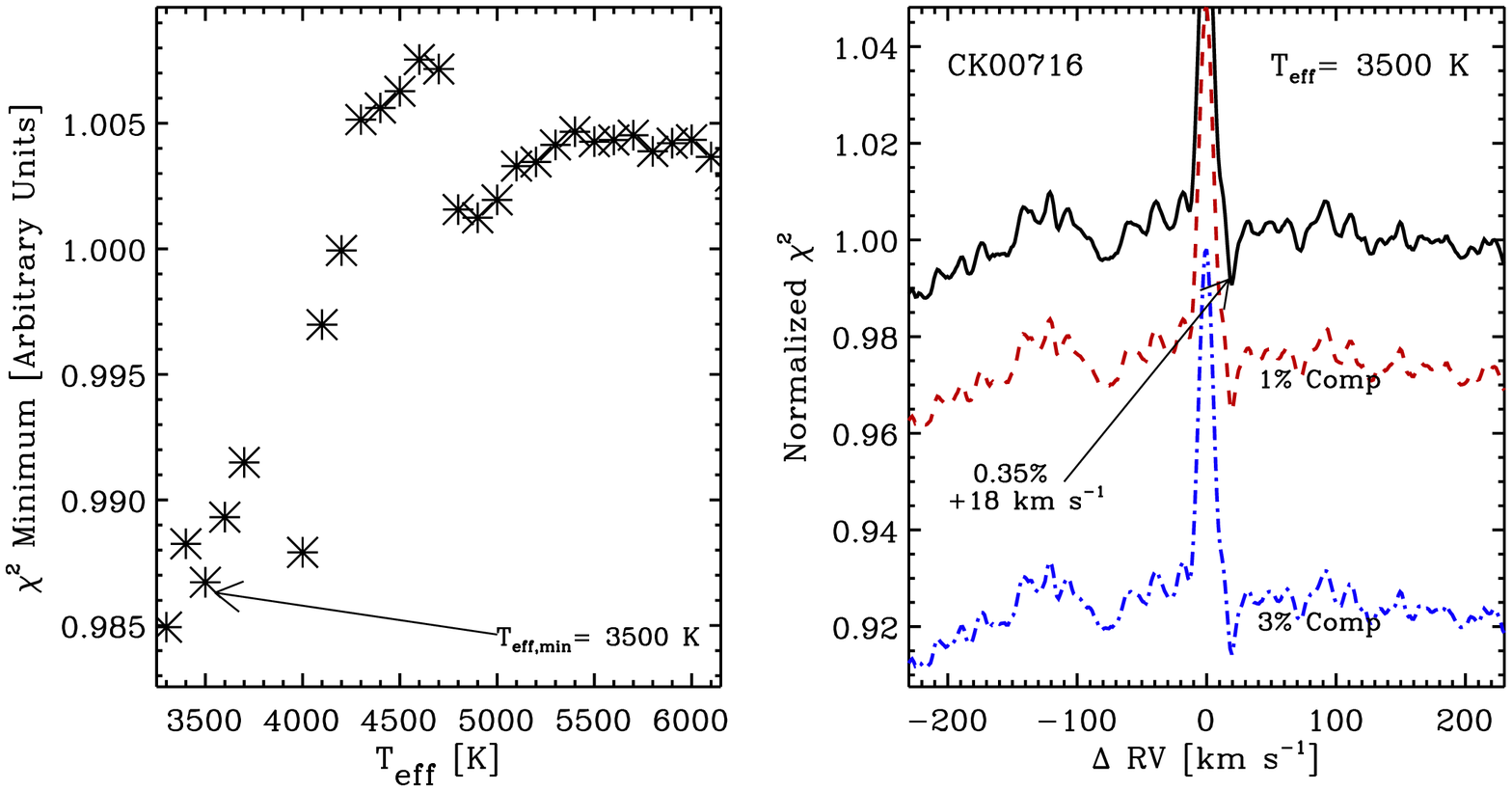}
	\caption{Final secondary star plot for KOI-716. Marginal detection. Same as Figure \ref{fig:complete_companion_plot}.}
	\label{fig:KOI-716}
\end{figure}

\begin{figure}[h]
	\plotone{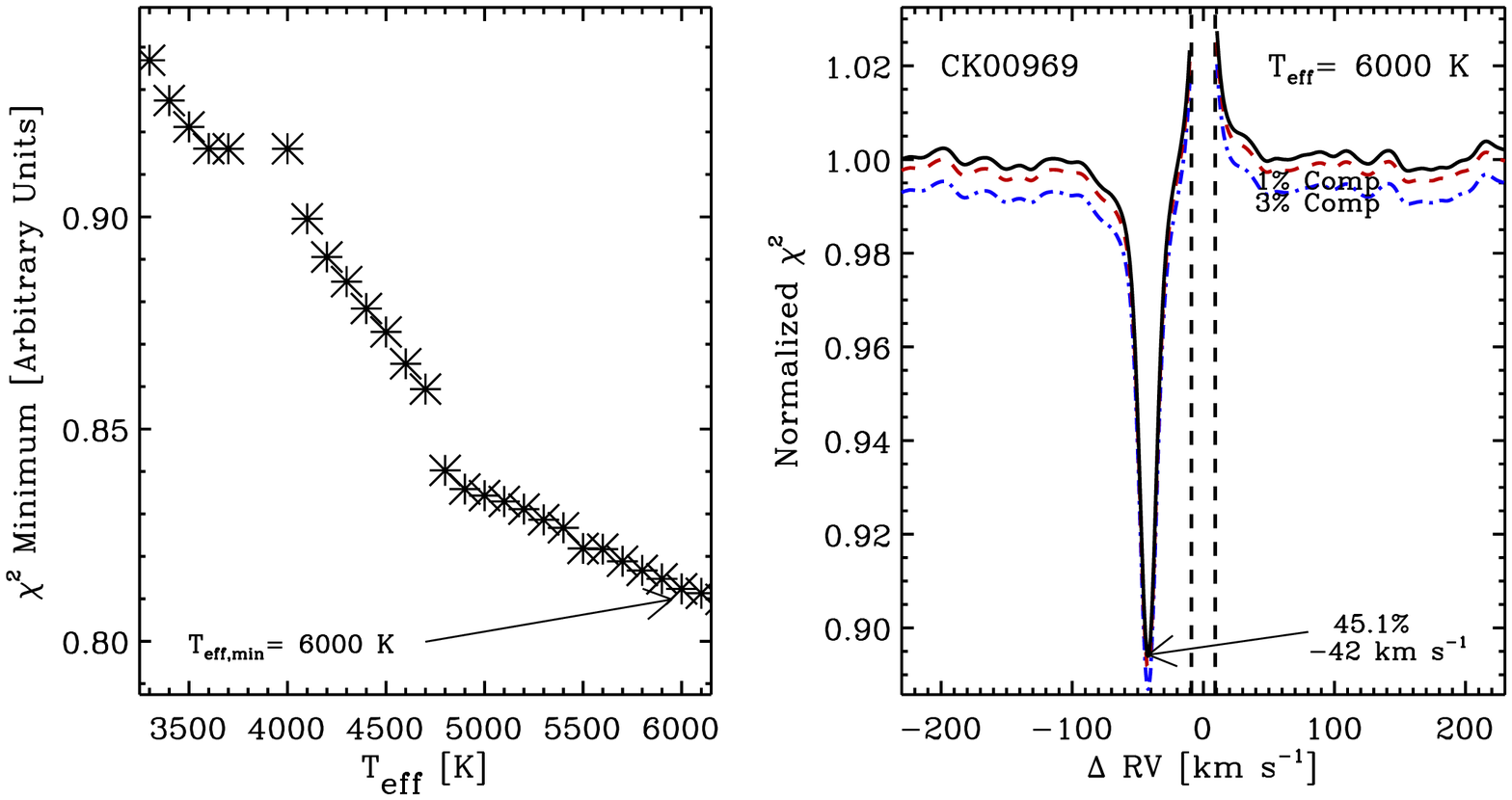}
	\caption{Final secondary star plot for KOI-969. Same as Figure \ref{fig:complete_companion_plot}.}
	\label{fig:KOI-969}
\end{figure}

\begin{figure}[h]
	\plotone{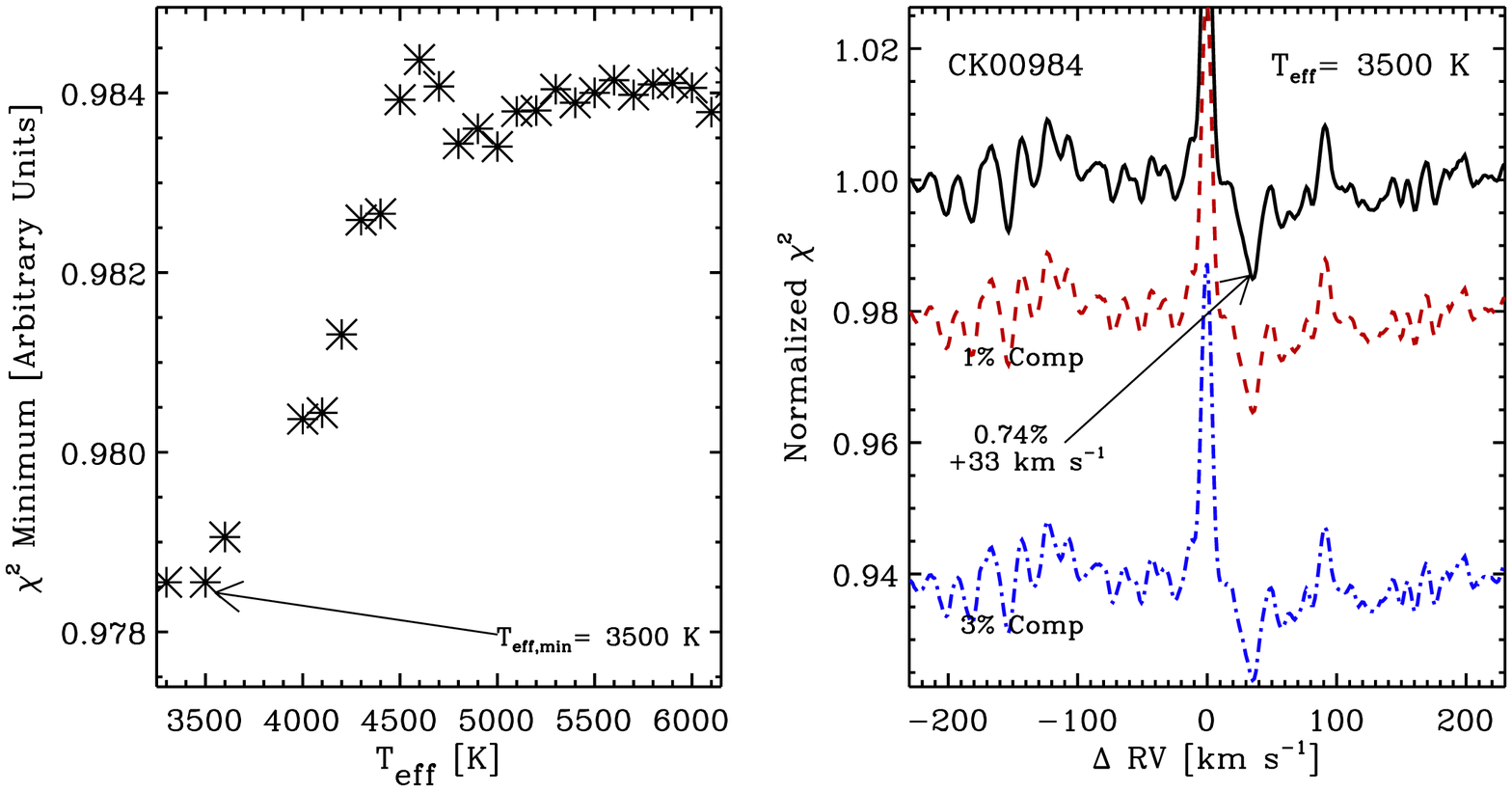}
	\caption{Final secondary star plot for KOI-984. Same as Figure \ref{fig:complete_companion_plot}.}
	\label{fig:KOI-984}
\end{figure}

\begin{figure}[h]
	\plotone{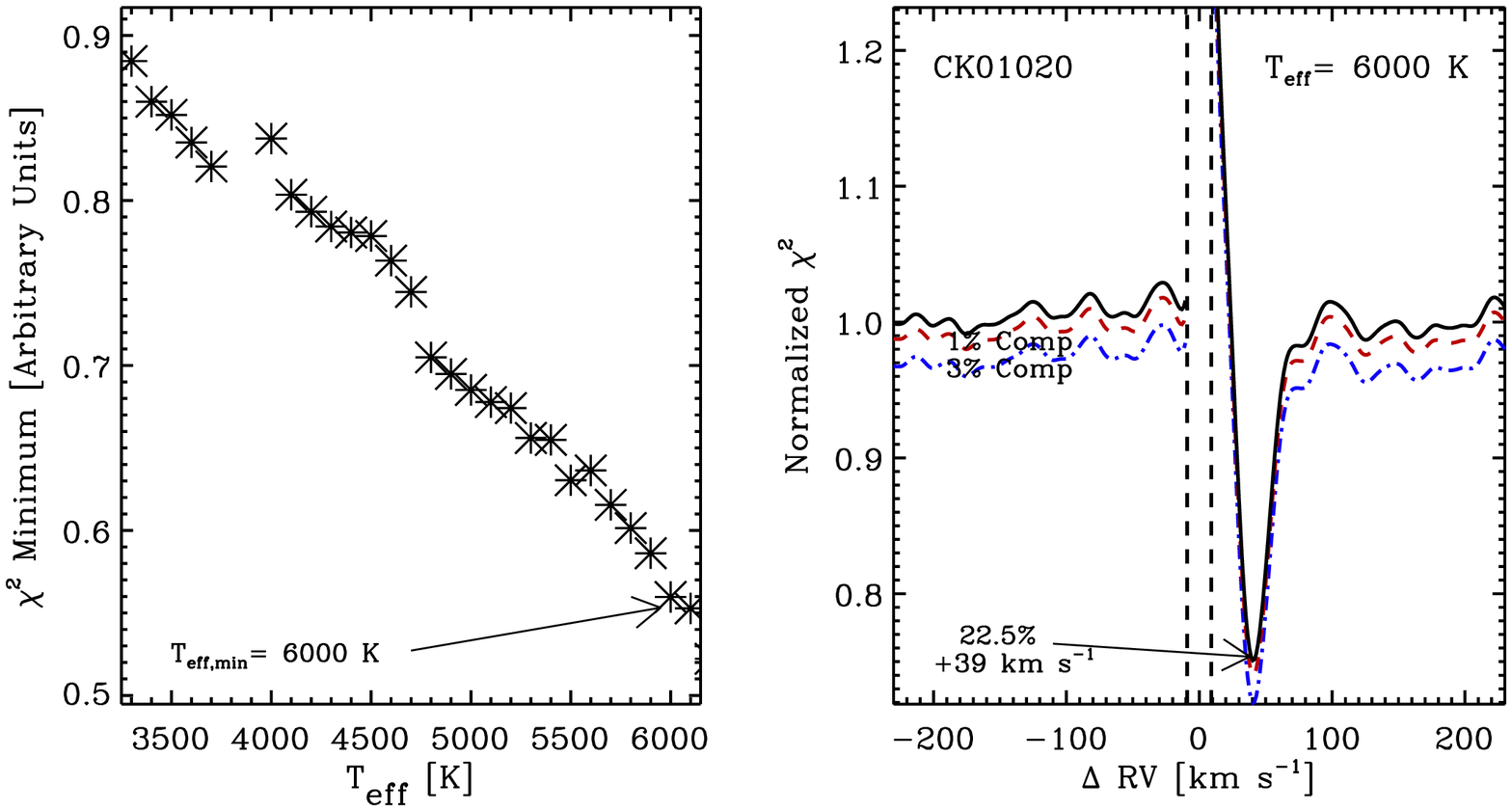}
	\caption{Final secondary star plot for KOI-1020. Same as Figure \ref{fig:complete_companion_plot}.}
	\label{fig:KOI-1020}
\end{figure}

\begin{figure}[h]
	\plotone{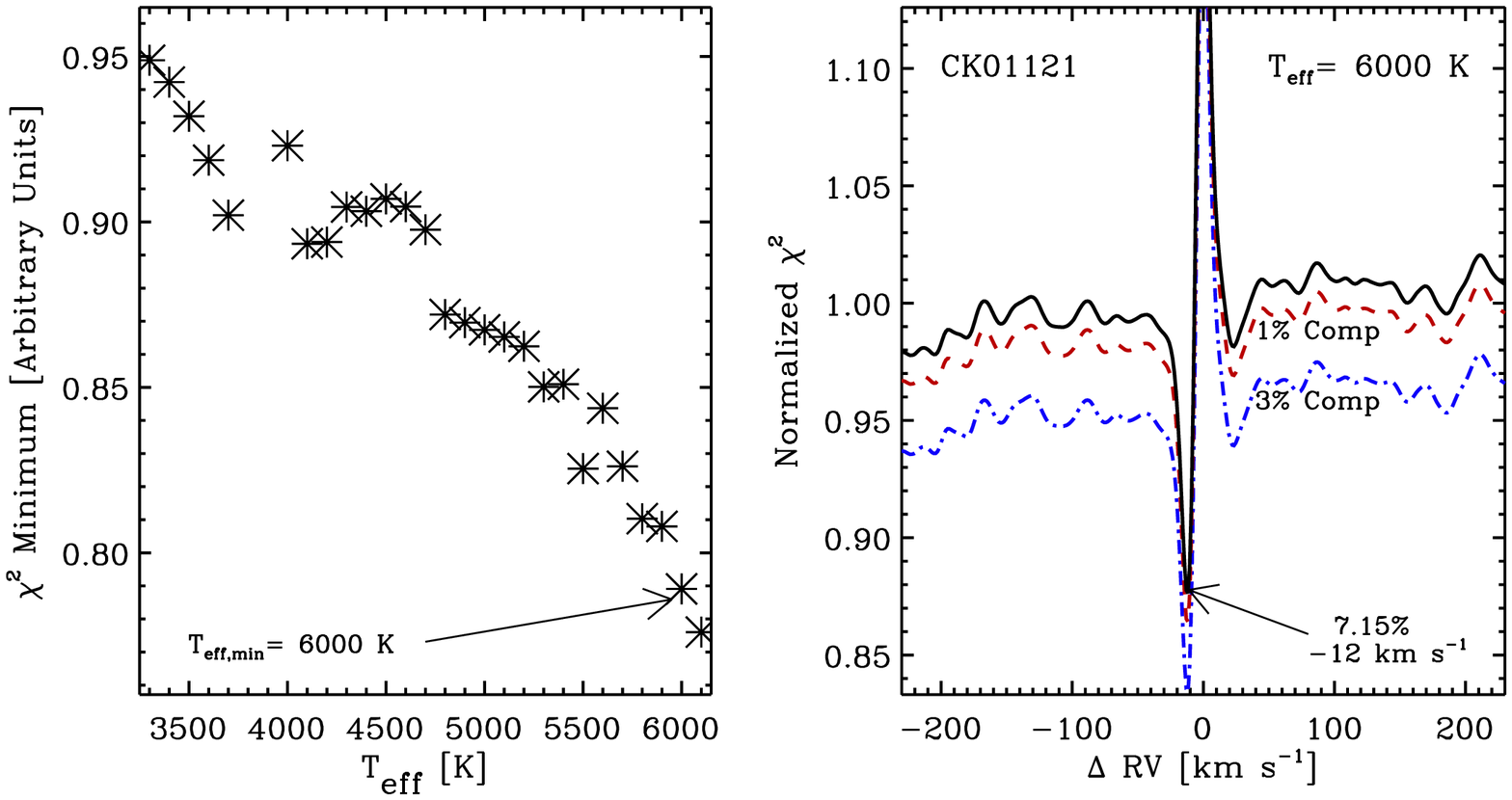}
	\caption{Final secondary star plot for KOI-1121. Same as Figure \ref{fig:complete_companion_plot}.}
	\label{fig:KOI-1121}
\end{figure}

\begin{figure}[h]
	\plotone{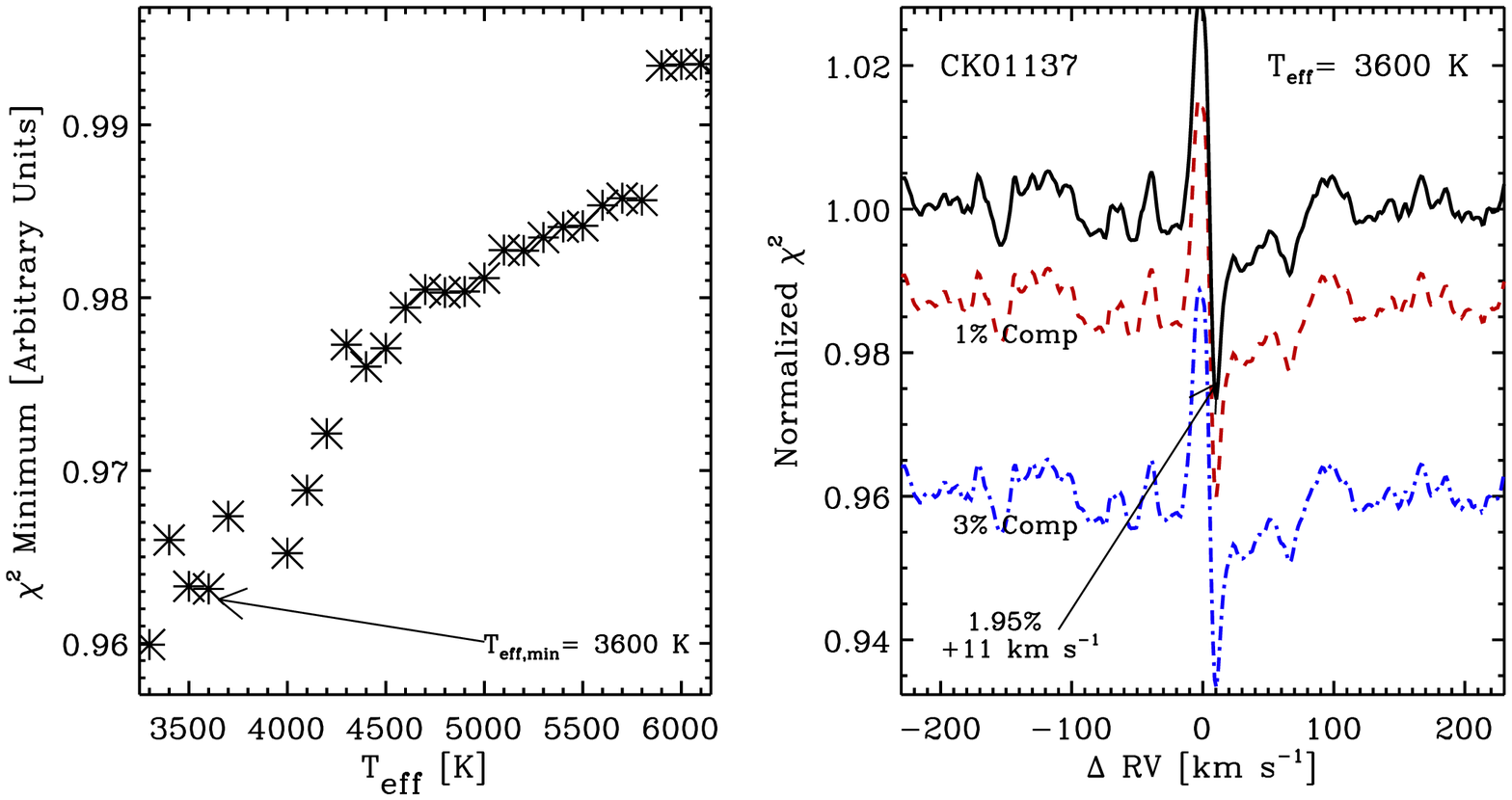}
	\caption{Final secondary star plot for KOI-1137. Same as Figure \ref{fig:complete_companion_plot}.}
	\label{fig:KOI-1137}
\end{figure}
\begin{figure}[h]
	\plotone{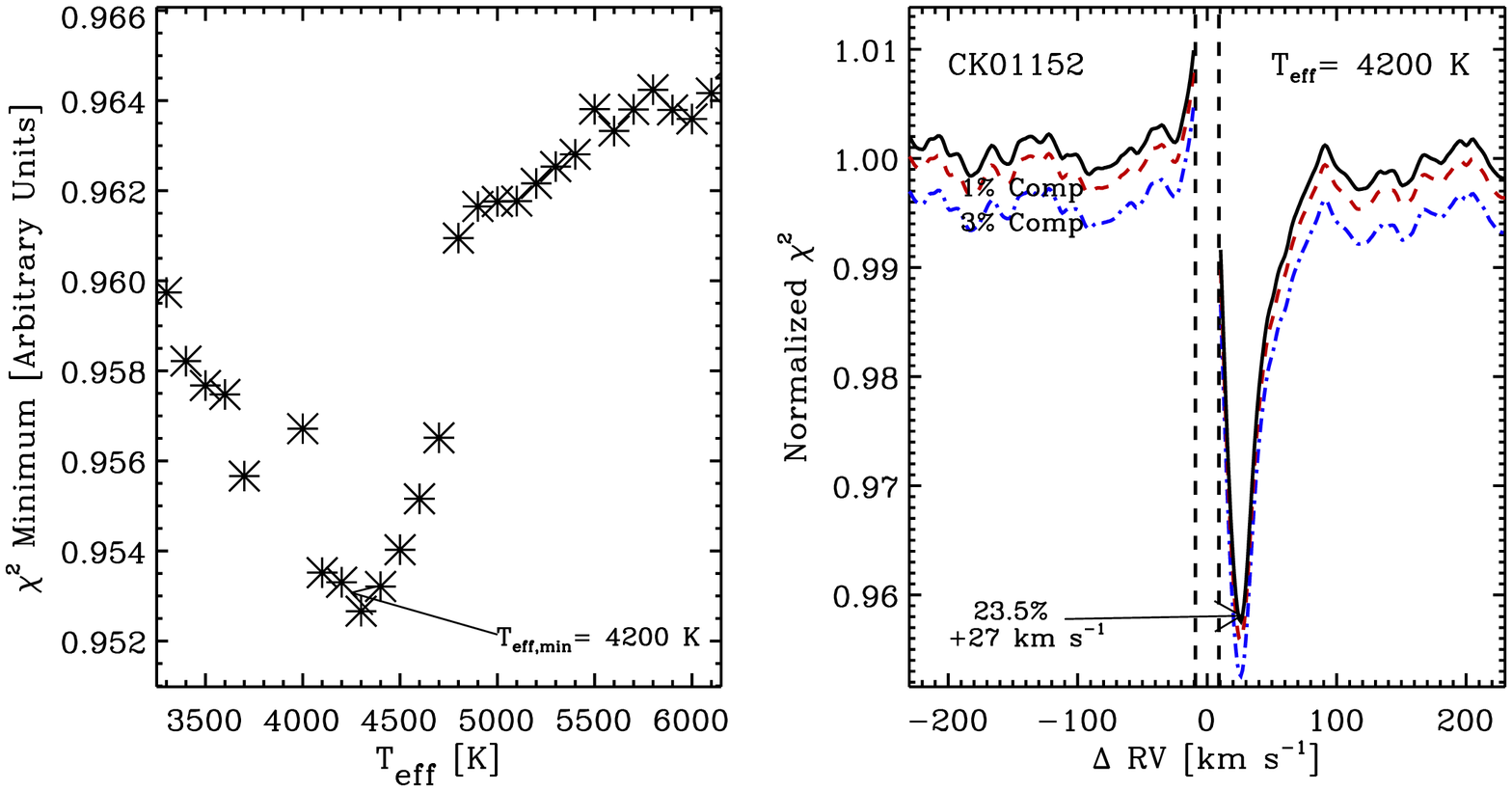}
	\caption{Final secondary star plot for KOI-1152. Same as Figure \ref{fig:complete_companion_plot}.}
	\label{fig:KOI-1152}
\end{figure}

\begin{figure}[h]
	\plotone{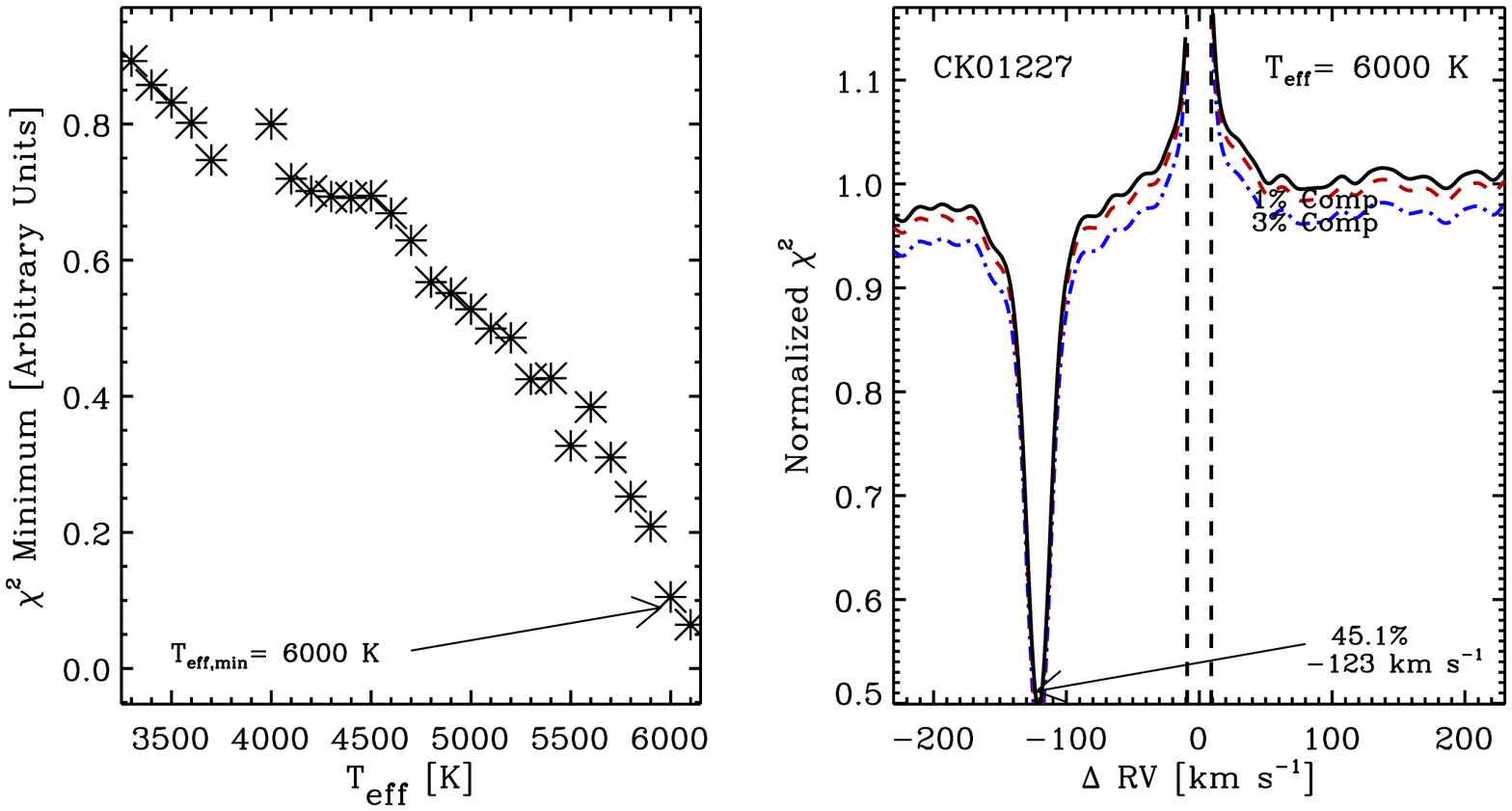}
	\caption{Final secondary star plot for KOI-1227. Same as Figure \ref{fig:complete_companion_plot}.}
	\label{fig:KOI-1227}
\end{figure}

\begin{figure}[h]
	\plotone{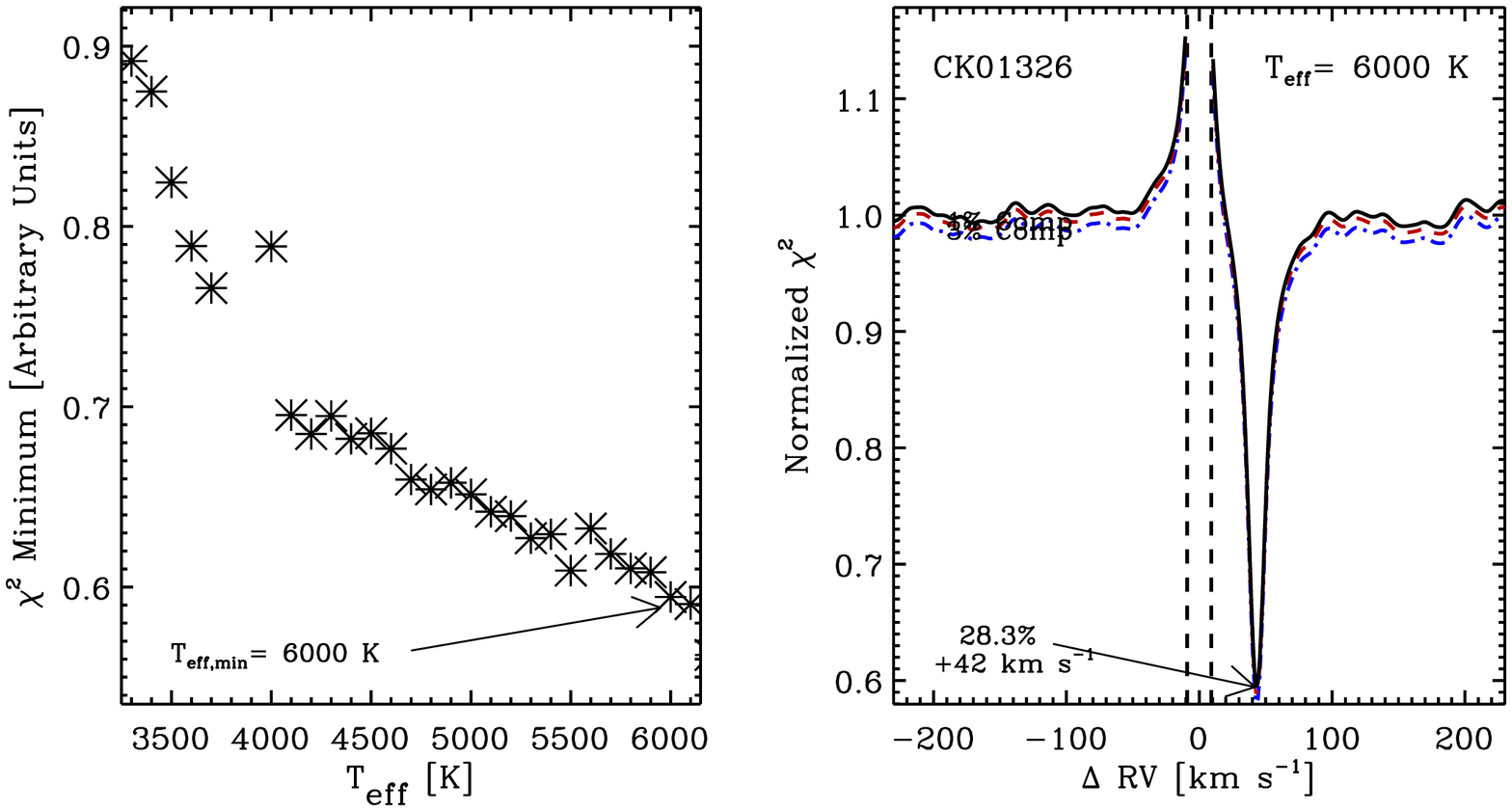}
	\caption{Final secondary star plot for KOI-1326. Same as Figure \ref{fig:complete_companion_plot}.}
	\label{fig:KOI-1326}
\end{figure}

\begin{figure}[h]
	\plotone{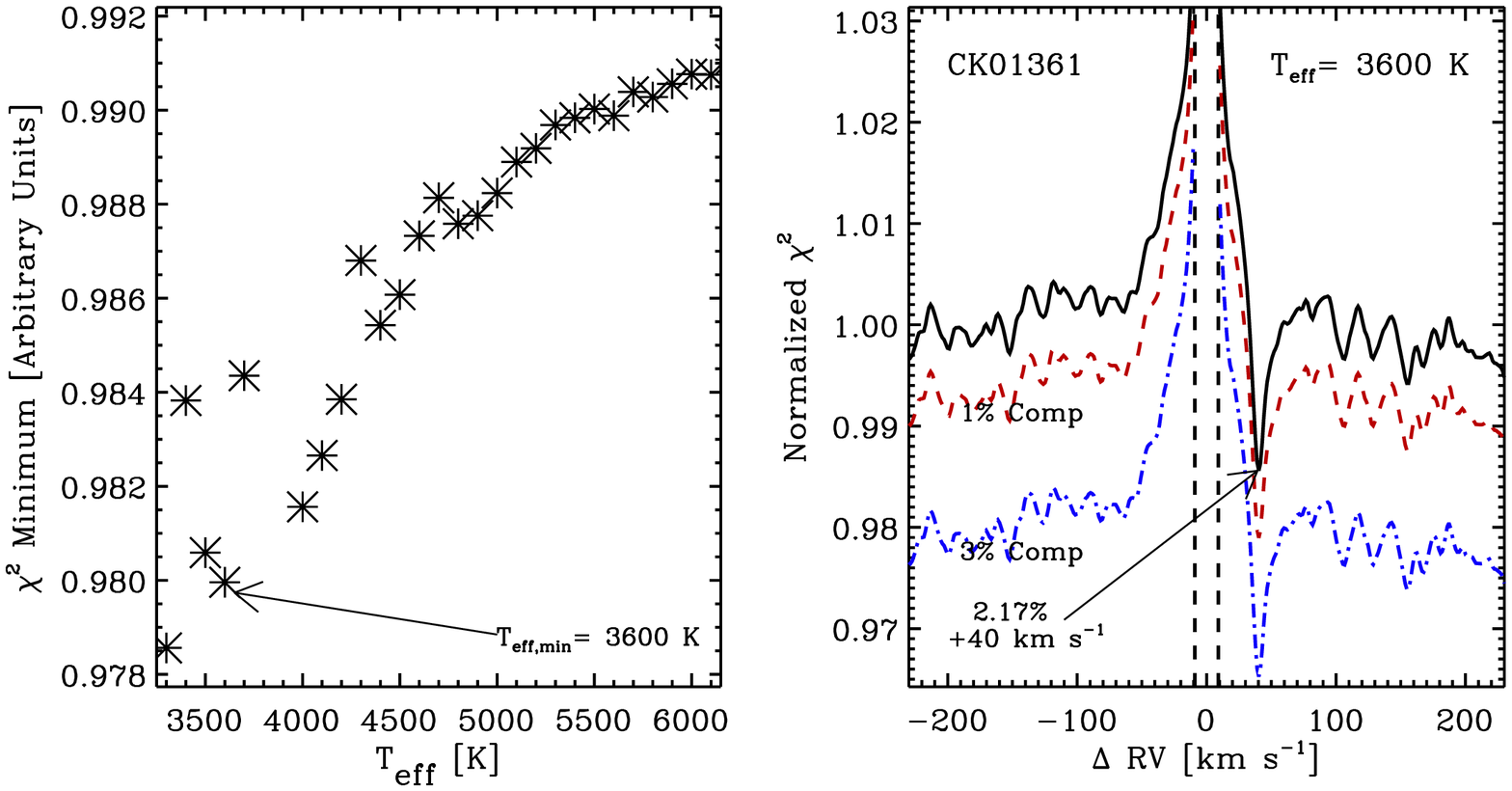}
	\caption{Final secondary star plot for KOI-1361. Same as Figure \ref{fig:complete_companion_plot}.}
	\label{fig:KOI-1361}
\end{figure}

\begin{figure}[h]
	\plotone{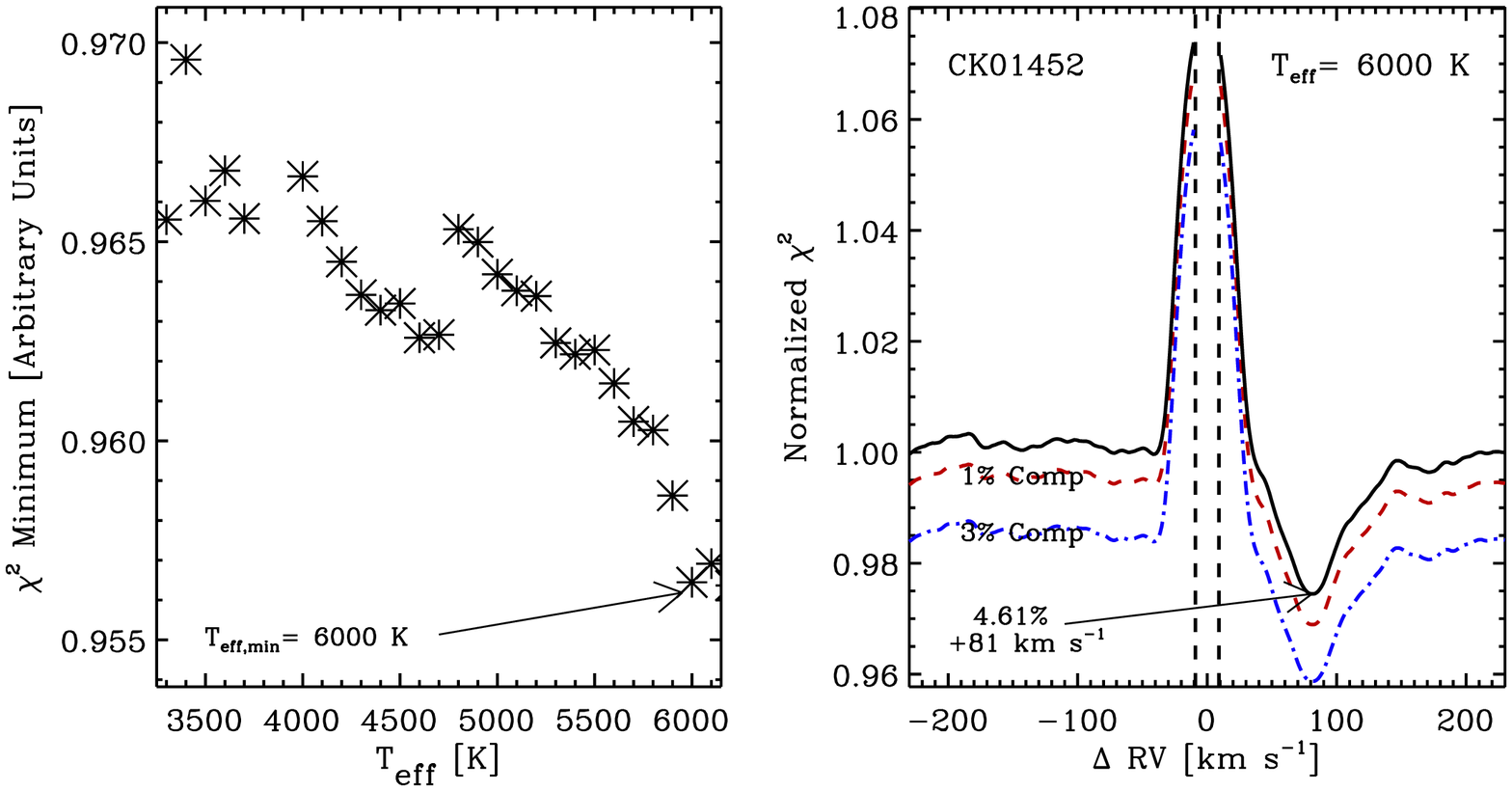}
	\caption{Final secondary star plot for KOI-1452. Same as Figure \ref{fig:complete_companion_plot}.}
	\label{fig:KOI-1452}
\end{figure}

\begin{figure}[h]
	\plotone{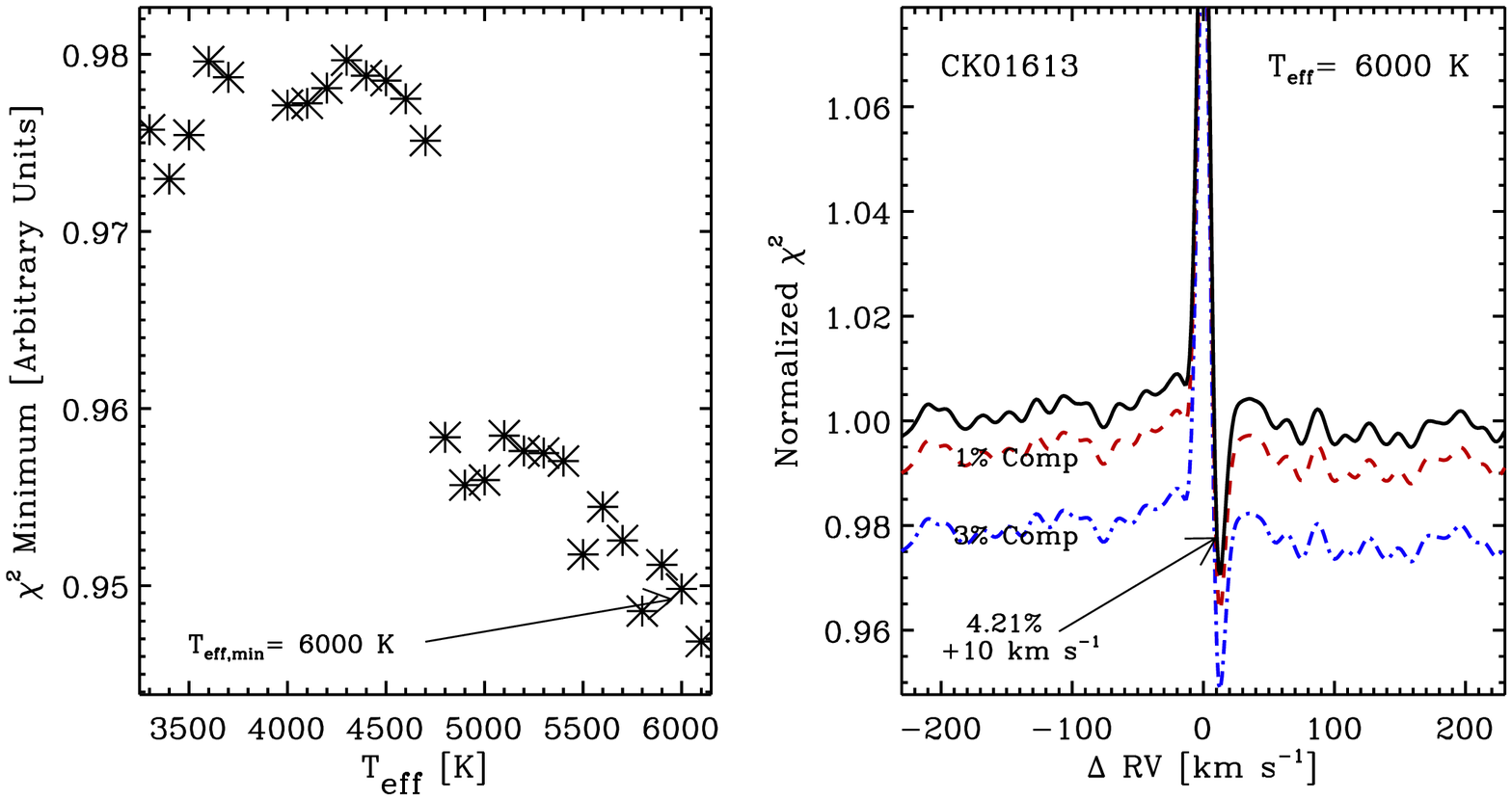}
	\caption{Final secondary star plot for KOI-1613. Same as Figure \ref{fig:complete_companion_plot}.}
	\label{fig:KOI-1613}
\end{figure}

\begin{figure}[h]
	\plotone{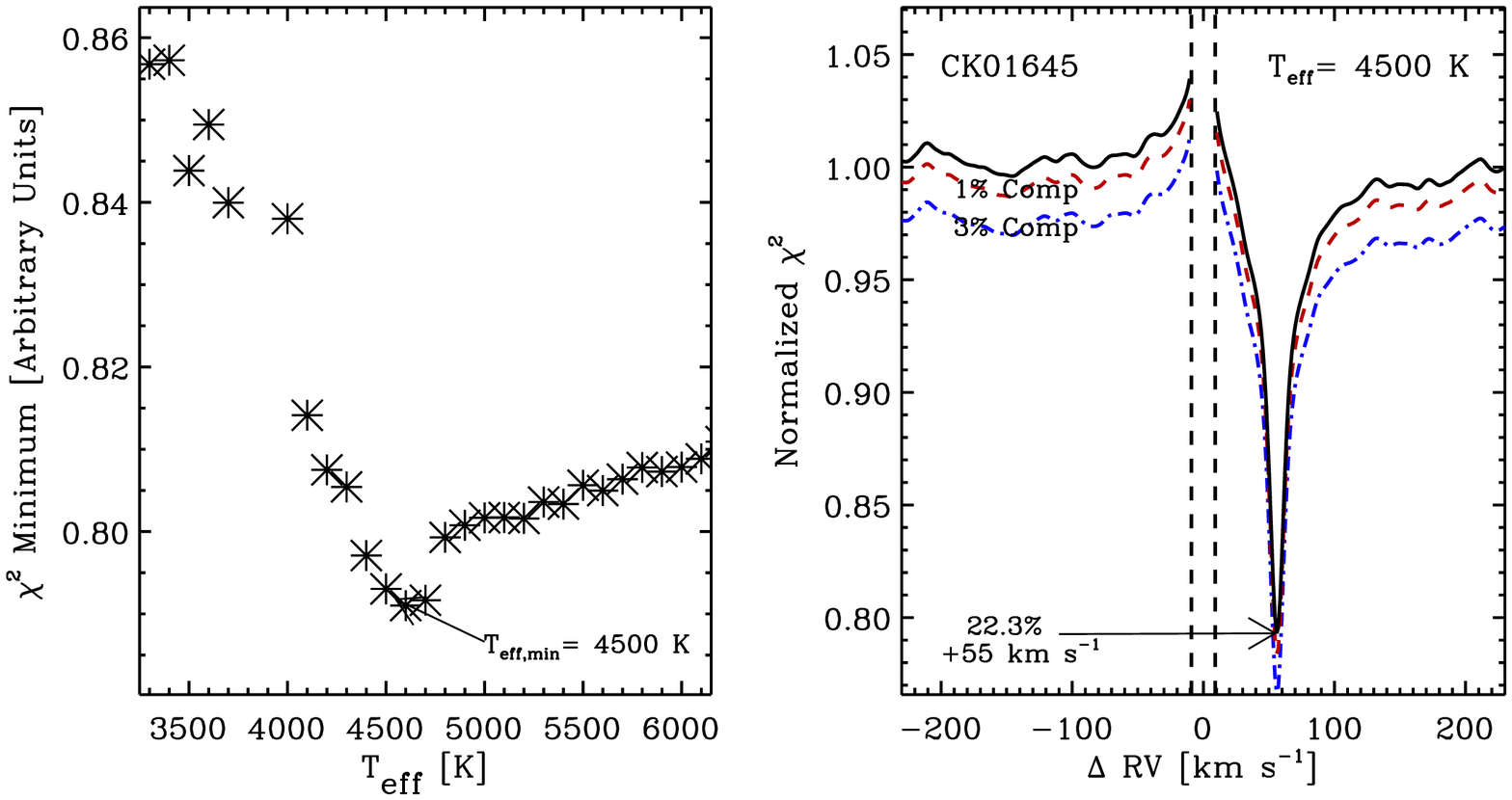}
	\caption{Final secondary star plot for KOI-1645. Same as Figure \ref{fig:complete_companion_plot}.}
	\label{fig:KOI-1645}
\end{figure}

\begin{figure}[h]
	\plotone{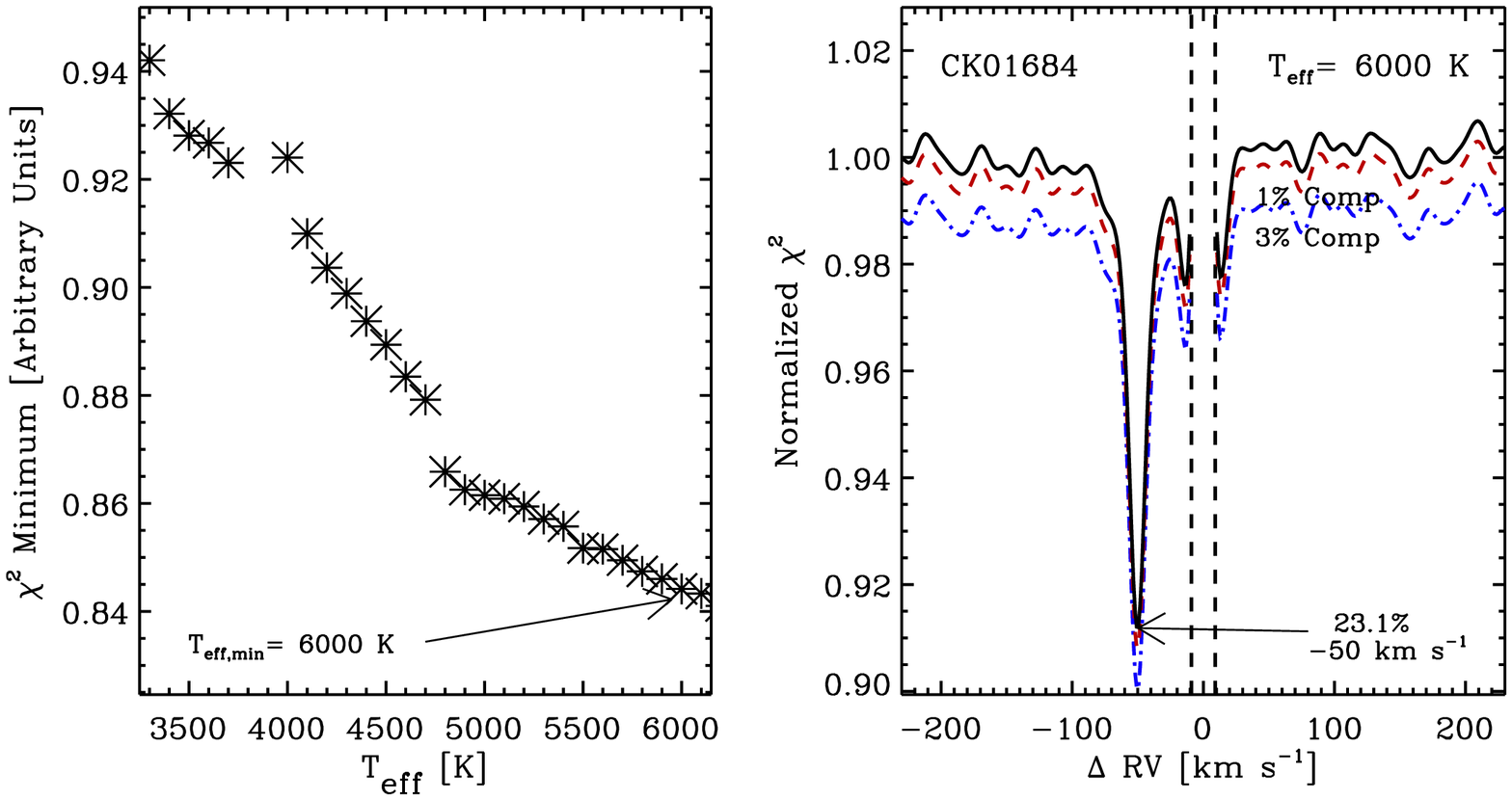}
	\caption{Final secondary star plot for KOI-1684. Same as Figure \ref{fig:complete_companion_plot}.}
	\label{fig:KOI-1684}
\end{figure}
\clearpage
\begin{figure}[h]
	\plotone{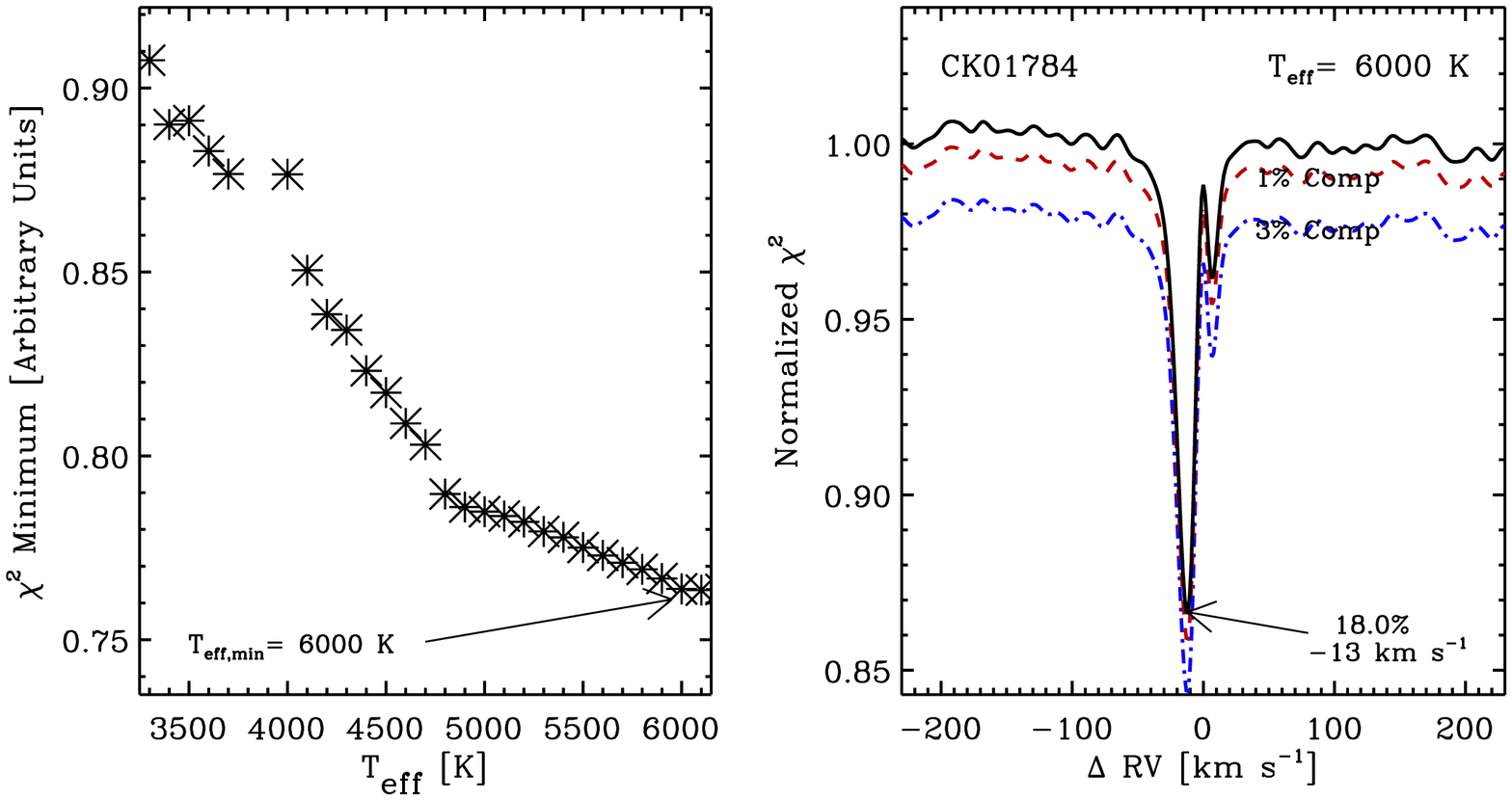}
	\caption{Final secondary star plot for KOI-1784. Same as Figure \ref{fig:complete_companion_plot}.}
	\label{fig:KOI-1784}
\end{figure}

\begin{figure}[h]
	\plotone{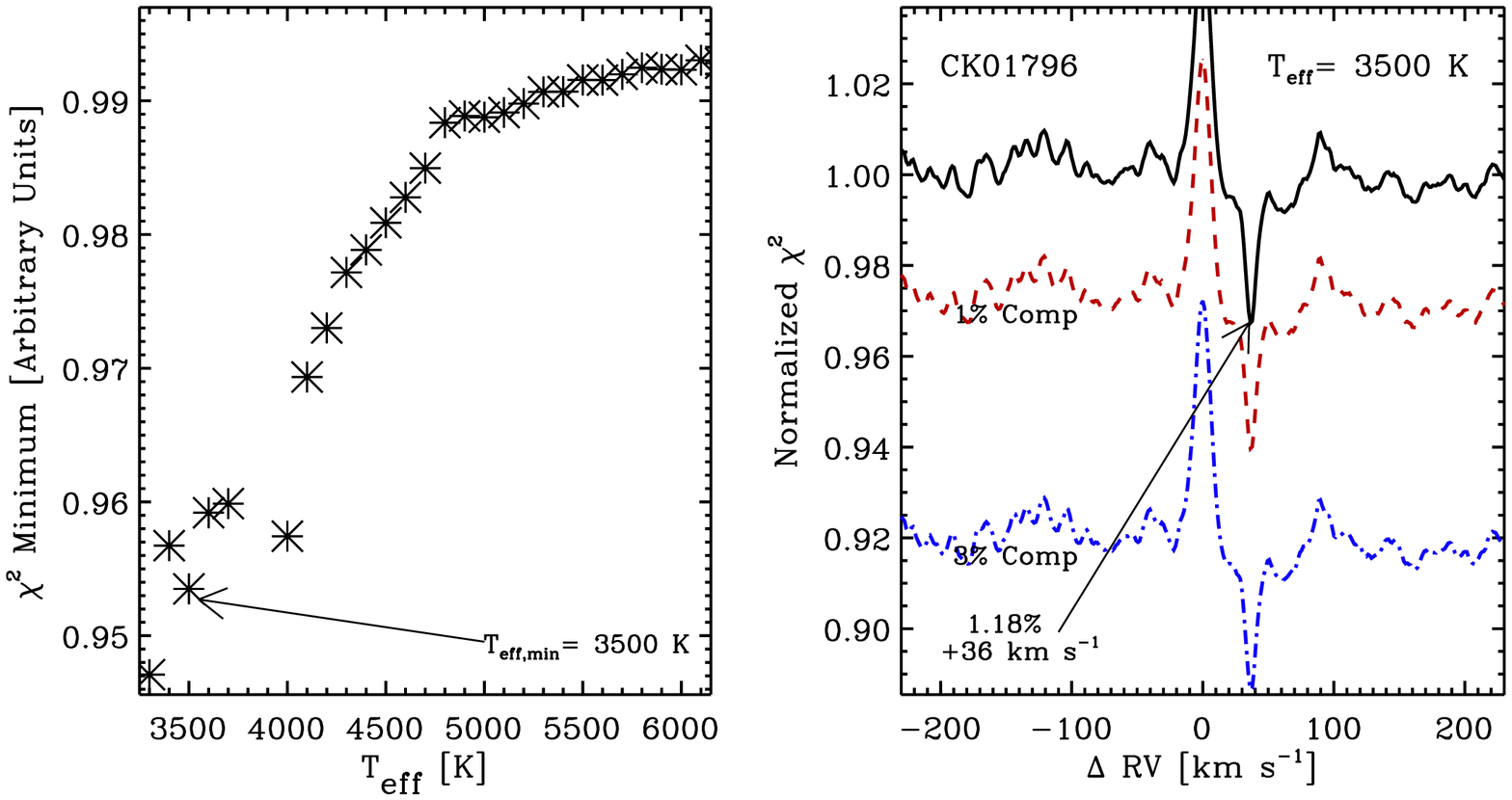}
	\caption{Final secondary star plot for KOI-1796. Same as Figure \ref{fig:complete_companion_plot}.}
	\label{fig:KOI-1798}
\end{figure}

\begin{figure}[h]
	\plotone{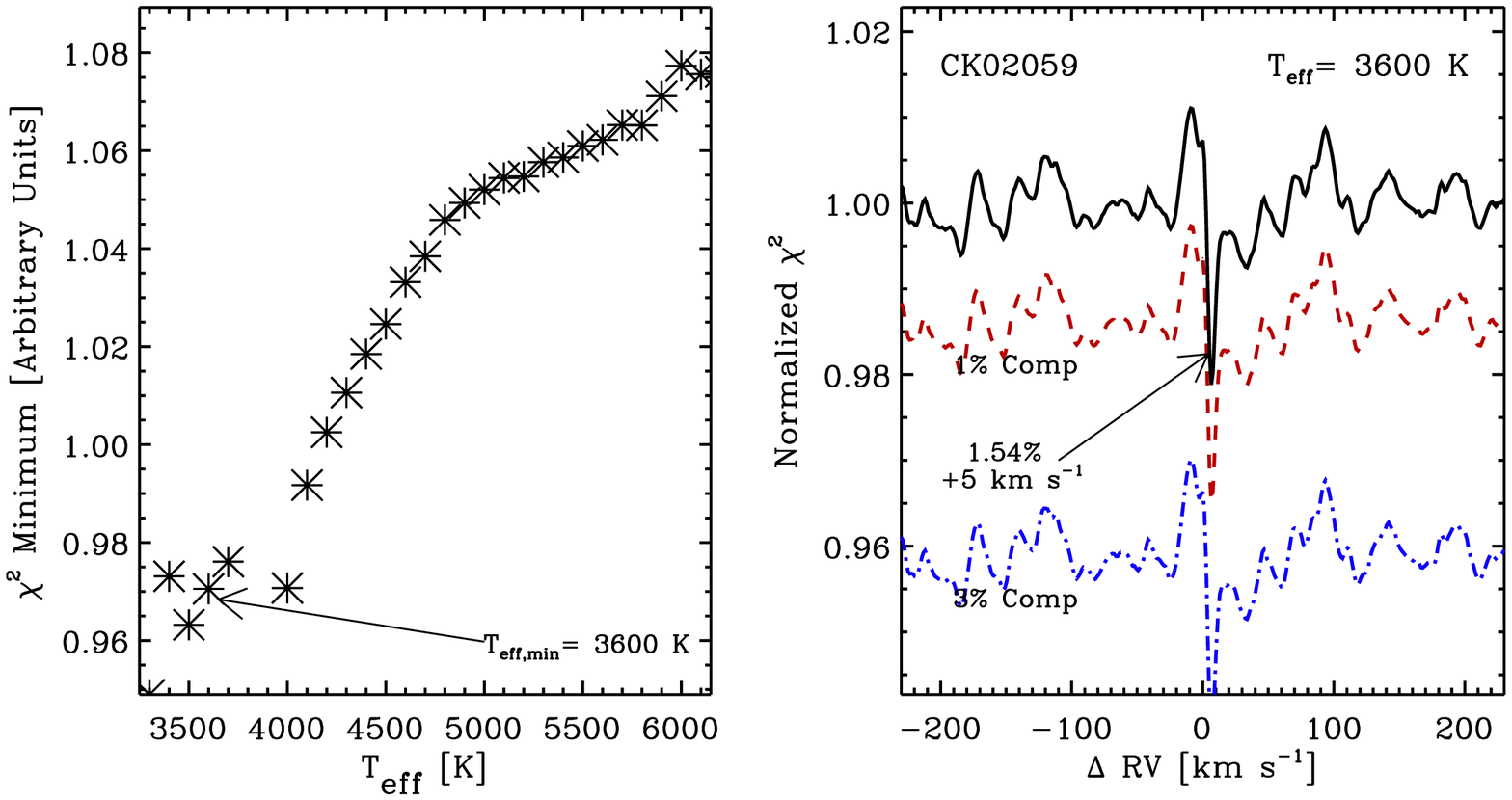}
	\caption{Final secondary star plot for KOI-2059. Same as Figure \ref{fig:complete_companion_plot}.}
	\label{fig:KOI-2059}
\end{figure}

\begin{figure}[h]
	\plotone{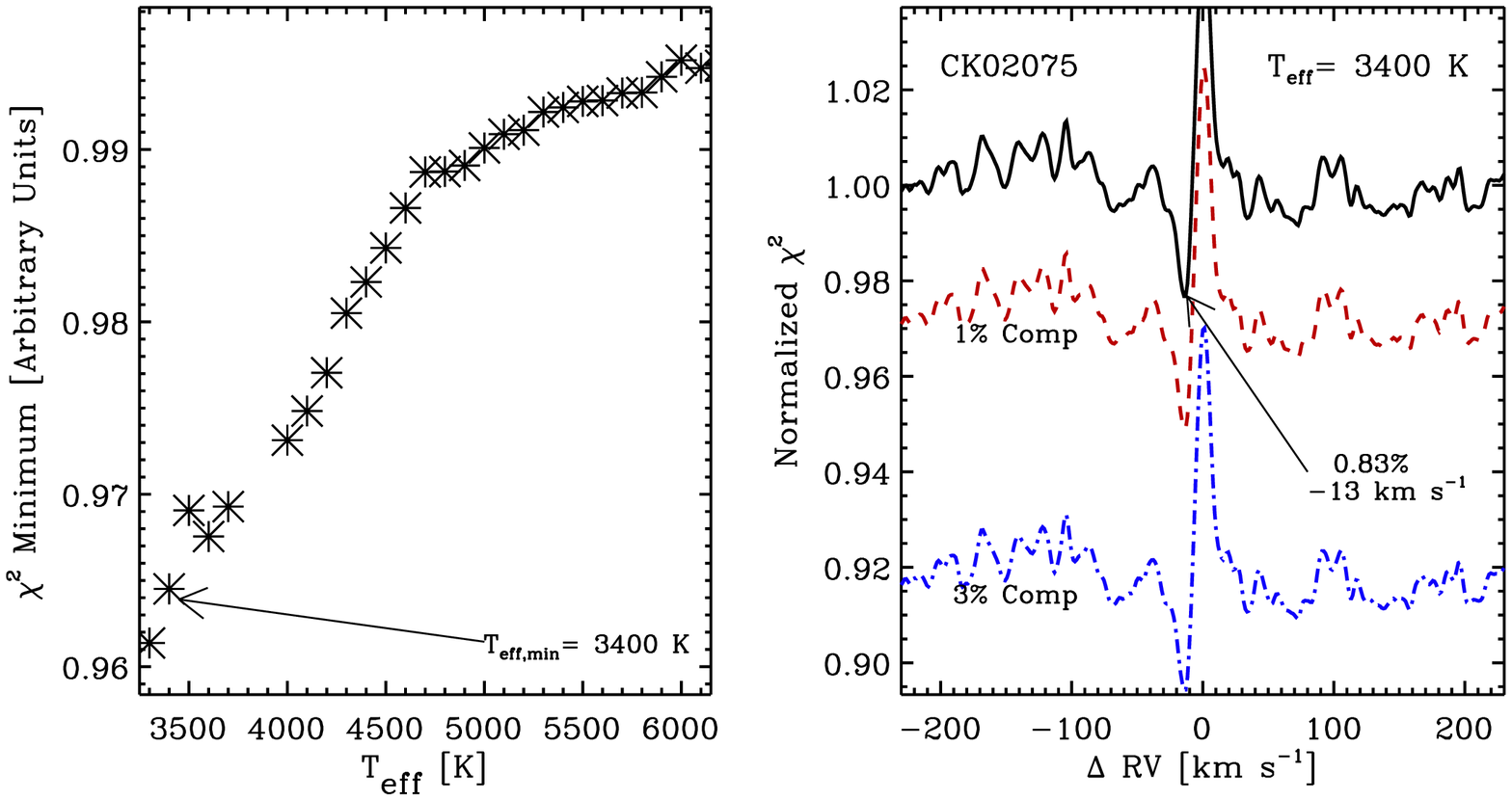}
	\caption{Final secondary star plot for KOI-2075. Same as Figure \ref{fig:complete_companion_plot}.}
	\label{fig:KOI-2075}
\end{figure}

\begin{figure}[h]
	\plotone{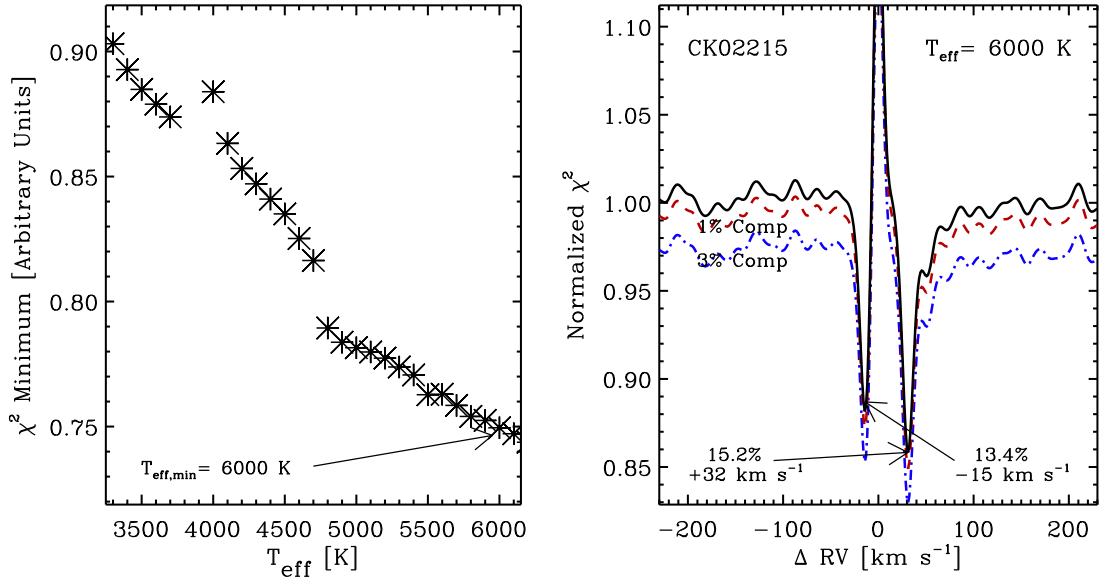}
	\caption{Final plot for KOI-2215. Same as Figure \ref{fig:complete_companion_plot}. Only the brightest, secondary star $\chi^2$ minima function is shown at left, while both additional stars are annotated with an arrow at right.}
	\label{fig:KOI-2215}
\end{figure}

\begin{figure}[h]
	\plotone{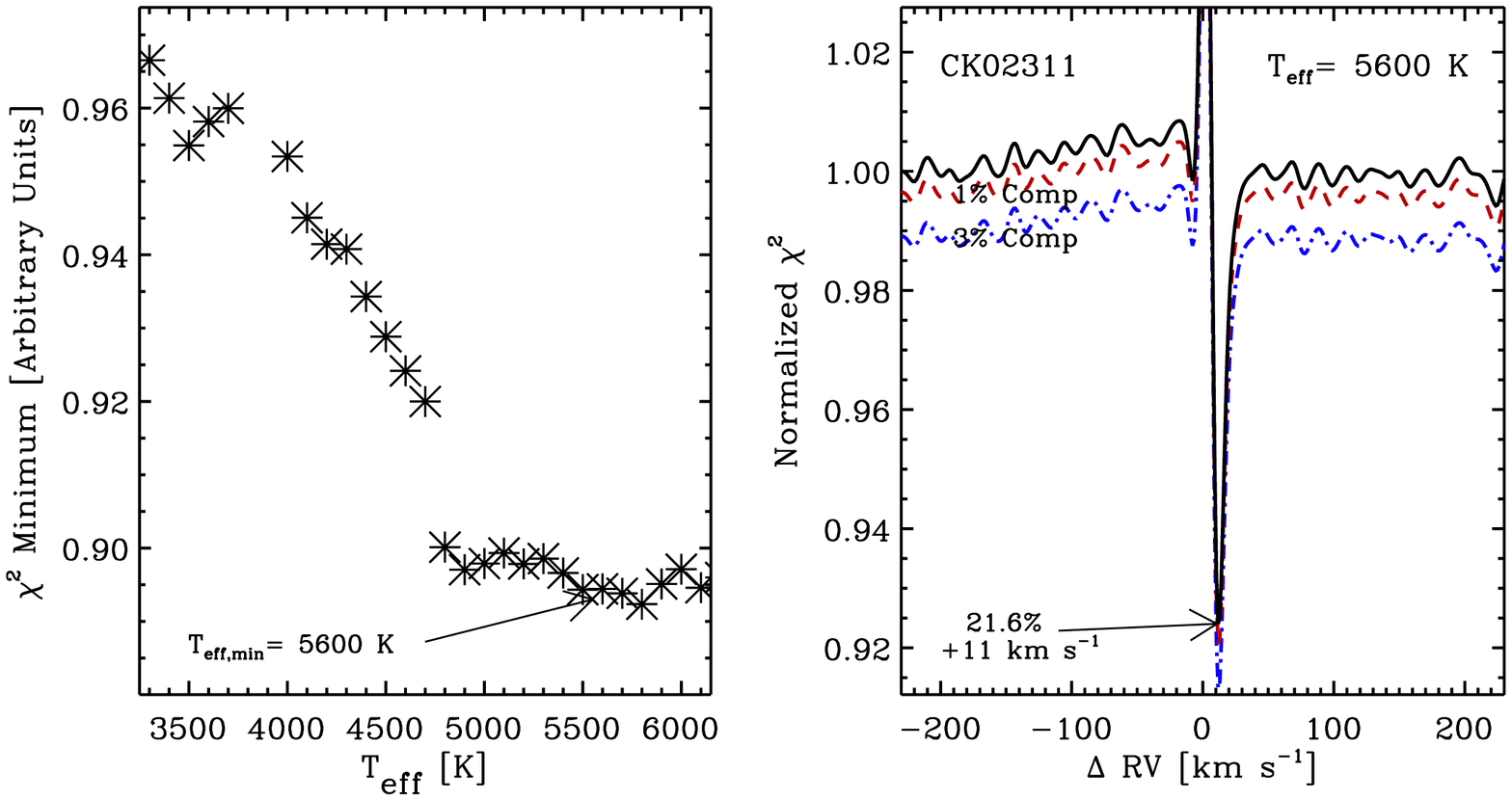}
	\caption{Final secondary star plot for KOI-2311. Same as Figure \ref{fig:complete_companion_plot}.}
	\label{fig:KOI-2311}
\end{figure}

\begin{figure}[h]
	\plotone{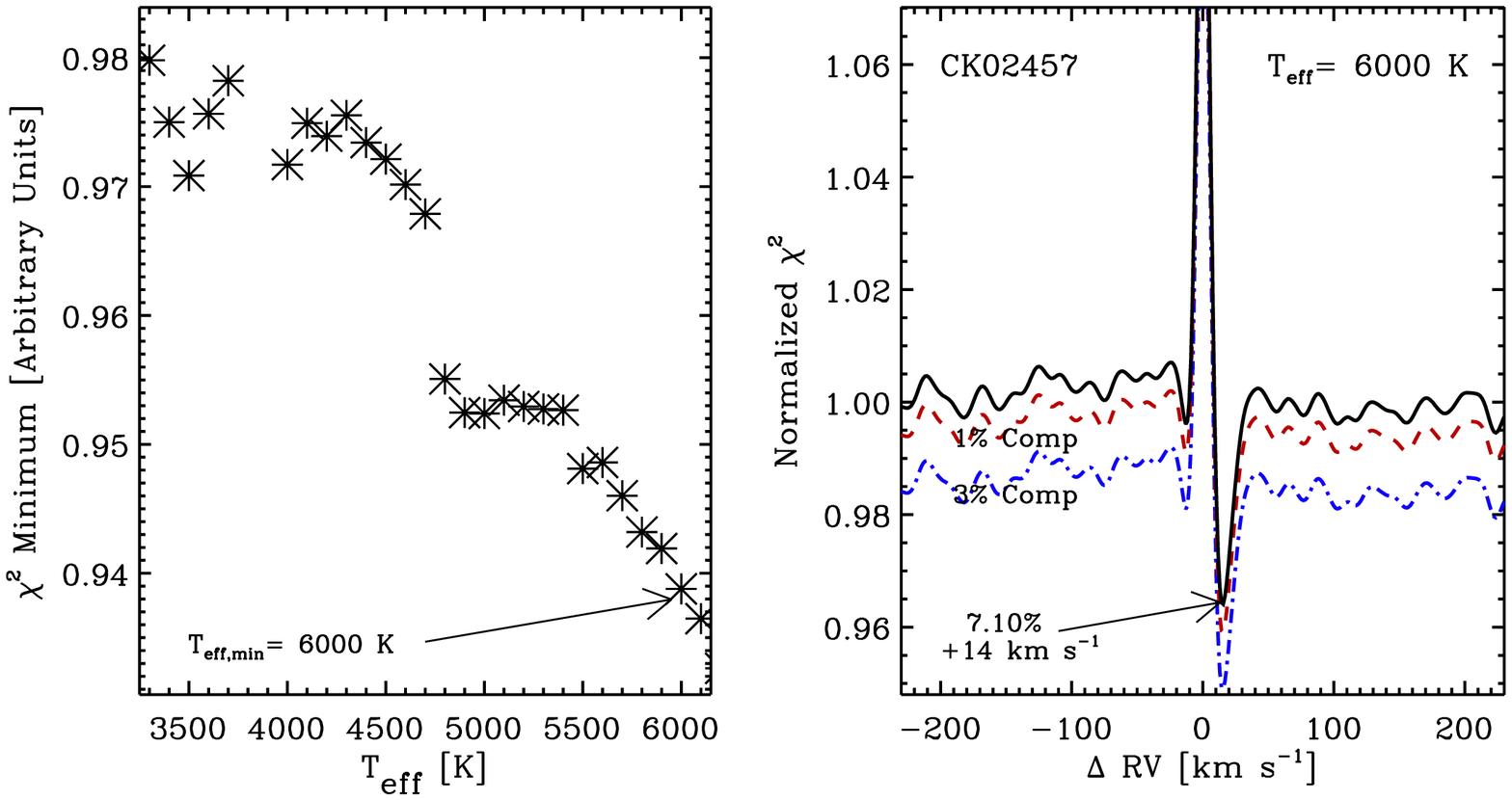}
	\caption{Final secondary star plot for KOI-2457. Same as Figure \ref{fig:complete_companion_plot}.}
	\label{fig:KOI-2457}
\end{figure}

\begin{figure}[h]
	\plotone{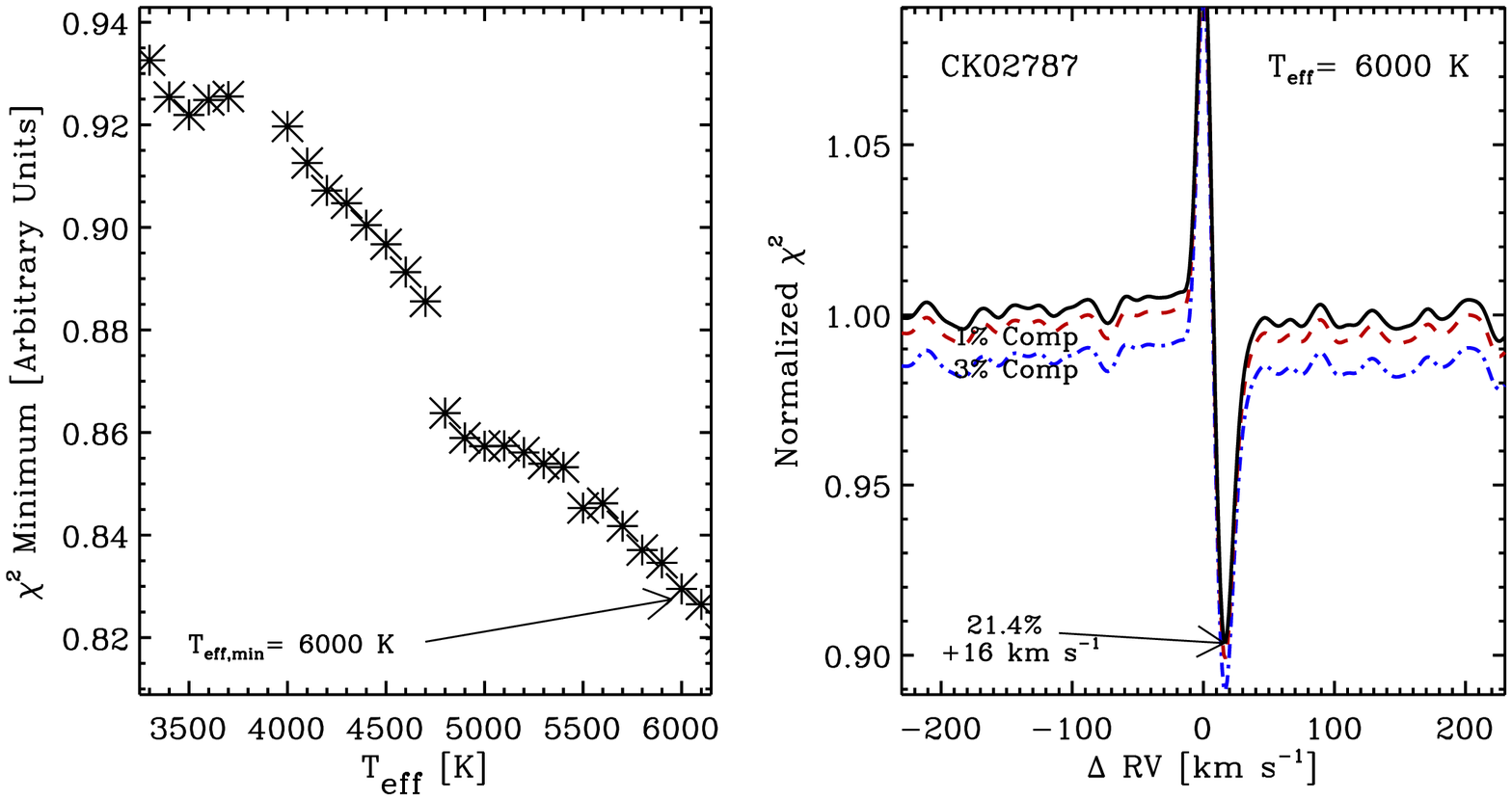}
	\caption{Final secondary star plot for KOI-2787. Same as Figure \ref{fig:complete_companion_plot}.}
	\label{fig:KOI-2787}
\end{figure}

\begin{figure}[h]
	\plotone{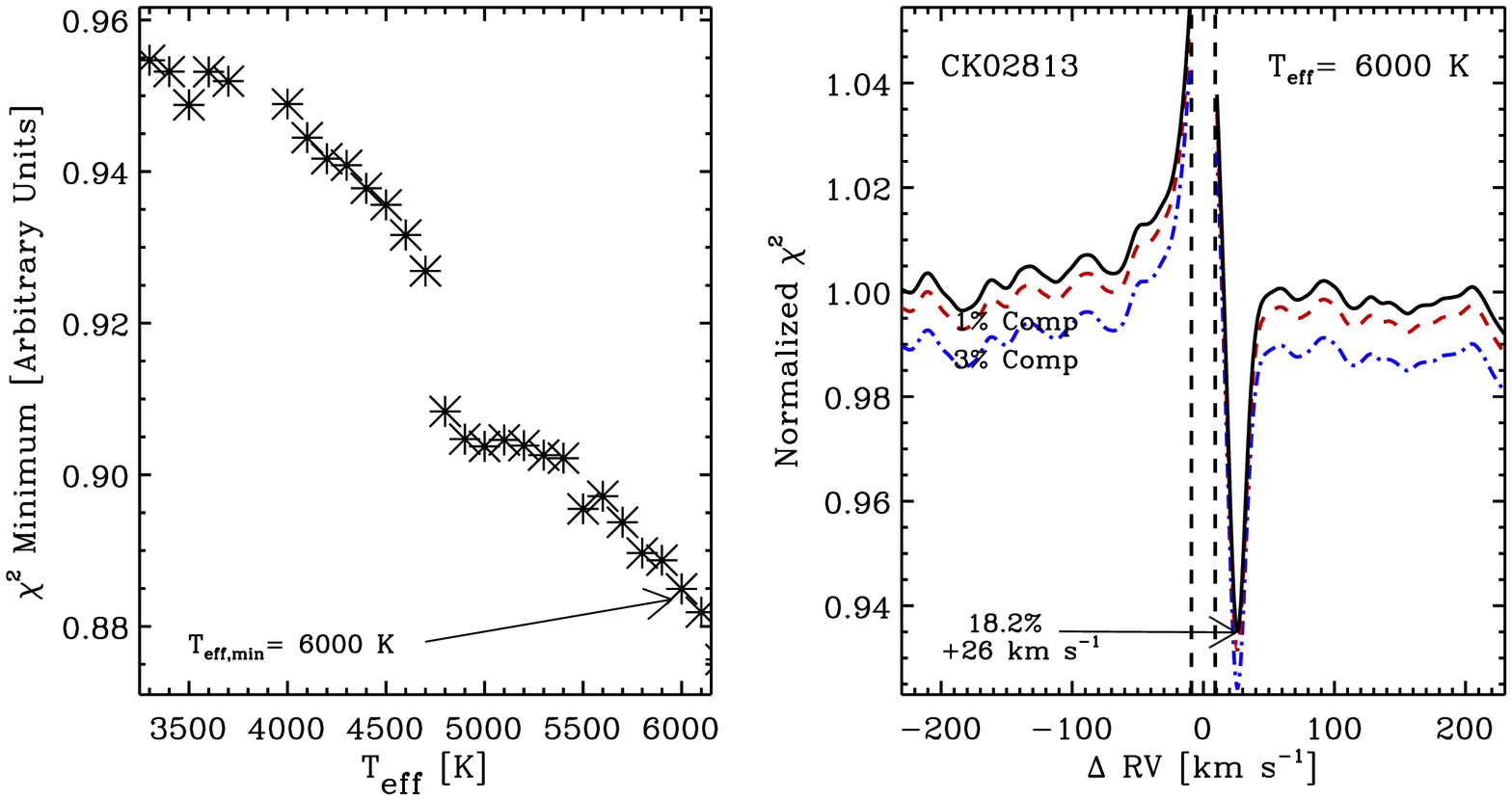}
	\caption{Final secondary star plot for KOI-2813. Same as Figure \ref{fig:complete_companion_plot}.}
	\label{fig:KOI-2813}
\end{figure}

\begin{figure}[H]
	\plotone{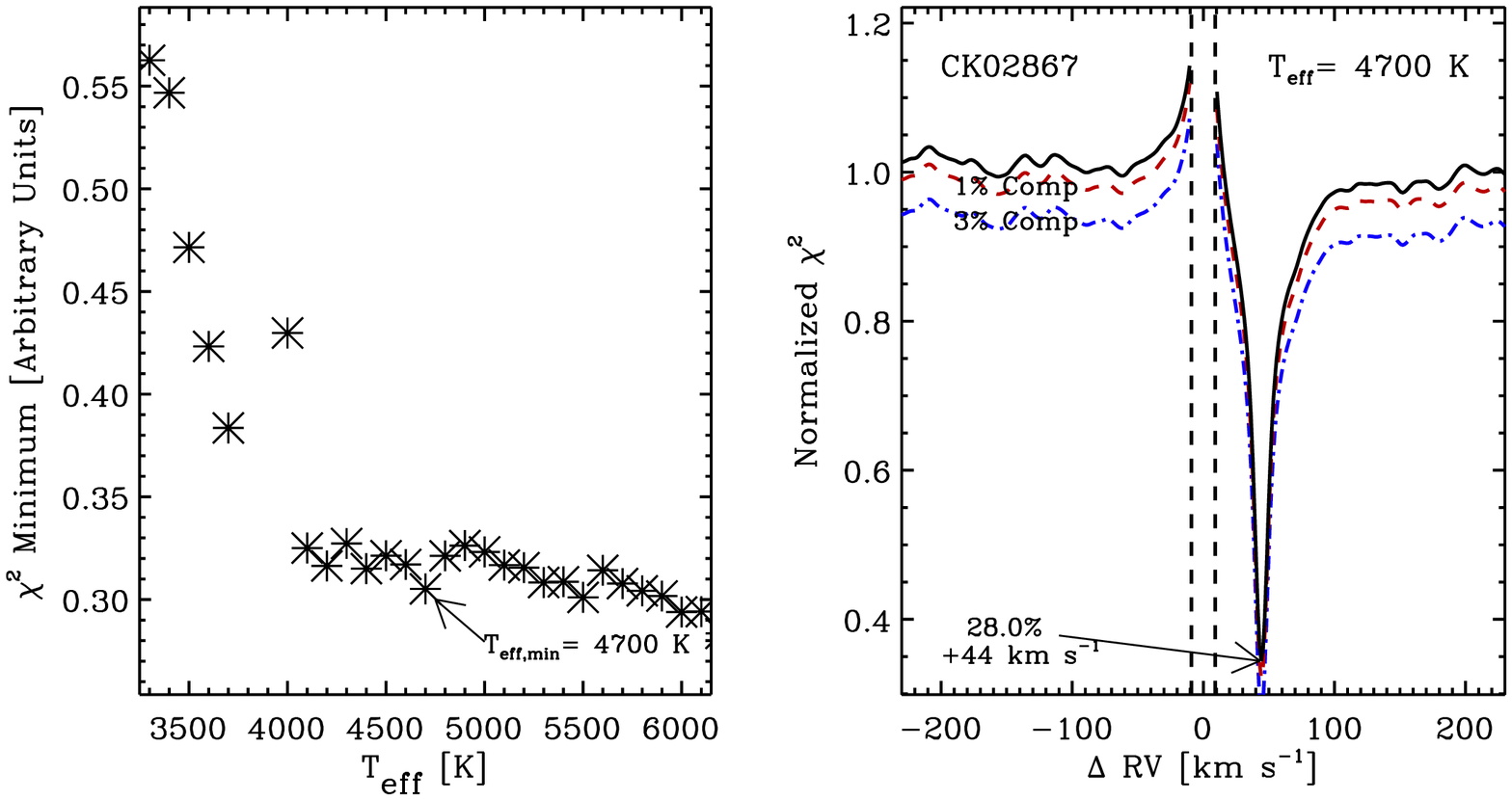}
	\caption{Final secondary star plot for KOI-2867. Same as Figure \ref{fig:complete_companion_plot}.}
	\label{fig:KOI-2867}
\end{figure}

\begin{figure}[h]
	\plotone{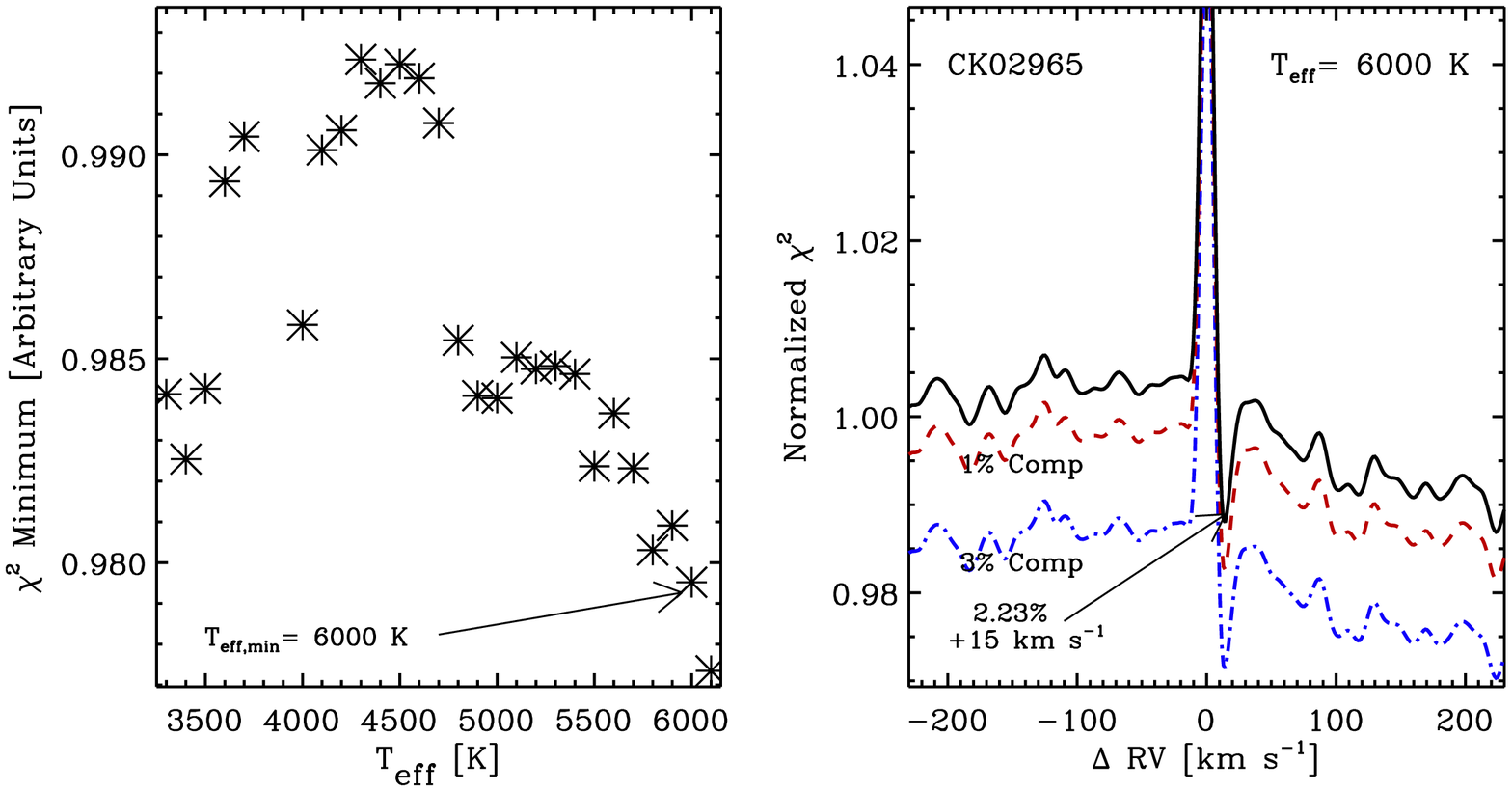}
	\caption{Final secondary star plot for KOI-2965. Same as Figure \ref{fig:complete_companion_plot}.}
	\label{fig:KOI-2965}
\end{figure}

\begin{figure}[h]
	\plotone{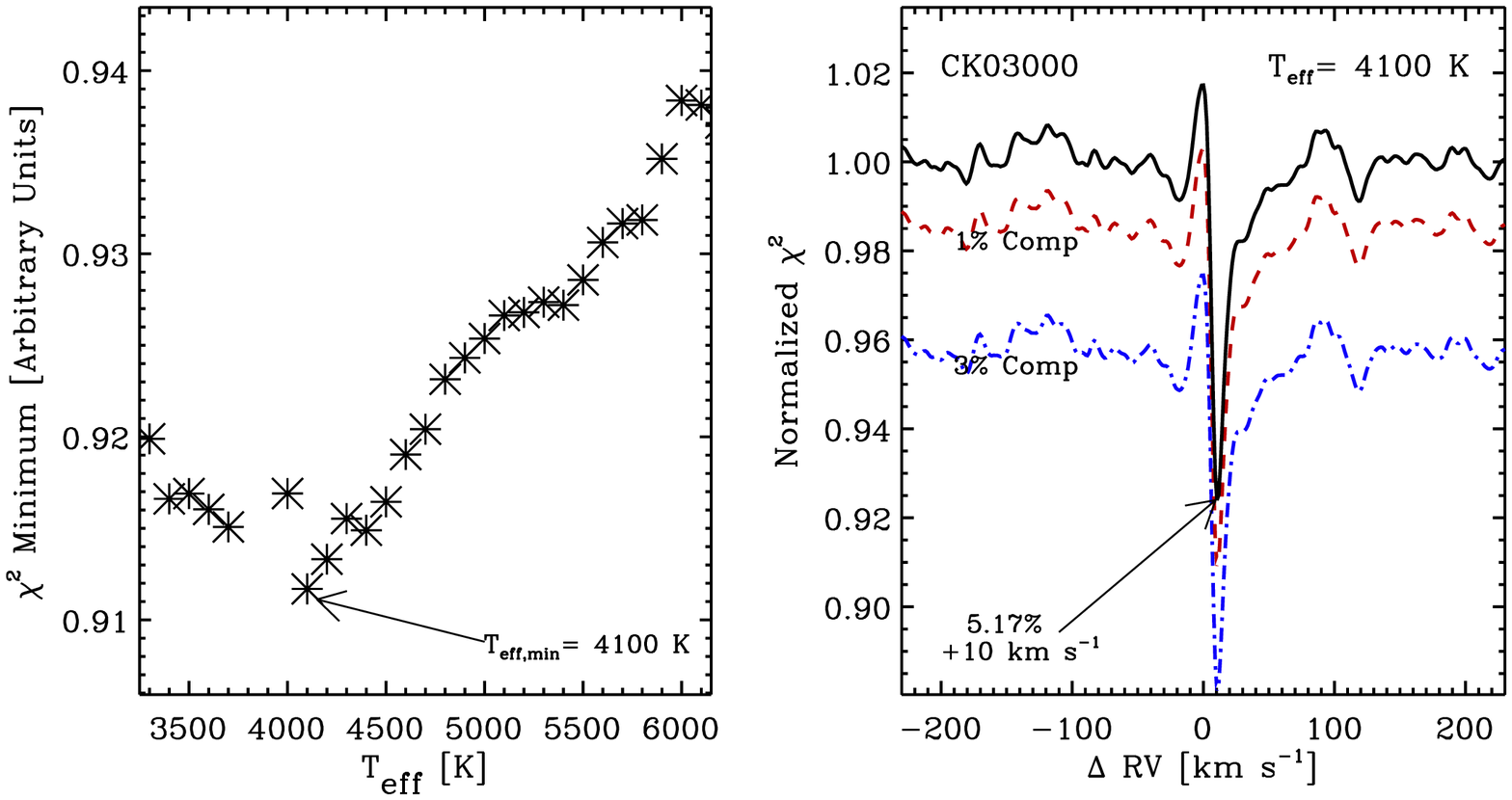}
	\caption{Final secondary star plot for KOI-3000. Same as Figure \ref{fig:complete_companion_plot}.}
	\label{fig:KOI-3000}
\end{figure}

\begin{figure}[h]
	\plotone{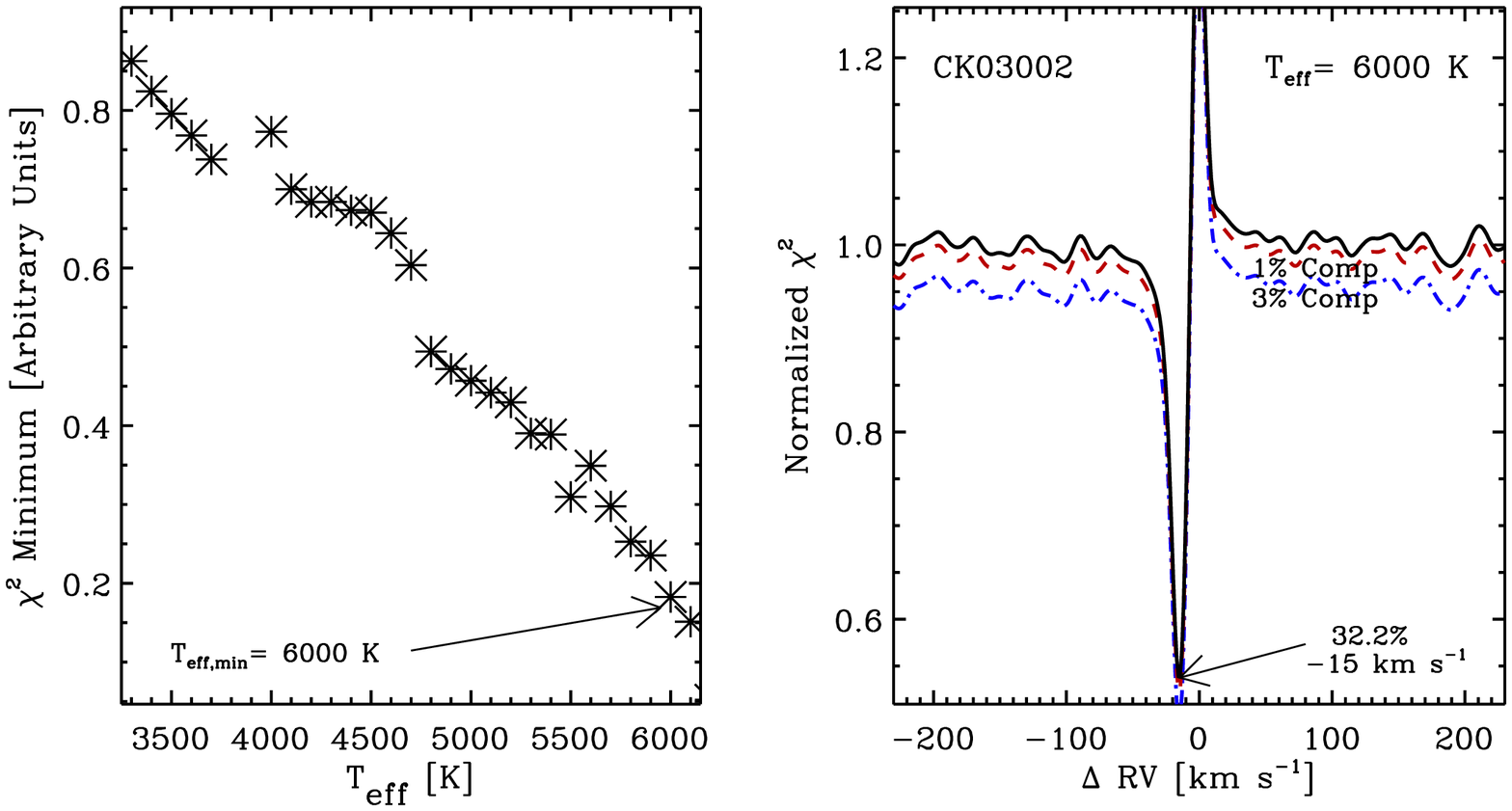}
	\caption{Final secondary star plot for KOI-3002. Same as Figure \ref{fig:complete_companion_plot}.}
	\label{fig:KOI-3002}
\end{figure}

\begin{figure}[h]
	\plotone{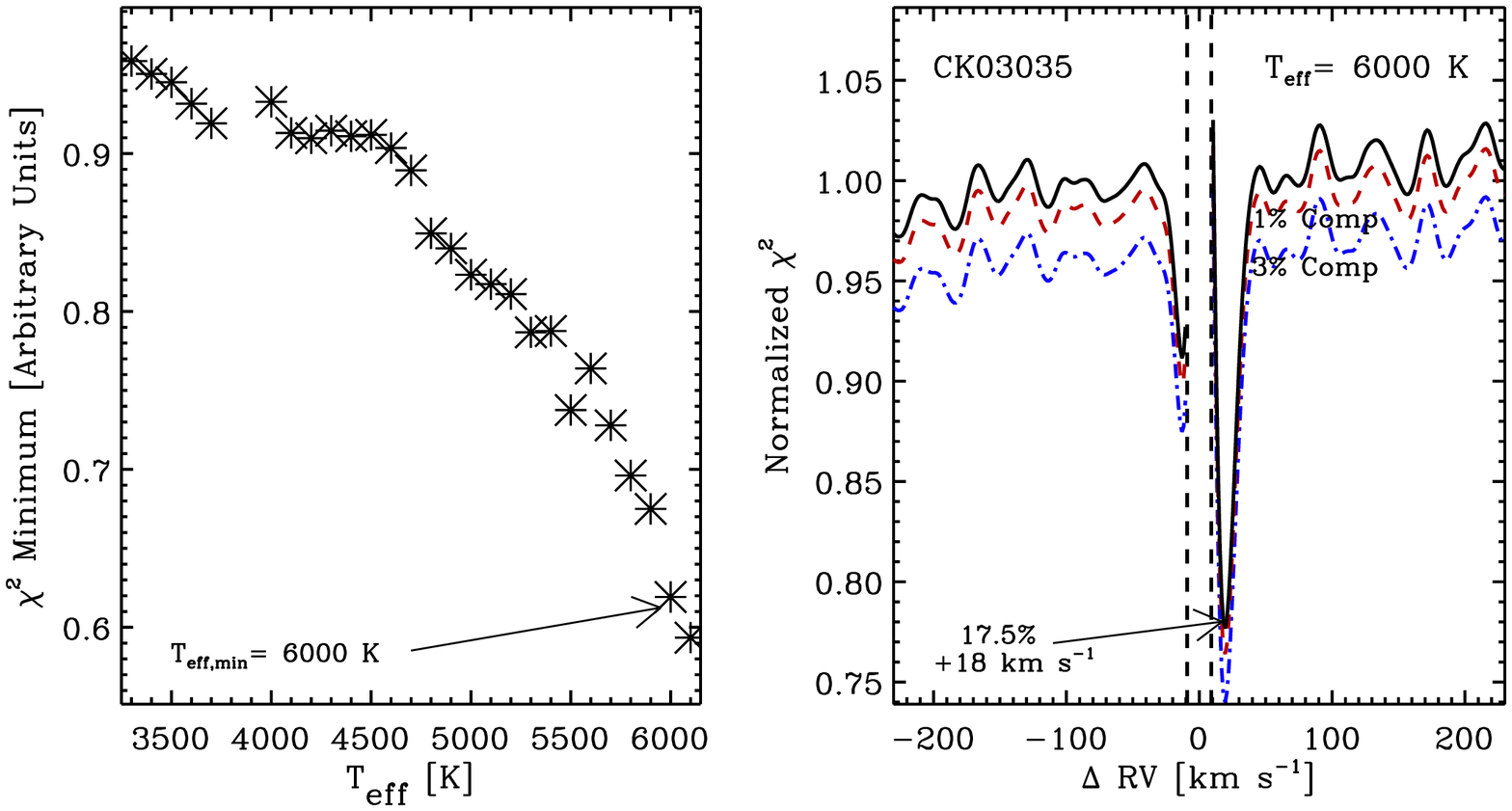}
	\caption{Final secondary star plot for KOI-3035. Same as Figure \ref{fig:complete_companion_plot}.}
	\label{fig:KOI-3035}
\end{figure}

\begin{figure}[h]
	\plotone{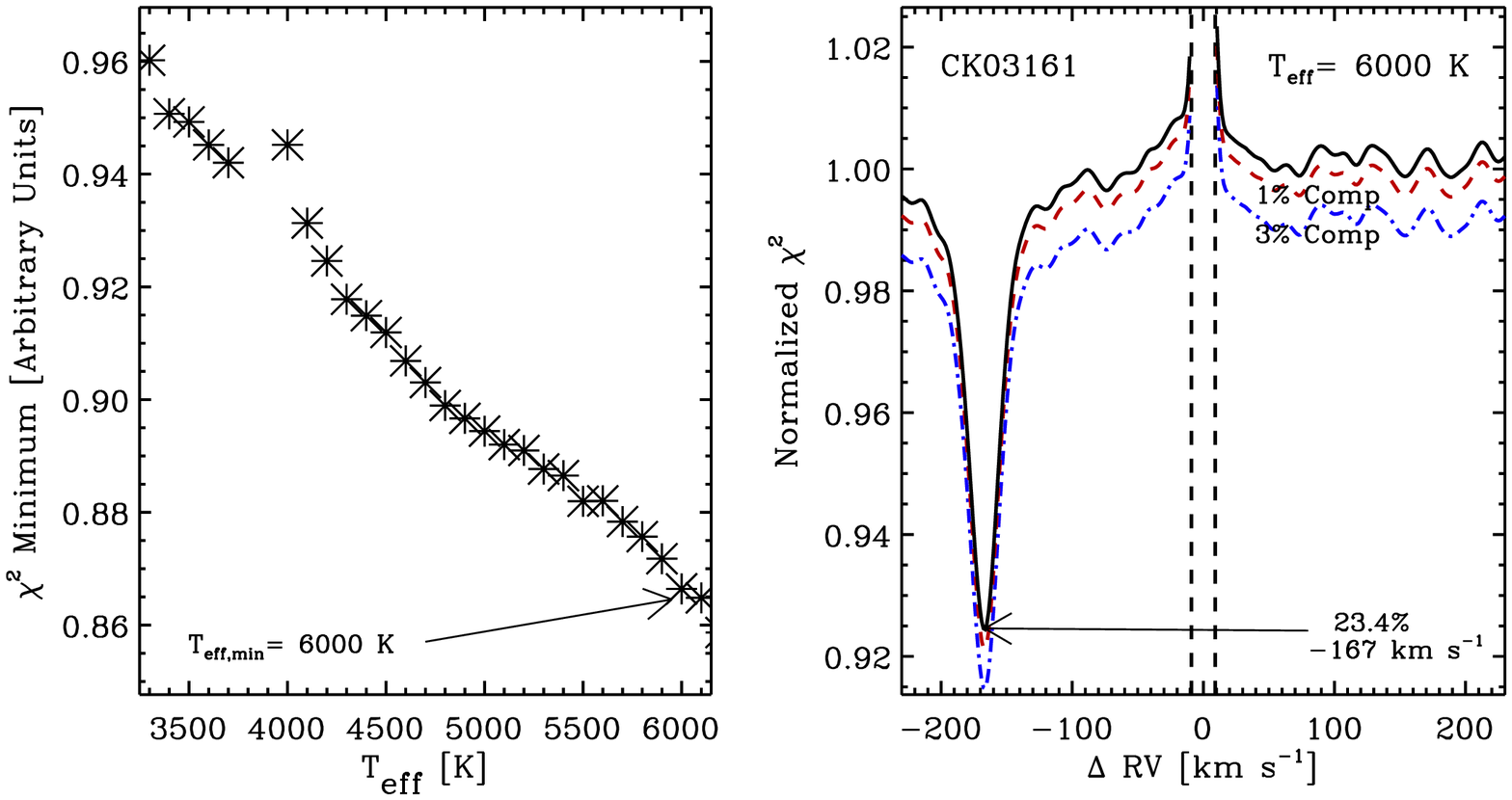}
	\caption{Final secondary star plot for KOI-3161. Same as Figure \ref{fig:complete_companion_plot}.}
	\label{fig:KOI-3161}
\end{figure}

\begin{figure}[h]
	\plotone{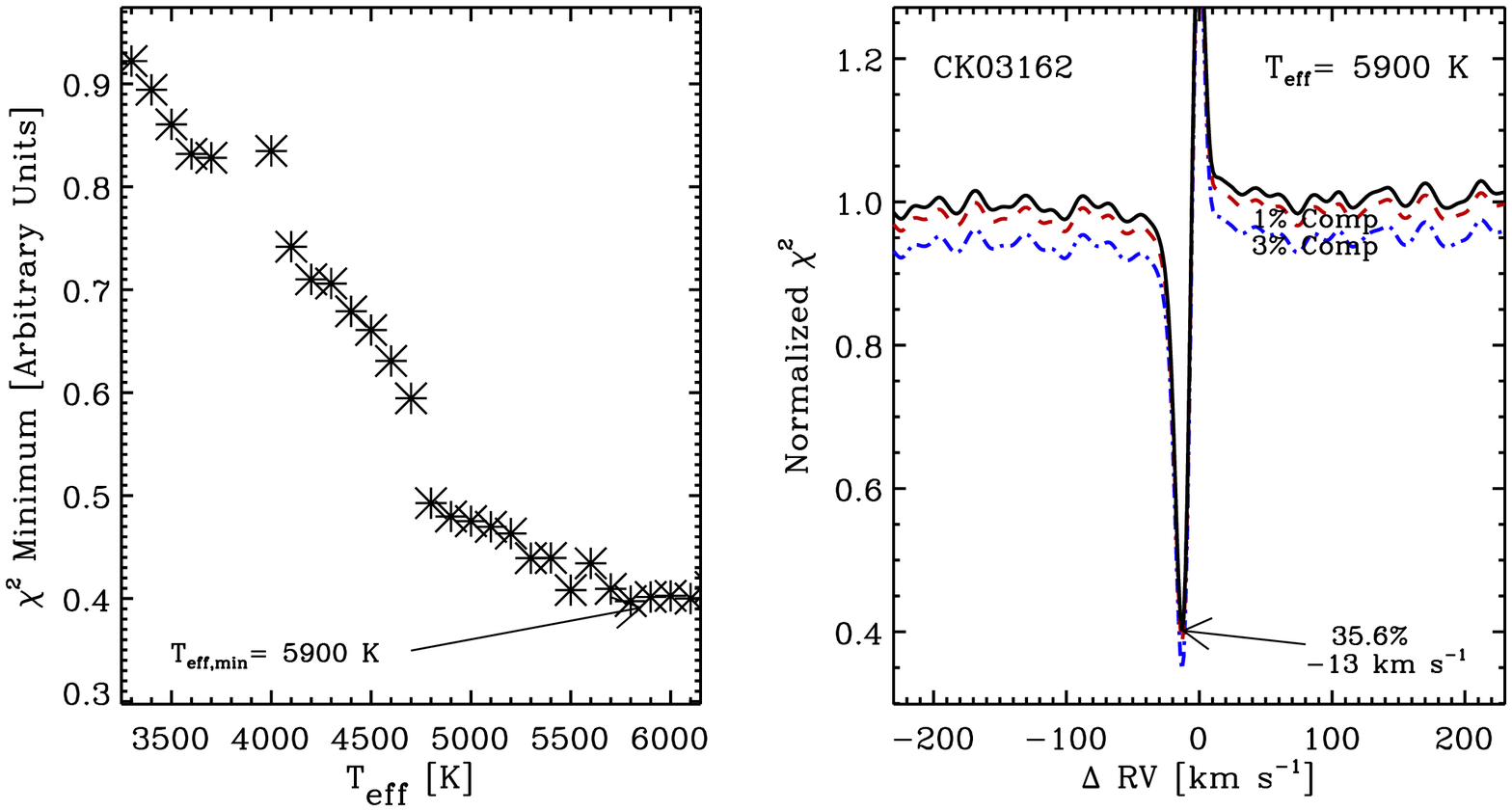}
	\caption{Final secondary star plot for KOI-3162. Same as Figure \ref{fig:complete_companion_plot}.}
	\label{fig:KOI-3162}
\end{figure}

\begin{figure}[h]
	\plotone{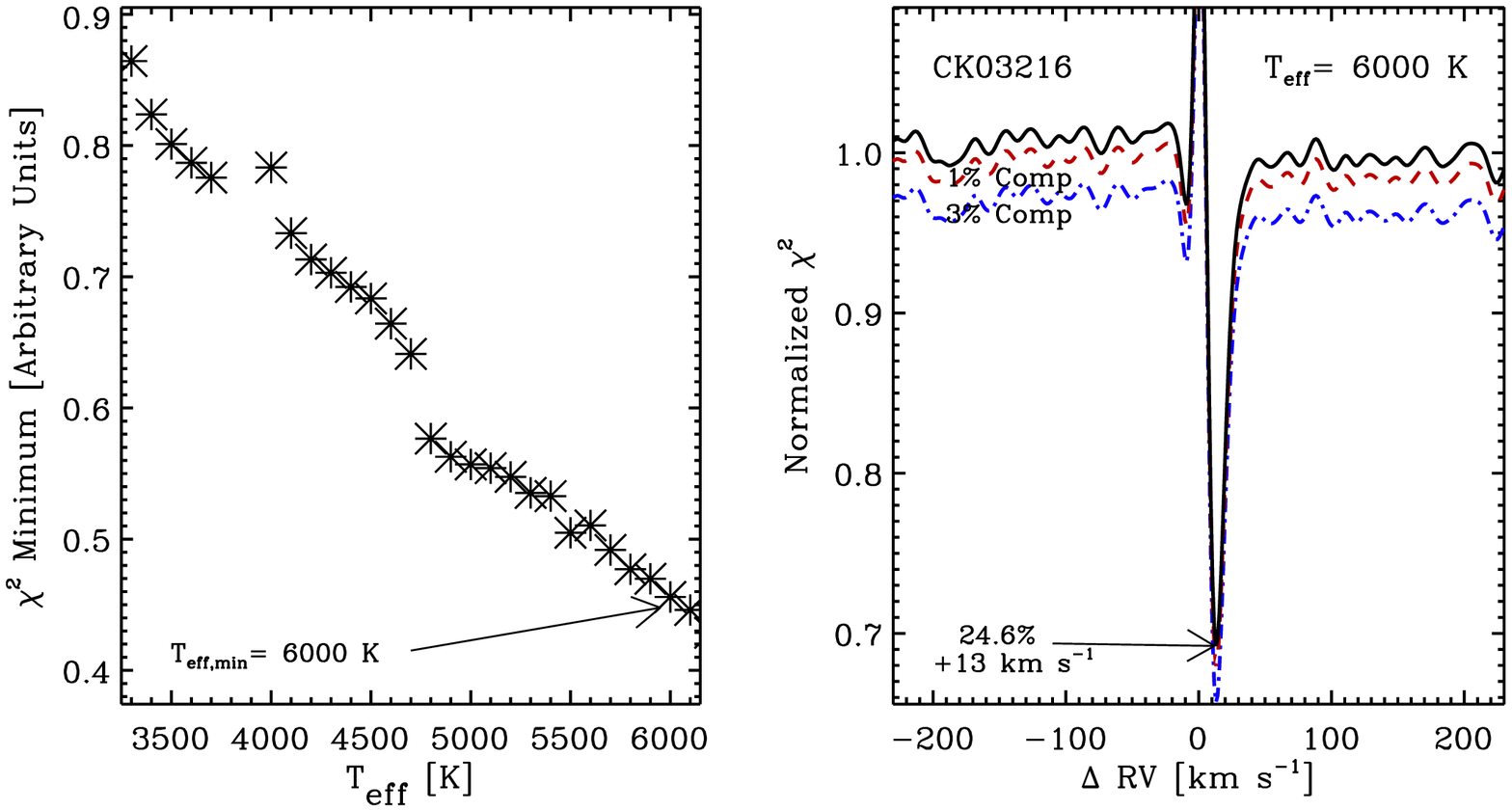}
	\caption{Final secondary star plot for KOI-3216. Same as Figure \ref{fig:complete_companion_plot}.}
	\label{fig:KOI-3216}
\end{figure}

\begin{figure}[h]
	\plotone{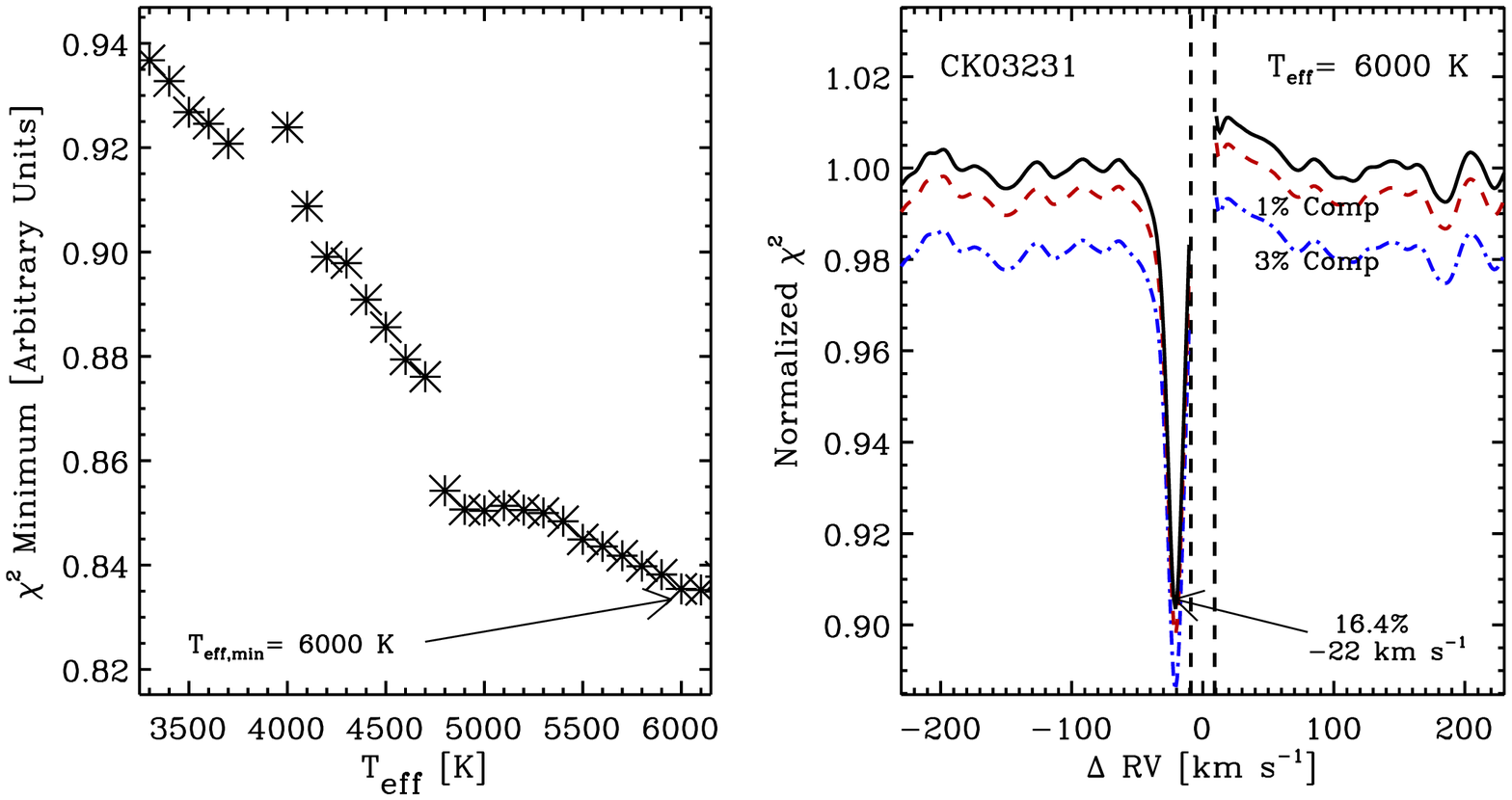}
	\caption{Final secondary star plot for KOI-3231. Same as Figure \ref{fig:complete_companion_plot}.}
	\label{fig:KOI-3231}
\end{figure}

\begin{figure}[h]
	\plotone{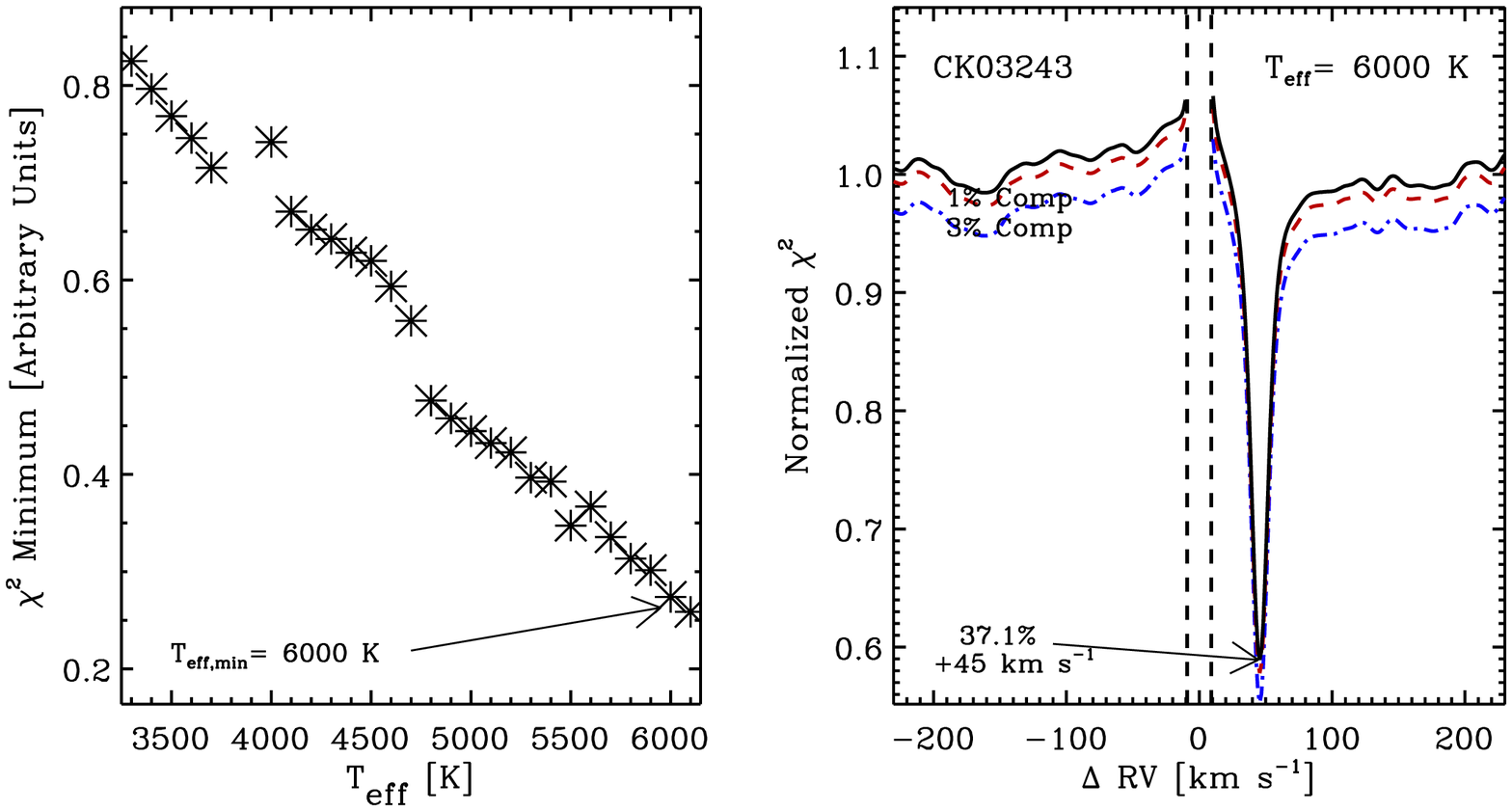}
	\caption{Final secondary star plot for KOI-3243. Same as Figure \ref{fig:complete_companion_plot}.}
	\label{fig:KOI-3243}
\end{figure}

\begin{figure}[h]
	\plotone{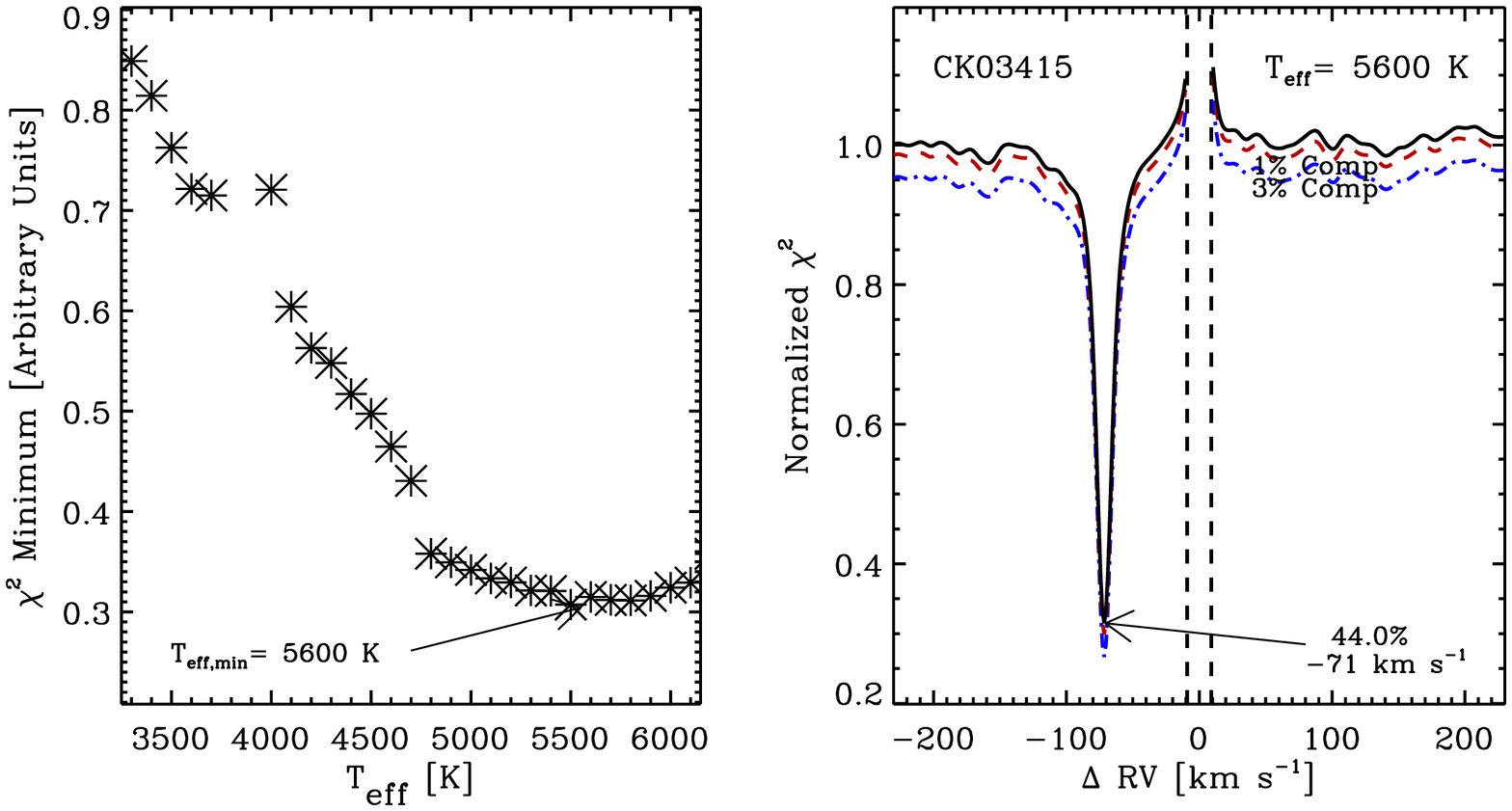}
	\caption{Final secondary star plot for KOI-3415. Same as Figure \ref{fig:complete_companion_plot}.}
	\label{fig:KOI-3415}
\end{figure}

\begin{figure}[h]
	\plotone{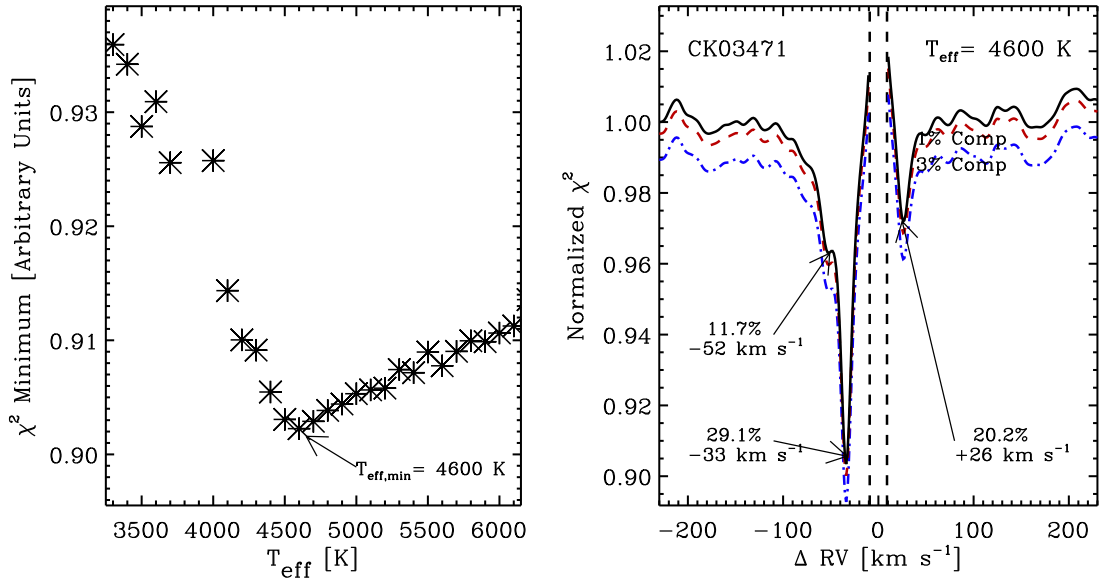}
	\caption{Final plot for KOI-3471. Same as Figure \ref{fig:complete_companion_plot}. At left, we show only the $\chi^2$ minima function for the brightest, secondary star. At left, we annotate all three detected additional stars to KOI-3471 with an arrow.}
	\label{fig:KOI-3471}
\end{figure}

\begin{figure}[h]
	\plotone{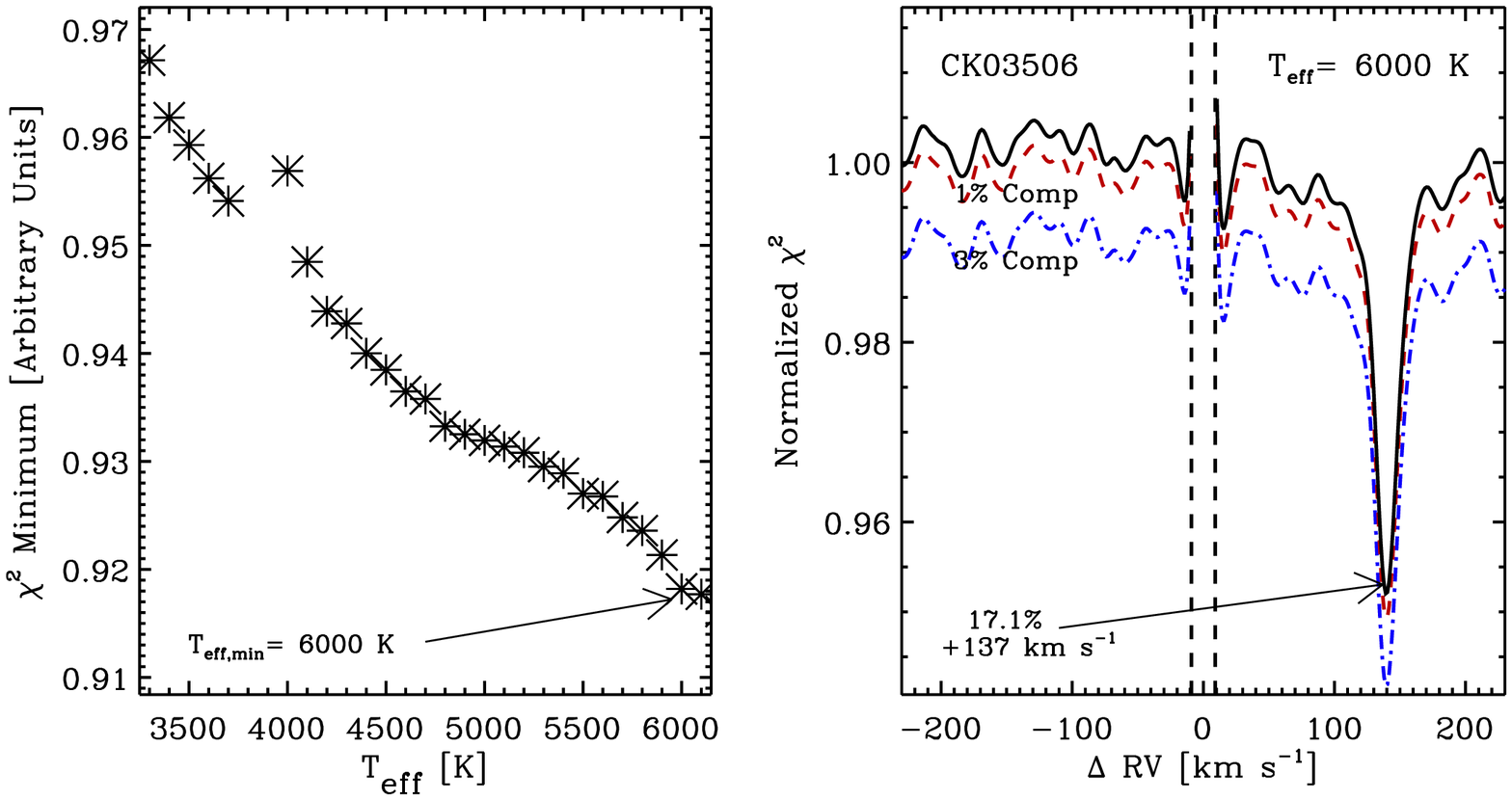}
	\caption{Final secondary star plot for KOI-3506. Same as Figure \ref{fig:complete_companion_plot}.}
	\label{fig:KOI-3506}
\end{figure}

\begin{figure}[h]
	\plotone{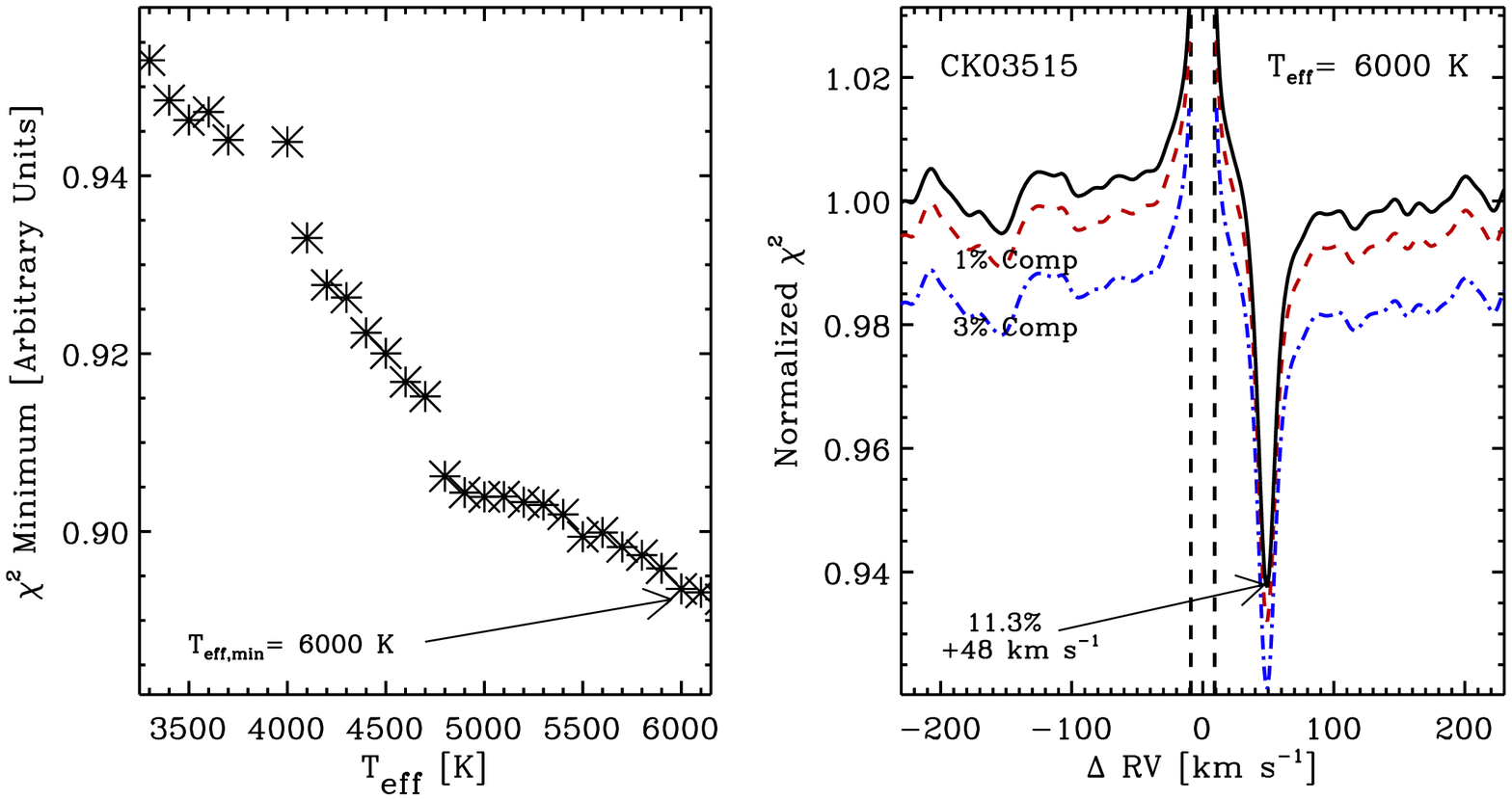}
	\caption{Final secondary star plot for KOI-3515. Same as Figure \ref{fig:complete_companion_plot}.}
	\label{fig:KOI-3515}
\end{figure}

\begin{figure}[h]
	\plotone{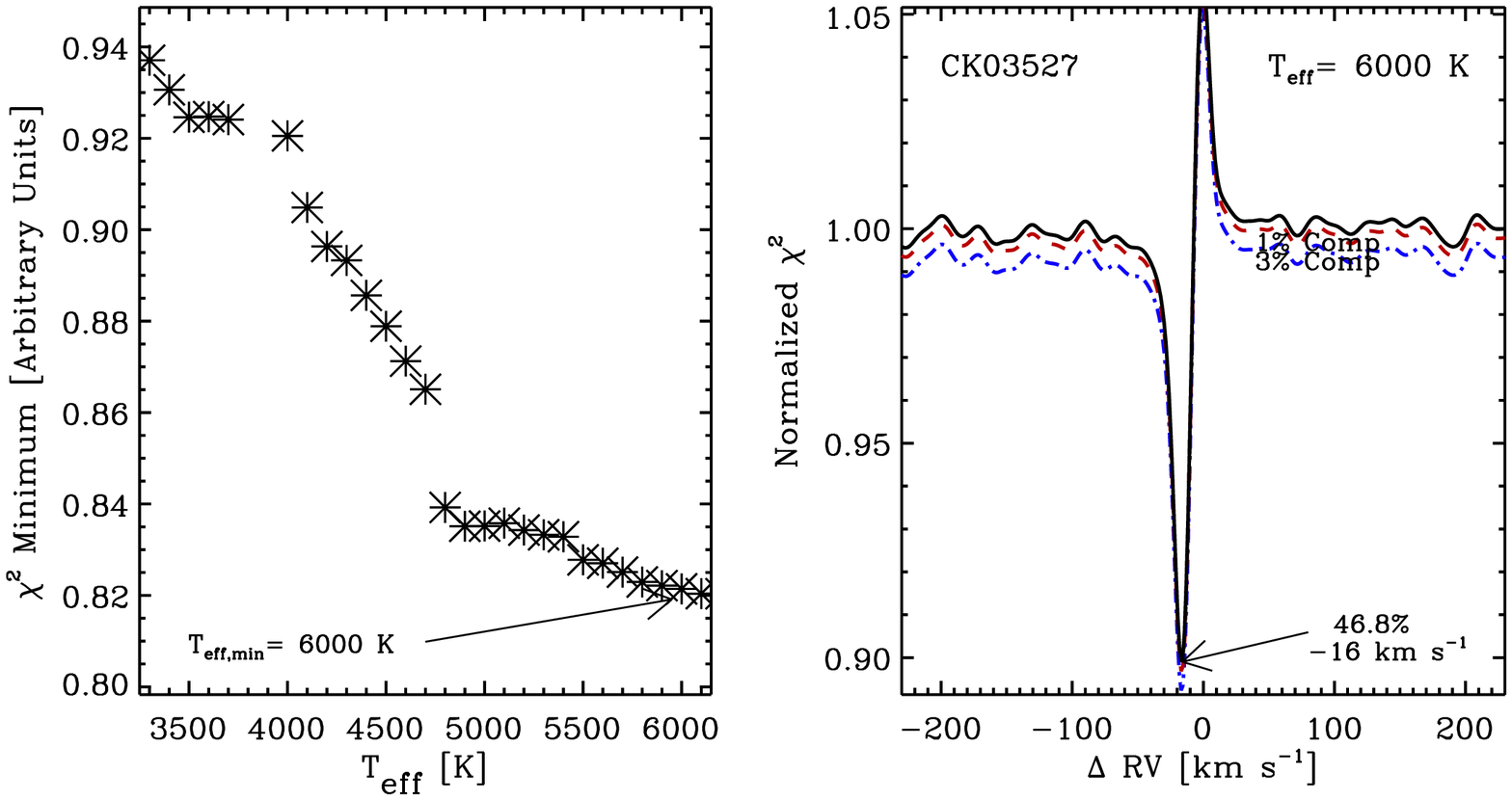}
	\caption{Final secondary star plot for KOI-3527. Same as Figure \ref{fig:complete_companion_plot}.}
	\label{fig:KOI-3527}
\end{figure}

\begin{figure}[h]
	\plotone{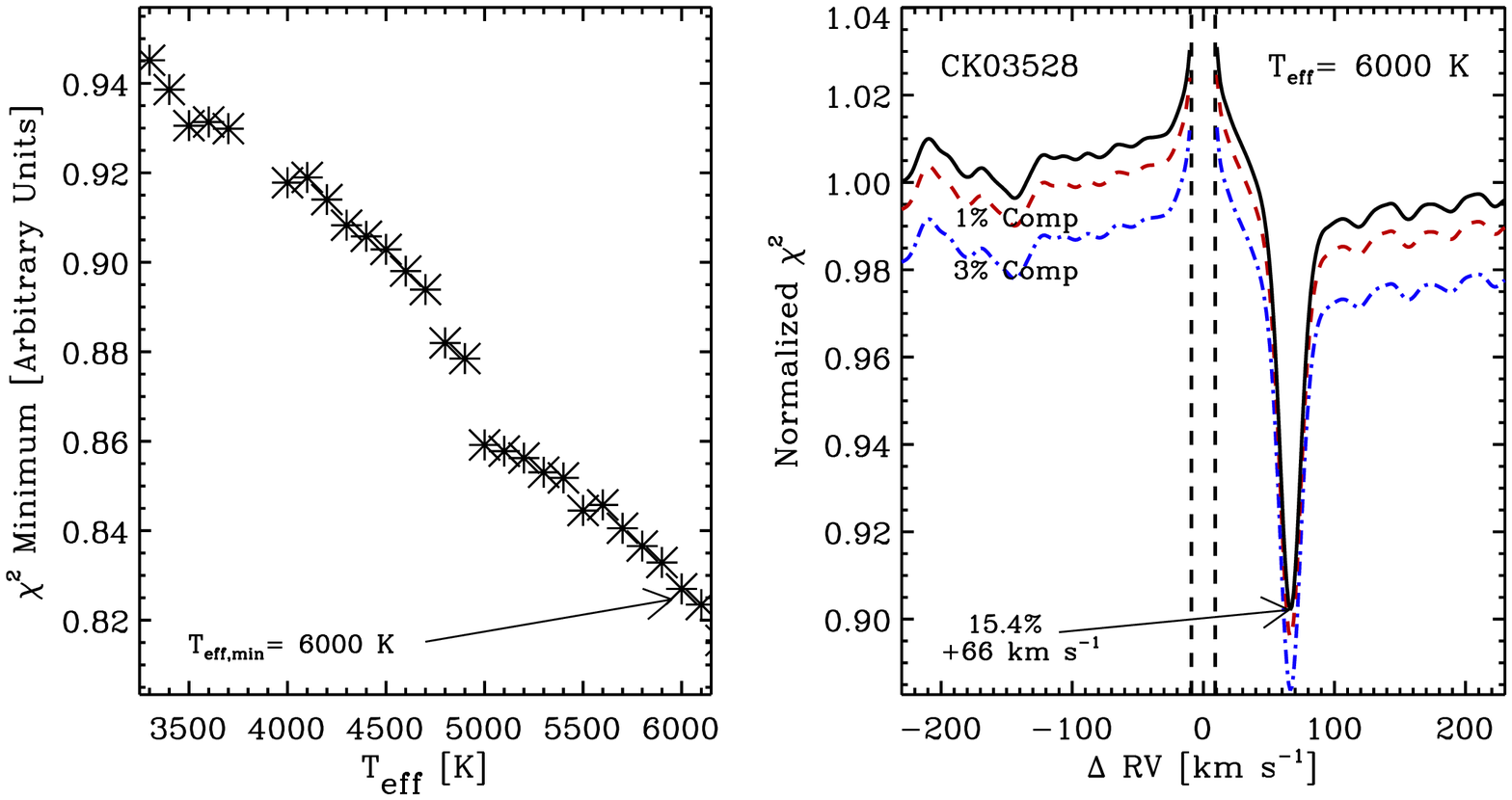}
	\caption{Final secondary star plot for KOI-3528. Same as Figure \ref{fig:complete_companion_plot}.}
	\label{fig:KOI-3528}
\end{figure}

\begin{figure}[h]
	\plotone{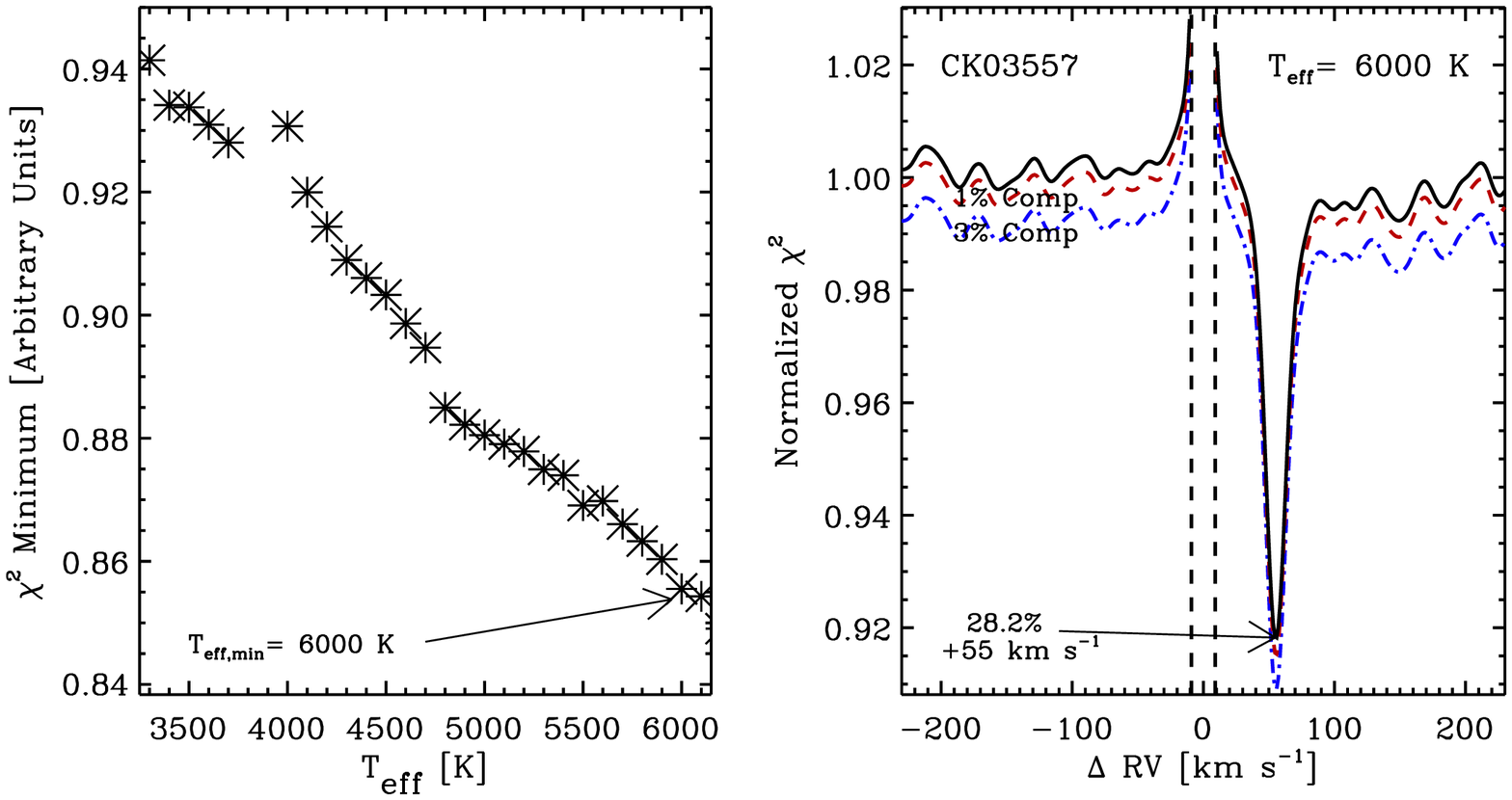}
	\caption{Final secondary star plot for KOI-3557. Same as Figure \ref{fig:complete_companion_plot}.}
	\label{fig:KOI-3557}
\end{figure}

\begin{figure}[h]
	\plotone{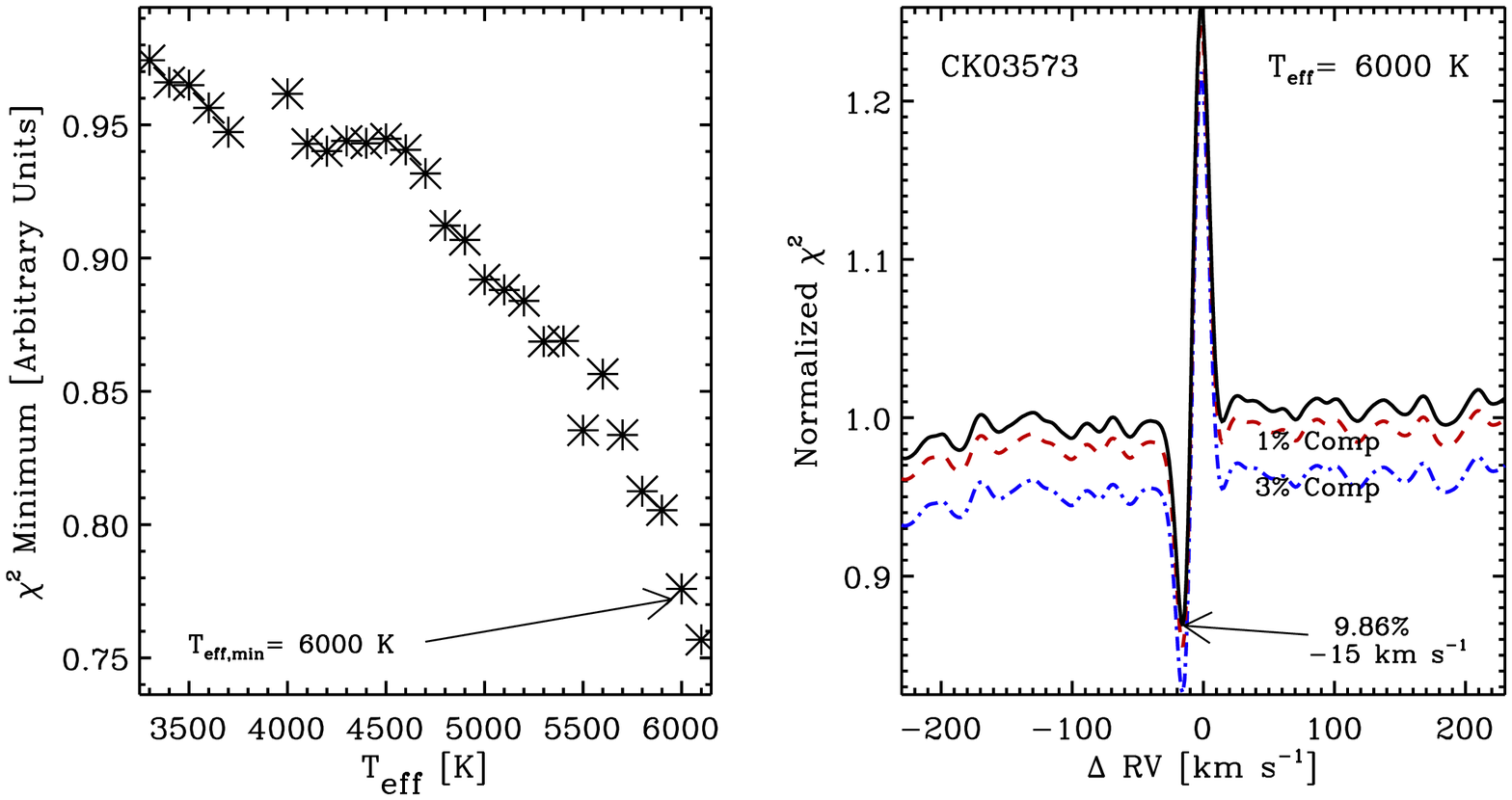}
	\caption{Final secondary star plot for KOI-3573. Same as Figure \ref{fig:complete_companion_plot}.}
	\label{fig:KOI-3573}
\end{figure}

\begin{figure}[h]
	\plotone{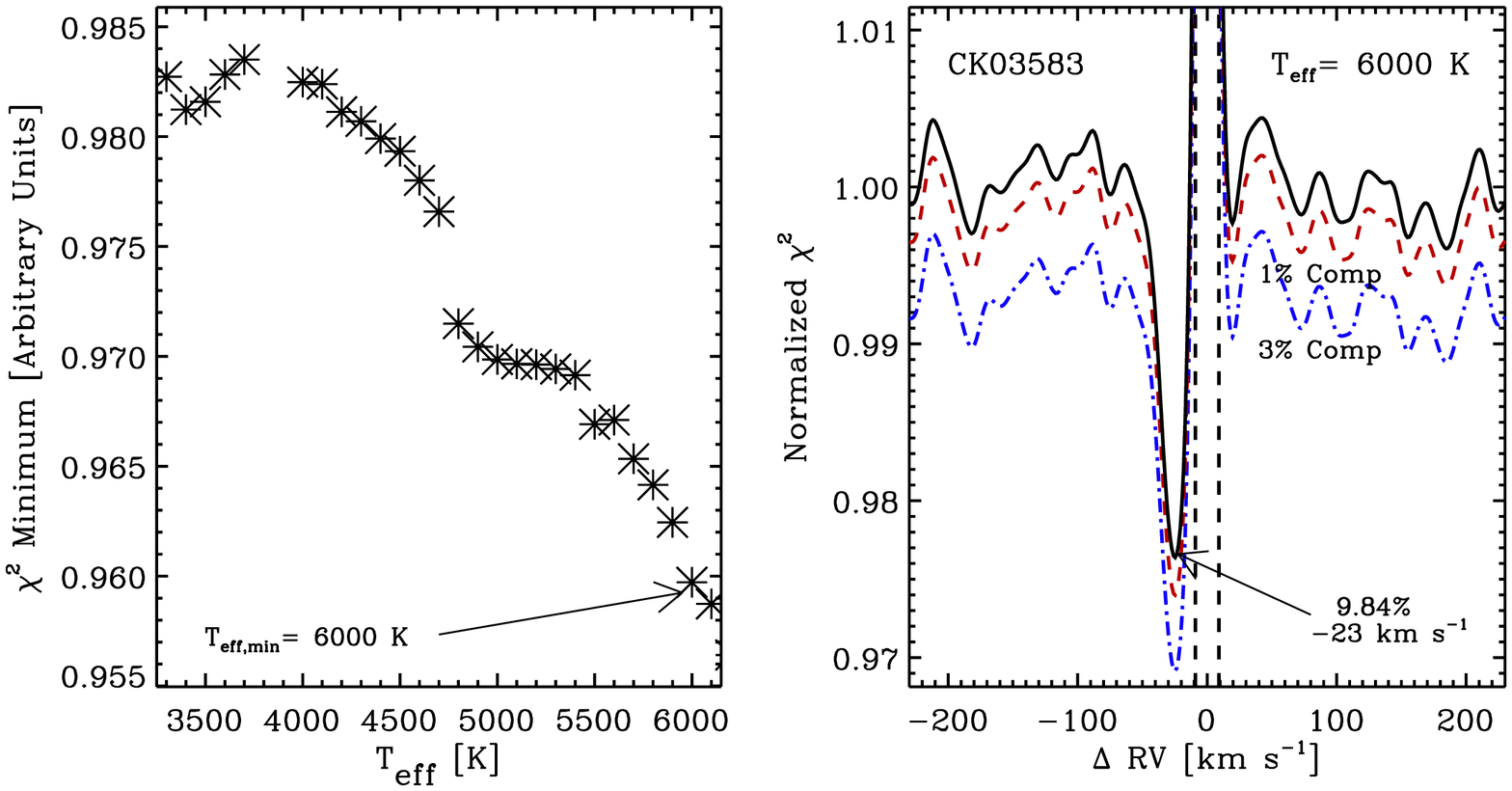}
	\caption{Final secondary star plot for KOI-3583. Same as Figure \ref{fig:complete_companion_plot}.}
	\label{fig:KOI-3583}
\end{figure}

\begin{figure}[h]
	\plotone{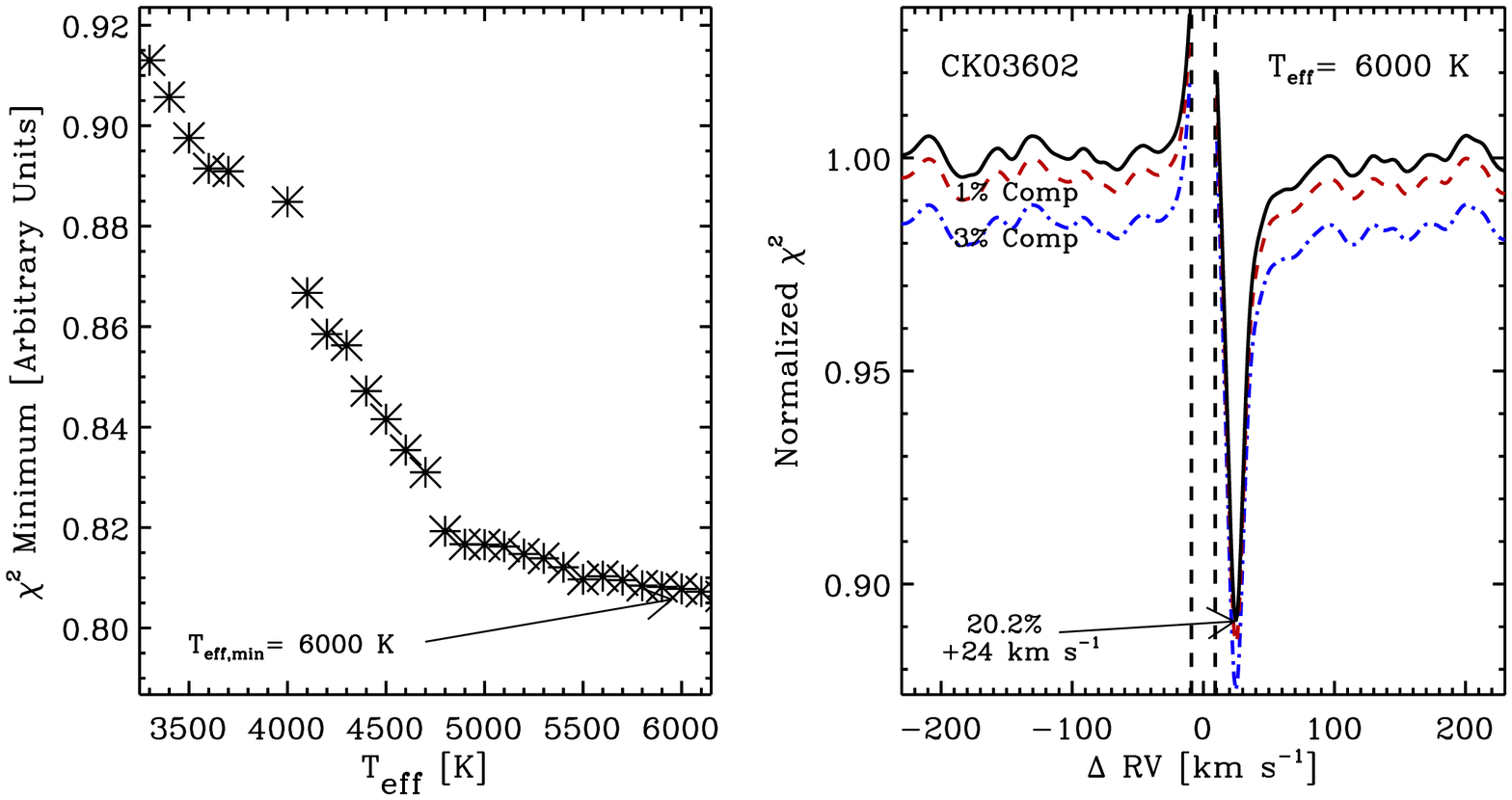}
	\caption{Final secondary star plot for KOI-3602. Same as Figure \ref{fig:complete_companion_plot}.}
	\label{fig:KOI-3602}
\end{figure}

\begin{figure}[h]
	\plotone{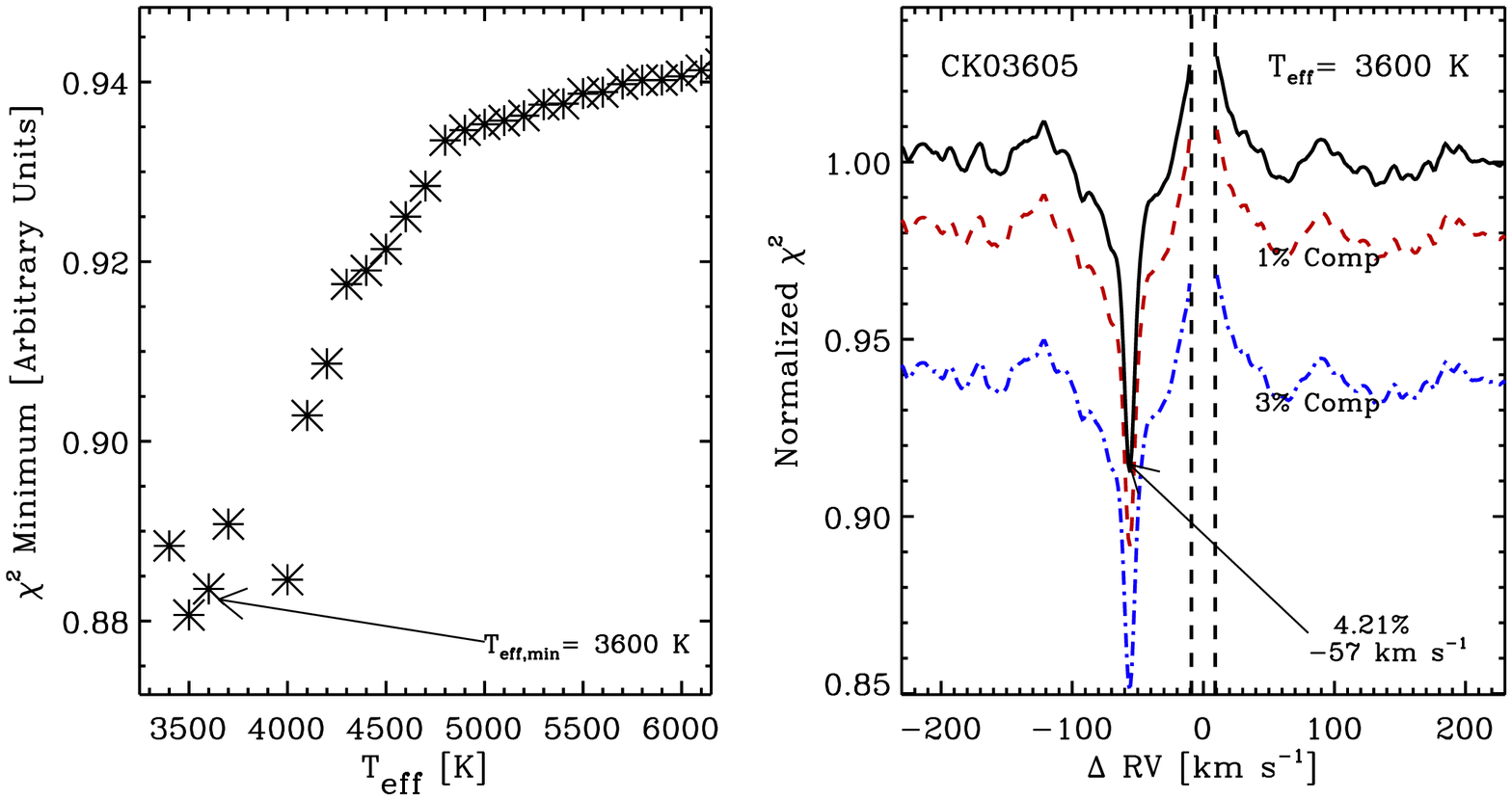}
	\caption{Final secondary star plot for KOI-3605. Same as Figure \ref{fig:complete_companion_plot}.}
	\label{fig:KOI-3605}
\end{figure}

\begin{figure}[h]
	\plotone{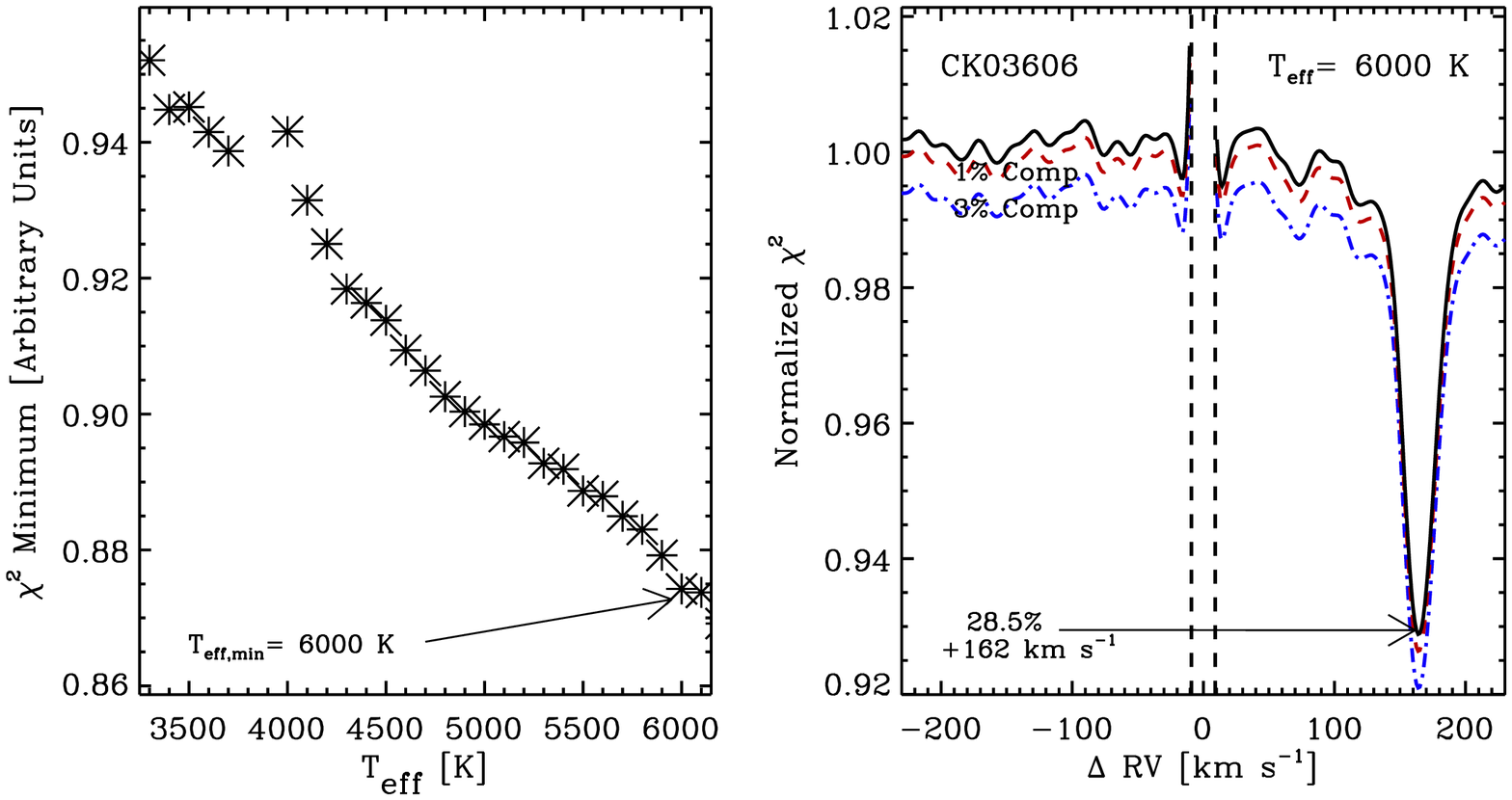}
	\caption{Final secondary star plot for KOI-3606. Same as Figure \ref{fig:complete_companion_plot}.}
	\label{fig:KOI-3606}
\end{figure}

\begin{figure}[h]
	\plotone{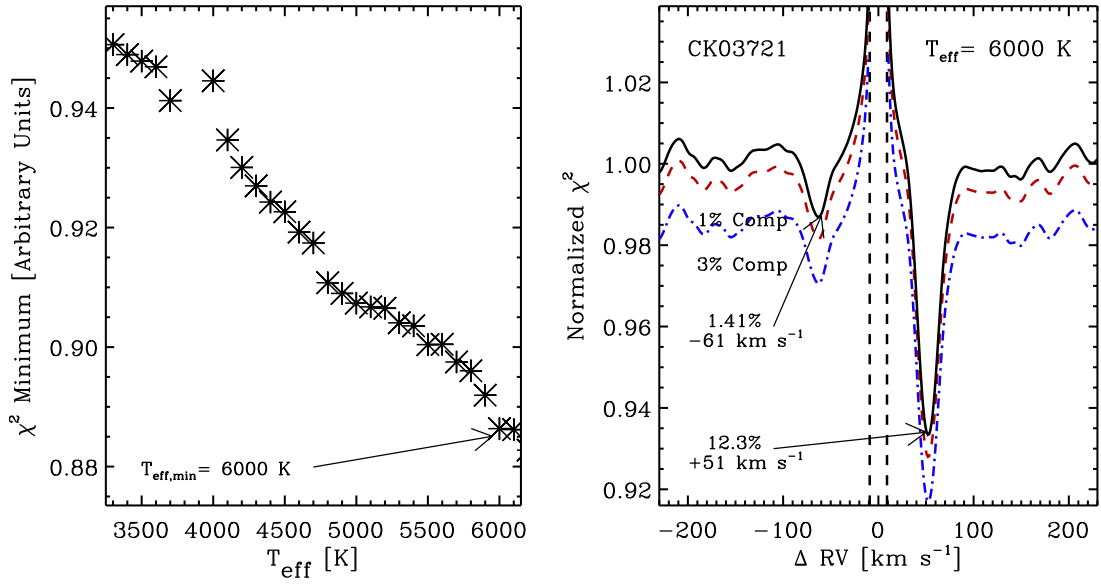}
	\caption{Final plot for KOI-3721. Same as Figure \ref{fig:complete_companion_plot}. At left, we only show the $\chi^2$ minima function for the brightest, secondary star. At right, we annotate both additional stars to KOI-3721 with an arrow.}
	\label{fig:KOI-3721}
\end{figure}

\begin{figure}[h]
	\plotone{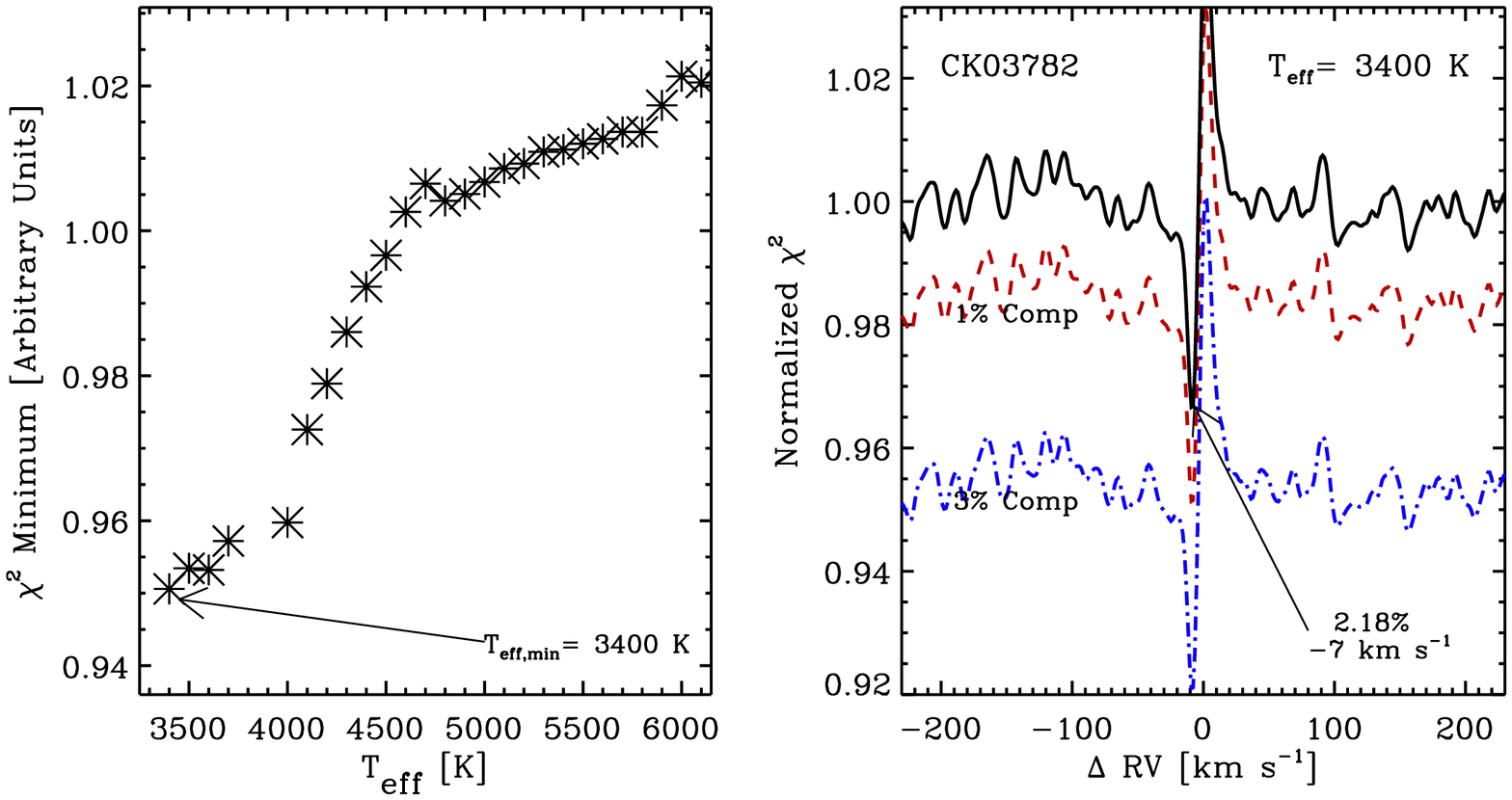}
	\caption{Final secondary star plot for KOI-3782. Same as Figure \ref{fig:complete_companion_plot}.}
	\label{fig:KOI-3782}
\end{figure}

\begin{figure}[h]
	\plotone{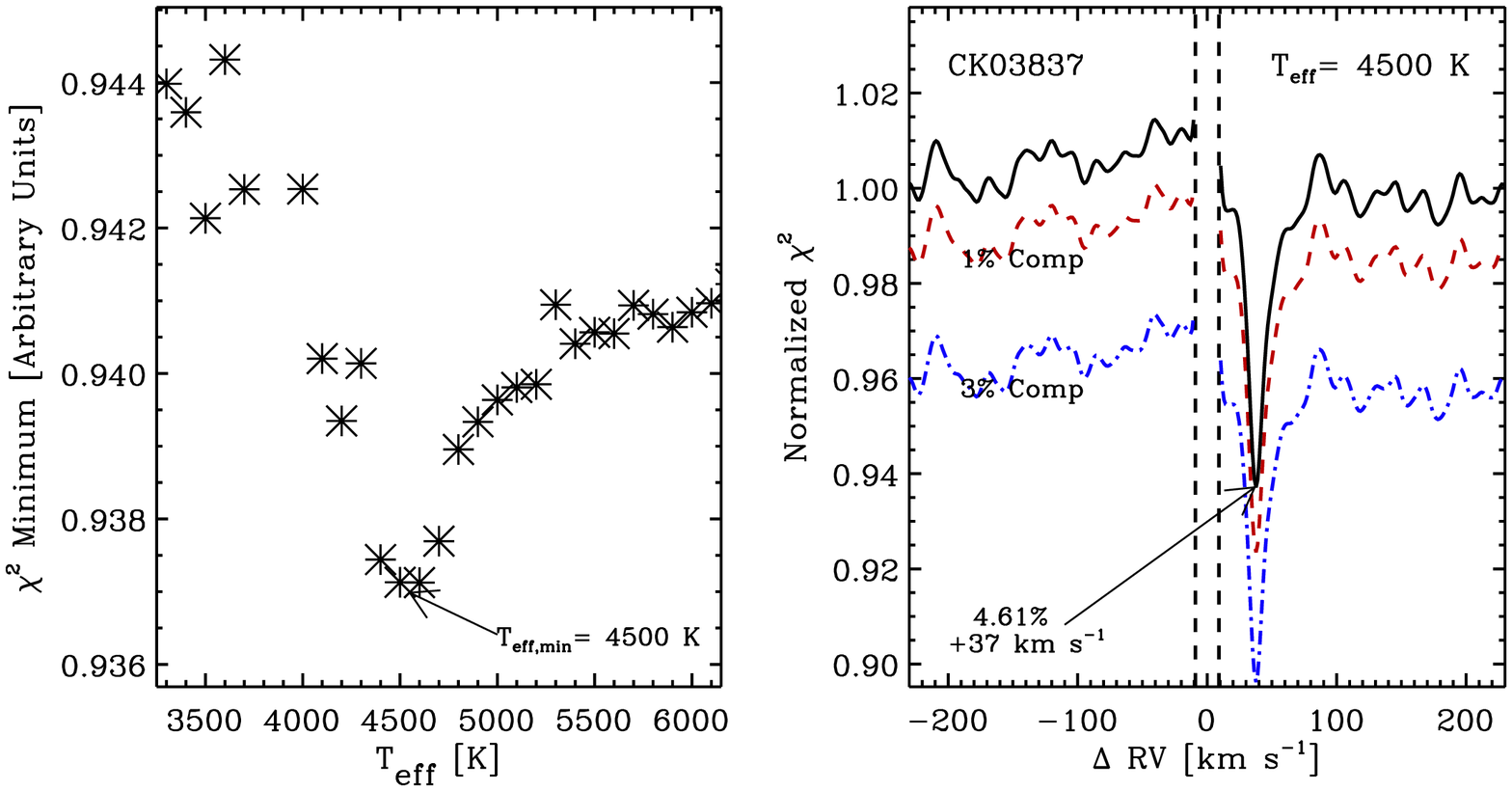}
	\caption{Final secondary star plot for KOI-3837. Same as Figure \ref{fig:complete_companion_plot}.}
	\label{fig:KOI-3837}
\end{figure}

\begin{figure}[h]
	\plotone{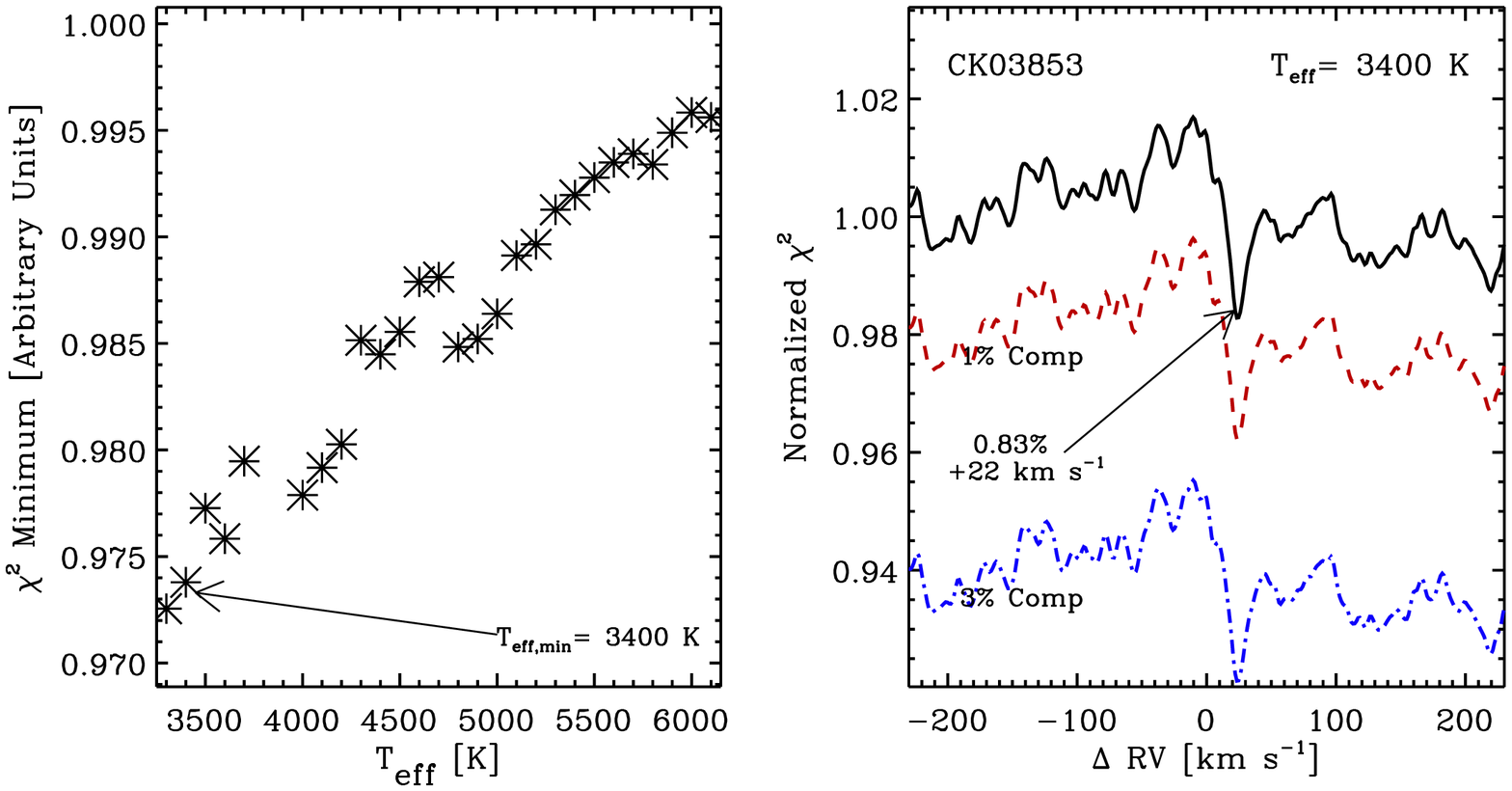}
	\caption{Final secondary star plot for KOI-3853. Same as Figure \ref{fig:complete_companion_plot}.}
	\label{fig:KOI-3853}
\end{figure}

\begin{figure}[h]
	\plotone{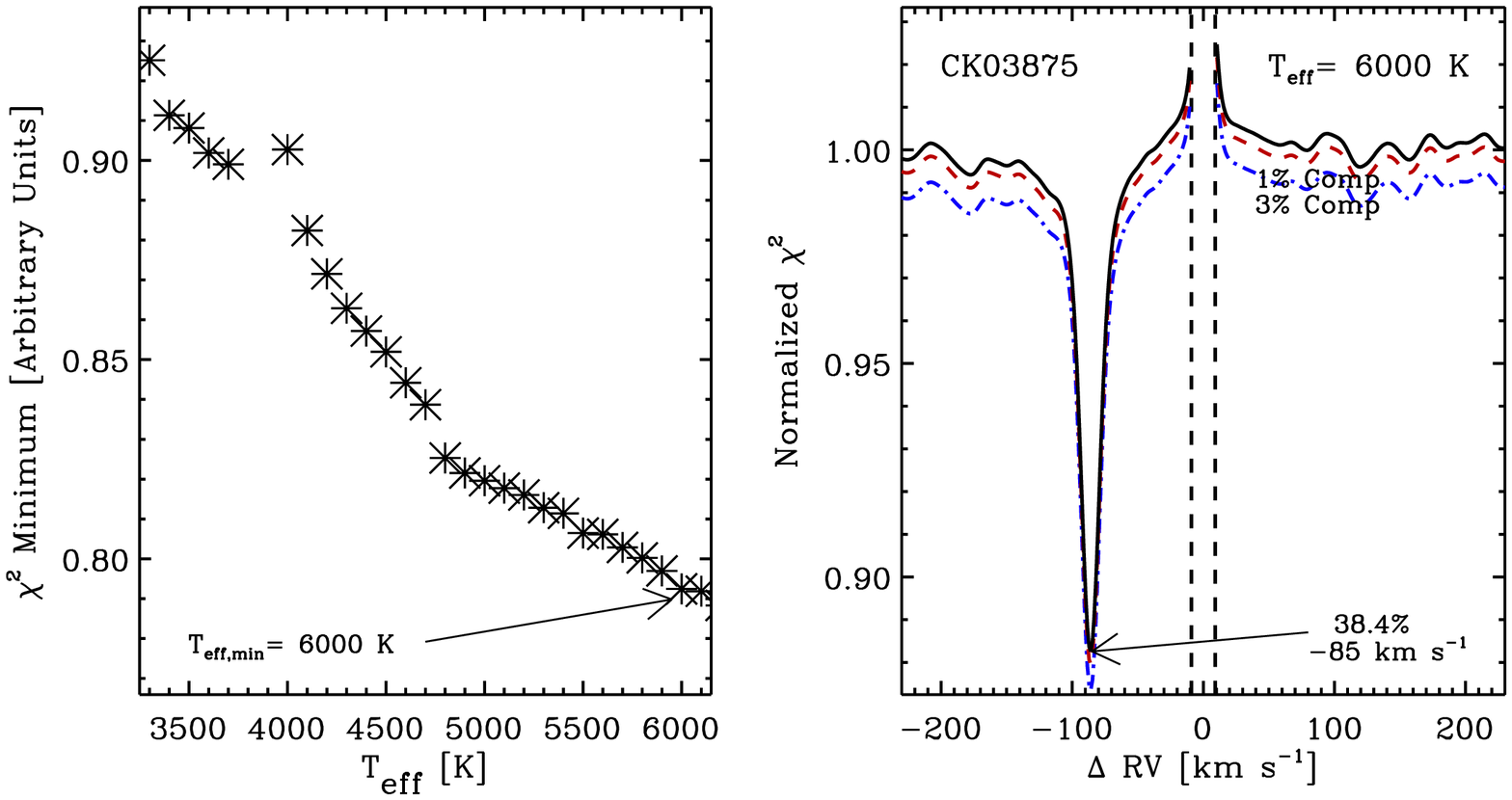}
	\caption{Final secondary star plot for KOI-3875. Same as Figure \ref{fig:complete_companion_plot}.}
	\label{fig:KOI-3875}
\end{figure}

\begin{figure}[h]
	\plotone{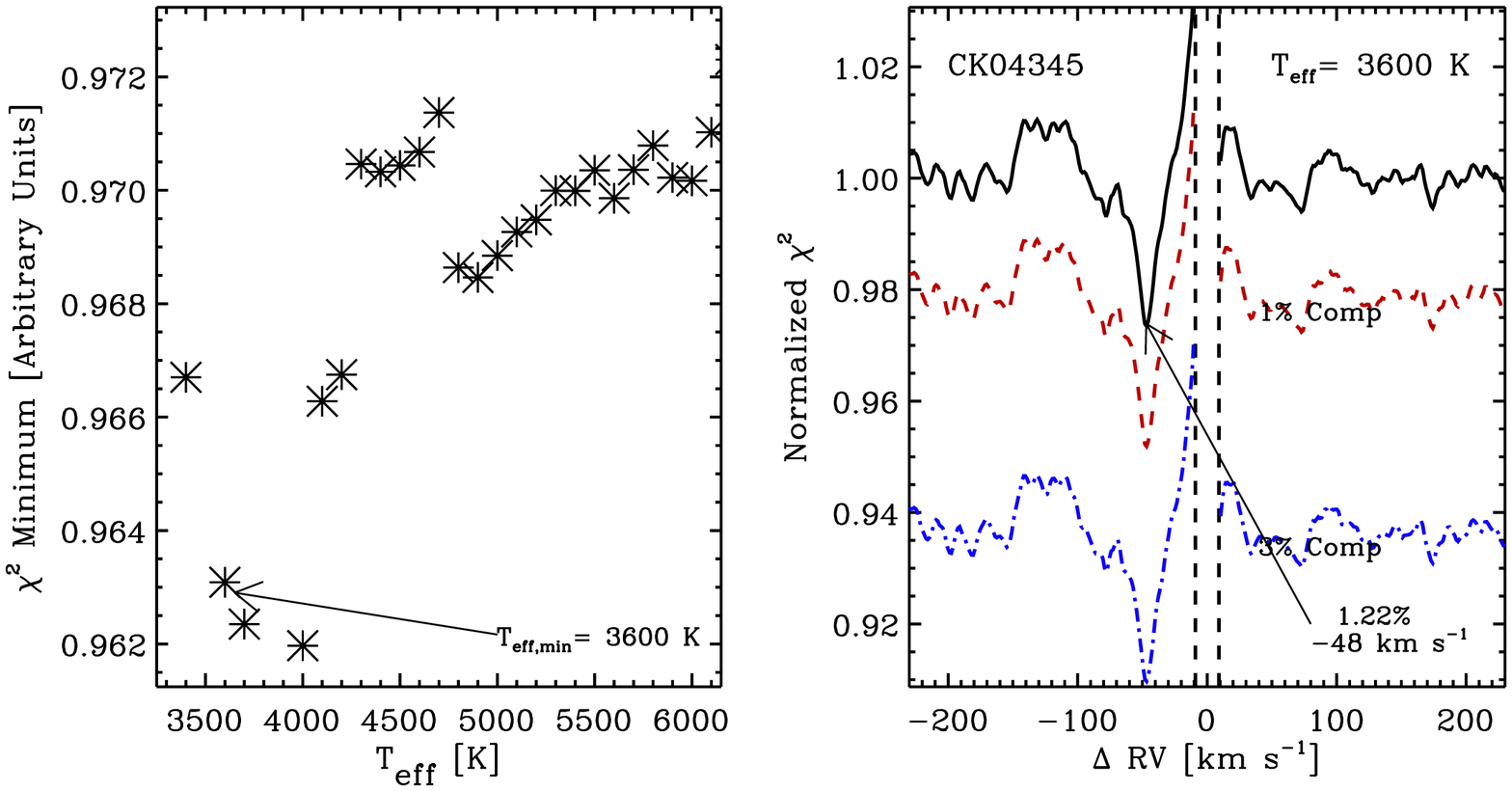}
	\caption{Final secondary star plot for KOI-4345. Same as Figure \ref{fig:complete_companion_plot}.}
	\label{fig:KOI-4345}
\end{figure}

\begin{figure}[h]
	\plotone{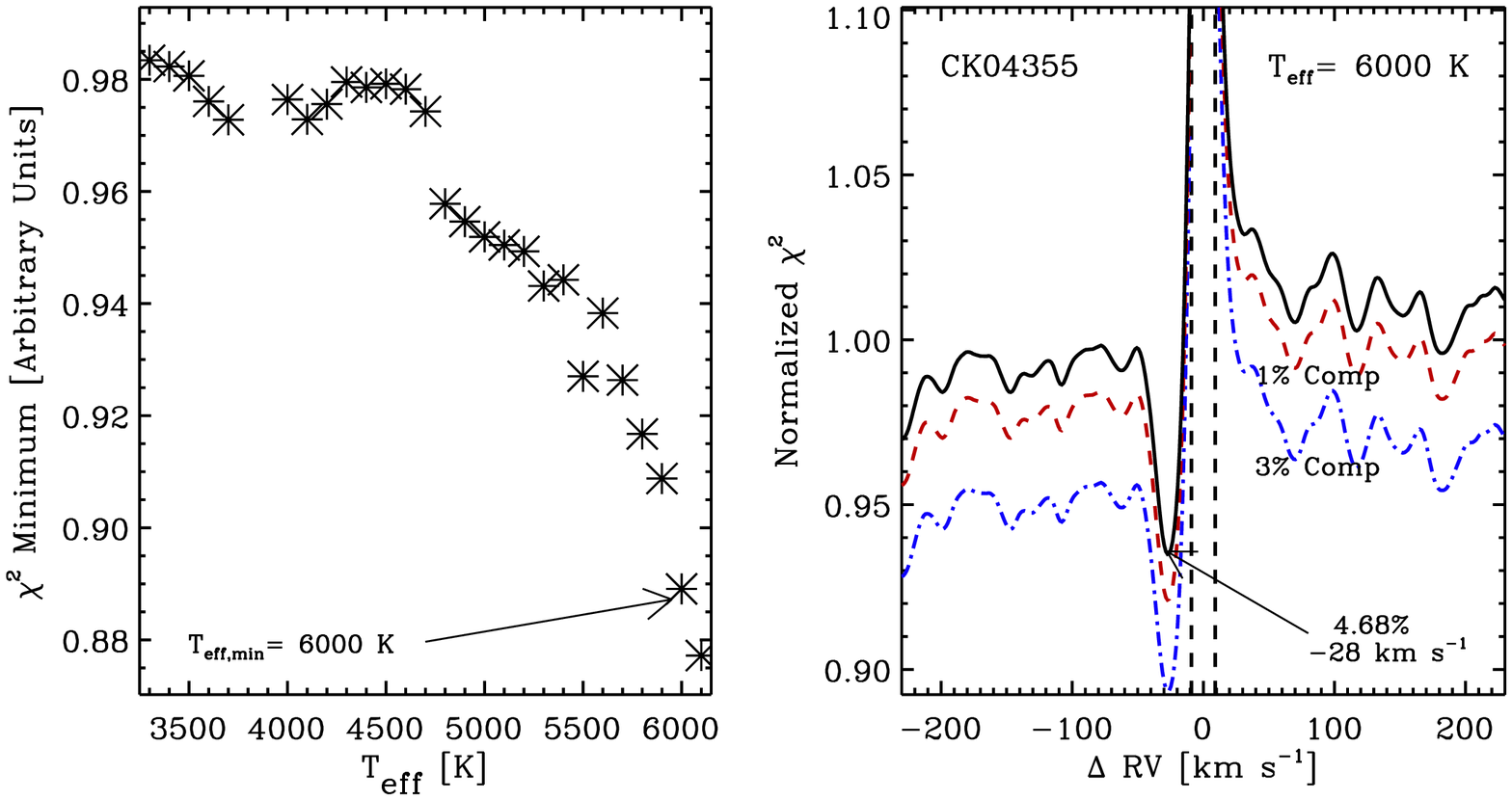}
	\caption{Final secondary star plot for KOI-4355. Same as Figure \ref{fig:complete_companion_plot}.}
	\label{fig:KOI-4355}
\end{figure}

\begin{figure}[h]
	\plotone{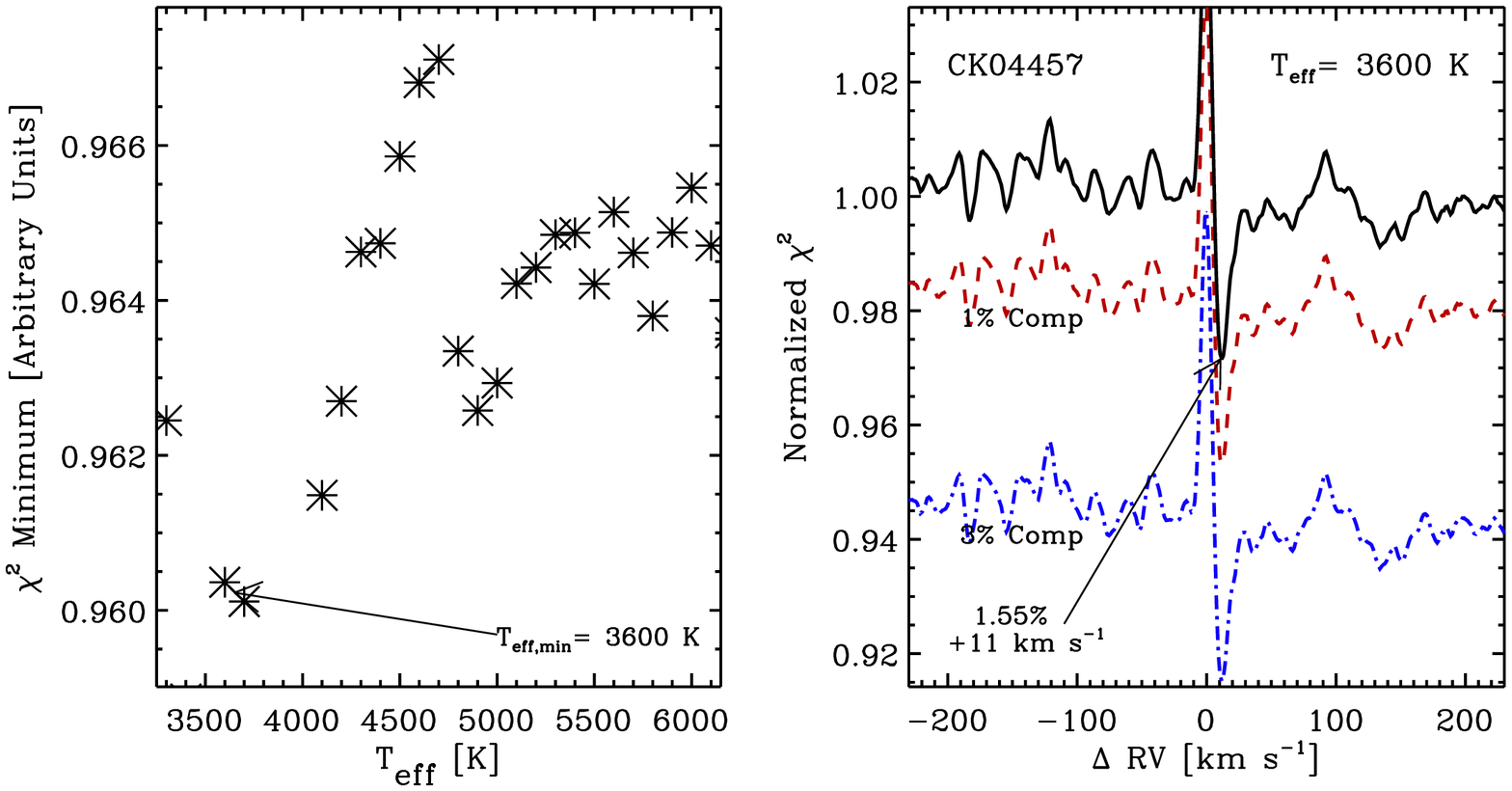}
	\caption{Final secondary star plot for KOI-4457. Same as Figure \ref{fig:complete_companion_plot}.}
	\label{fig:KOI-4457}
\end{figure}

\begin{figure}[h]
	\plotone{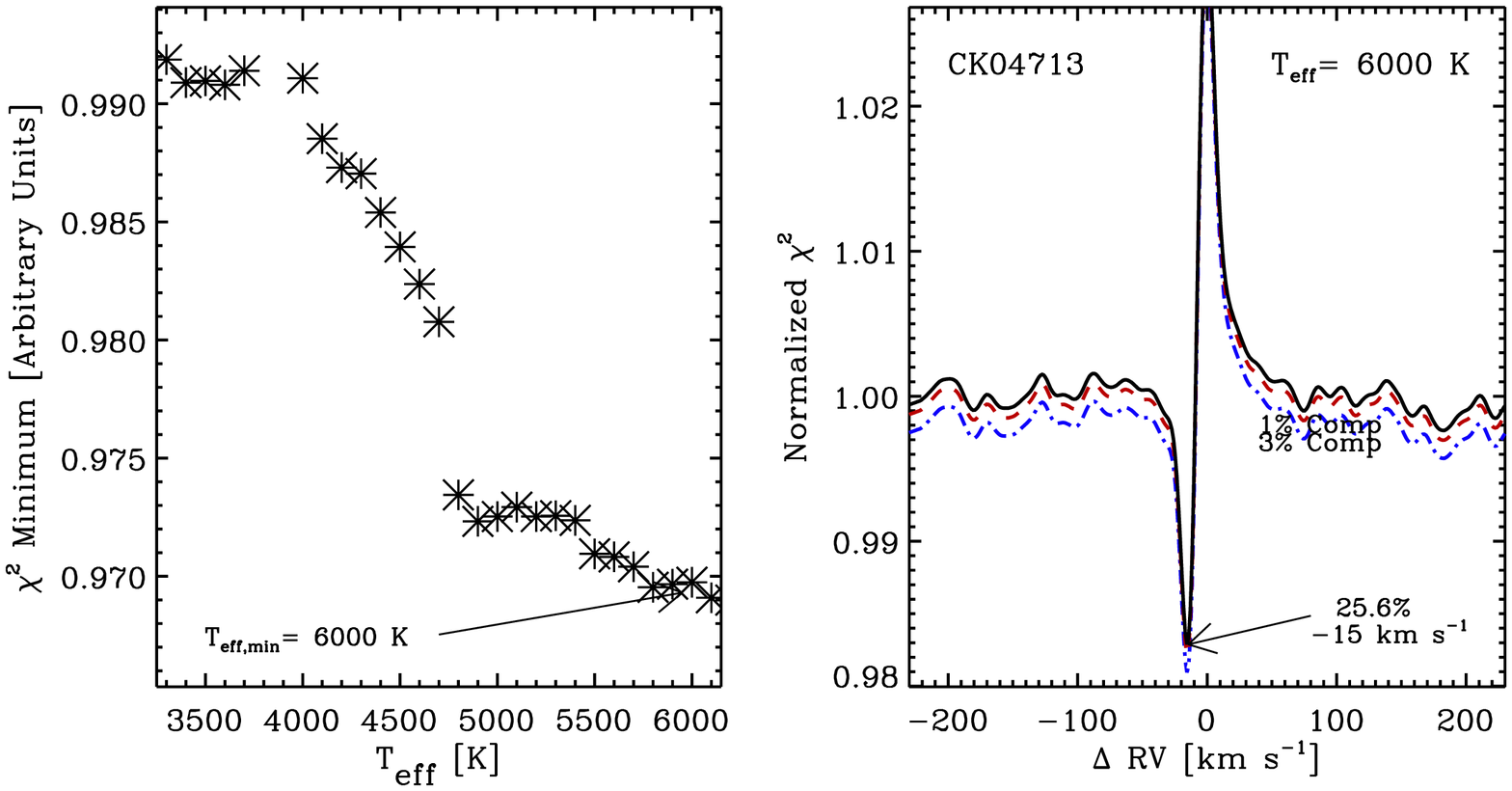}
	\caption{Final secondary star plot for KOI-4713. Same as Figure \ref{fig:complete_companion_plot}.}
	\label{fig:KOI-4713}
\end{figure}

\begin{figure}[h]
	\plotone{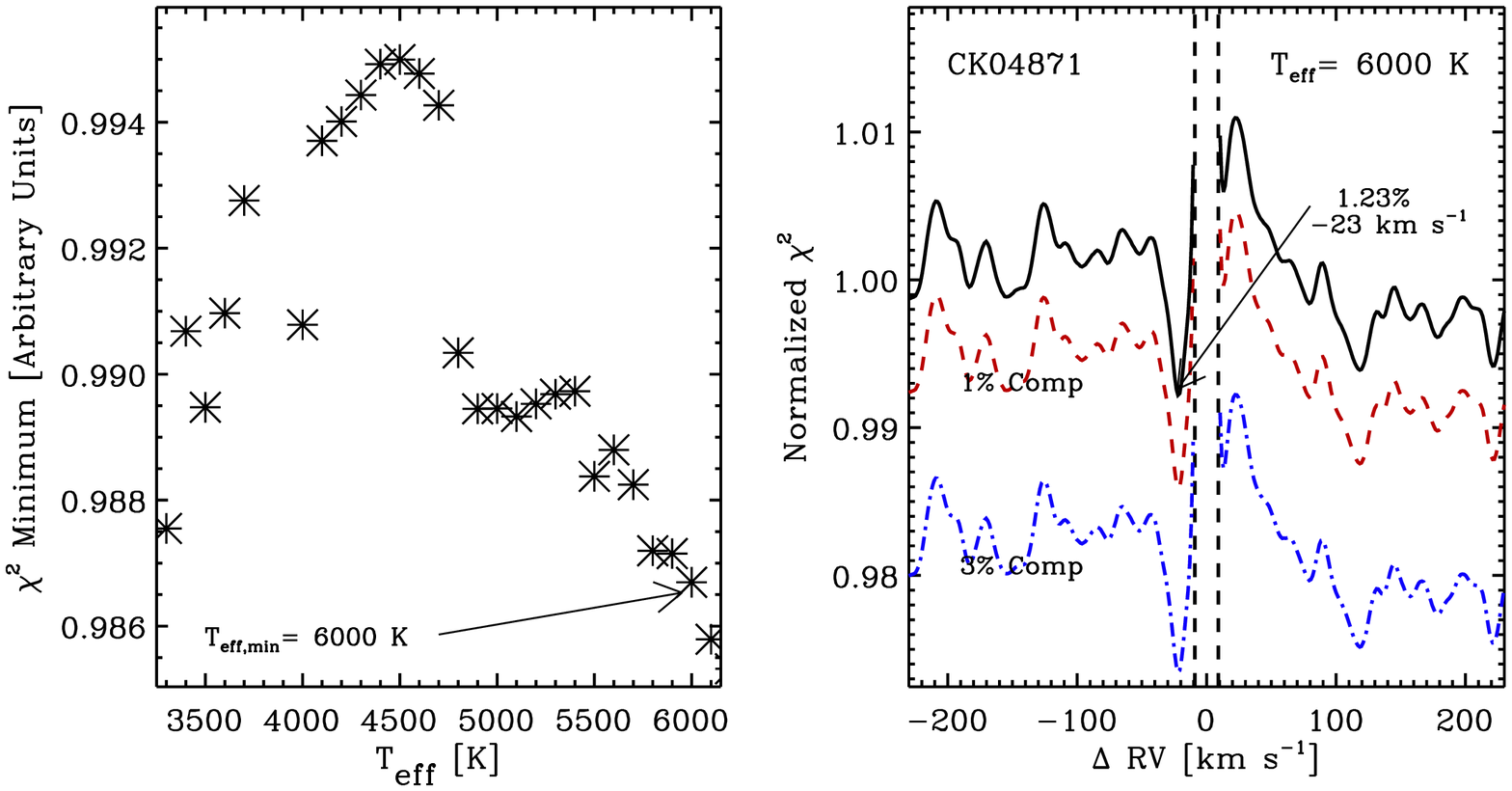}
	\caption{Final secondary star plot for KOI-4871. Same as Figure \ref{fig:complete_companion_plot}.}
	\label{fig:KOI-4871}
\end{figure}

\bibliographystyle{apj}
\bibliography{ms} 

\enddocument